\documentclass[a4paper,11pt]{article}
\pdfoutput=1
\usepackage{jheppub}
\usepackage{multirow}
\usepackage{graphicx}
\usepackage{amssymb, amsmath}
\usepackage{amsfonts}
\usepackage[utf8]{inputenc}
\usepackage{bbold}
\usepackage{bm}
\usepackage{xcolor}
\usepackage{soul}
\usepackage{float}
\usepackage{slashed}
\usepackage{braket}
\usepackage{subcaption}

\newcommand{\ii}{\mathrm{i}}
\newcommand{\rme}{\mathrm{e}}

\newcommand{\vev}[1]{{\left\langle #1 \right\rangle}}
\newcommand{\comm}[2]{\left[#1,#2\right]}

\DeclareMathOperator{\Tr}{Tr}
\DeclareMathOperator{\tr}{tr}

\makeatletter
\newcommand*{\letterdef@}{}
\newcommand*{\letterdef}[3]{%
	\def\letterdef@##1{\expandafter\newcommand\csname #1\endcsname{#2{##1}}}%
	\@tfor\@tempa :=#3\do{\expandafter\letterdef@\expandafter{\@tempa}}}
\makeatother
\letterdef{c#1} {\mathcal}{ABCDEFGHIJKLMNOPQRSTUVWXYZ} 
\letterdef{rm#1}{\mathrm} {dDmM} 

\newdimen\tableauside\tableauside=1.0ex
\newdimen\tableaurule\tableaurule=0.4pt
\newdimen\tableaustep
\def\phantomhrule#1{\hbox{\vbox to0pt{\hrule height\tableaurule
			width#1\vss}}}
\def\phantomvrule#1{\vbox{\hbox to0pt{\vrule width\tableaurule
			height#1\hss}}}
\def\sqr{\vbox{%
		\phantomhrule\tableaustep
		\hbox{\phantomvrule\tableaustep\kern\tableaustep\phantomvrule\tableaustep}%
		\hbox{\vbox{\phantomhrule\tableauside}\kern-\tableaurule}}}
\def\squares#1{\hbox{\count0=#1\noindent\loop\sqr
		\advance\count0 by-1 \ifnum\count0>0\repeat}}
\def\tableau#1{\vcenter{\offinterlineskip
		\tableaustep=\tableauside\advance\tableaustep by-\tableaurule
		\kern\normallineskip\hbox
		{\kern\normallineskip\vbox
			{\gettableau#1 0 }%
			\kern\normallineskip\kern\tableaurule}%
		\kern\normallineskip\kern\tableaurule}}
\def\gettableau#1 {\ifnum#1=0\let\next=\null\else
	\squares{#1}\let\next=\gettableau\fi\next}
\newcommand{\parenth}[1]{\left( #1 \right)}

\renewcommand{\a}{\alpha}
\renewcommand{\b}{\beta}

\title{\boldmath Chiral correlators in
	$\cN=2$ superconformal quivers}

\author[a]{Francesco Galvagno,}
\affiliation[a]{Institut f\"ur Theoretische Physik, ETH Z\"urich
\\
	Wolfgang-Pauli-Strasse 27, 8093 Z\"urich, Switzerland
\\}
\emailAdd{fgalvagno@phys.ethz.ch}

\author[b]{Michelangelo Preti}
\affiliation[b]{Nordita, KTH Royal Institute of Technology and Stockholm University \\ Roslagstullsbacken 23, SE-106 91 Stockholm, Sweden
\\}
\emailAdd{michelangelo.preti@gmail.com}

\abstract{We consider a family of $\cN=2$ superconformal field theories in four dimensions, defined as $\mathbb{Z}_q$ orbifolds of $\cN=4$ Super Yang-Mills theory. We compute the chiral/anti-chiral correlation functions at a perturbative level, using both the matrix model approach arising from supersymmetric localisation on the four-sphere and explicit field theory calculations on the flat space using the $\cN=1$ superspace formalism. We implement a highly efficient algorithm to produce a large number of results for finite values of $N$, exploiting the symmetries of the quiver to reduce the complexity of the mixing between the operators. Finally the interplay with the field theory calculations allows to isolate special observables which deviate from $\cN=4$ only at high orders in perturbation theory. }

\keywords{SUSY localisation, $\mathcal{N}=2$ SYM theories, multi-matrix model, chiral operator}
\preprint{NORDITA 2020-080}

\begin{document}
	\maketitle

\section{Introduction and summary of results}
\label{sec:intro}
The maximally supersymmetric theory in four dimensions, namely $\cN=4$ SYM has played an still plays a central role in our understanding of gauge theories. Furthermore, it provides one of the most successful realisation of the AdS/CFT correspondence, as well as the most favourable playground for obtaining exact results. A perfect example is  the resummation of the infinite series of perturbative corrections of the expectation value of the circular Wilson loops by means of a matrix-model \cite{Erickson:2000af,Berenstein:1998ij,Drukker:2000rr,Semenoff:2001xp}. Indeed, obtaining the exact result for the circular Wilson loop, it was also possible to match its strong coupling behaviour predicted by string theory in the AdS/CFT framework. This matrix-model originates from the supersymmetric localisation of the partition function on the sphere $S^4$ \cite{Pestun:2007rz} and it provides a recipe to compute not only supersymmetric Wilson loops but a large class of observables involving also local operators (see for instance \cite{Drukker:2007qr,Pestun:2009nn,Giombi:2009ds,Giombi:2009ek,Giombi:2012ep,Bonini:2014vta,Bonini:2015fng}).  
 
However, when we drop the amount of supersymmetries to $\cN=2$, some of the $\cN=4$ features get lost, but it is still possible to produce exact results. The most famous $\cN=2$ example is  superconformal QCD (SCQCD) with a gauge group $SU(N)$ coupled to $2N$ hypermultiplets. The power of supersymmetry is enough to localise the partition function on the four sphere, reducing the computation of the path integral to a matrix model \cite{Pestun:2007rz}. Many observables can be computed in this case, especially at the level of perturbation theory: the Wilson loop vacuum expectation values \cite{Erickson:2000af,Drukker:2000rr,Passerini:2011fe,Bourgine:2011ie}, chiral/antichiral correlators \cite{Baggio:2014sna,Gerchkovitz:2016gxx,Baggio:2016skg,Rodriguez-Gomez:2016cem,Rodriguez-Gomez:2016ijh,Billo:2017glv,Billo:2019fbi} and correlators of chiral operators and Wilson loops \cite{Semenoff:2001xp,Billo:2018oog}, also in special regimes such as the large charge \cite{Bourget:2018obm,Beccaria:2018xxl,Beccaria:2018owt,Beccaria:2020azj}. However, going beyond the weak coupling limit appears as a hard task, as a signal of the fact that SCQCD has no longer a simple string theory dual.
A special class of $\cN=2$ theories has been developed in \cite{Fiol:2015mrp,Bourget:2018fhe,Billo:2019job,Beccaria:2020hgy} by keeping a single $SU(N)$ gauge group and playing with the matter content of the theory. These special theories admit a dual holographic description as orientifold projections of the AdS$_5\times S^5$ geometry, and they enjoy many cancellation properties at the level of perturbation theory as well as the possibility of a resummation of the perturbative series \cite{Beccaria:2020hgy}.
 
In the present work we concentrate on another special family of $\cN=2$ superconformal theories (SCFTs) denoted as $A_{q-1}$, the so called necklace quiver theories with $q$ nodes. Each node represents a $SU(N)$ gauge group and they are connected by bifundamental hypermultiplets. There are many reasons for studying these theories. 
First, they possess a holographic dual as a type IIB string theory on a AdS$_5\times (S^5/\mathbb{Z}_q)$ \cite{Kachru:1998ys,Gukov:1998kk,Lee:1998bxa} and can be used for testing AdS/CFT correspondence.
Moreover, they are considered as the ideal framework where to introduce an integrability approach also in $\cN=2$ contexts \cite{Gadde:2009dj,Gadde:2010zi,Pomoni:2011jj,Pomoni:2013poa,Pomoni:2019oib}. 
Furthermore, necklace quivers can even play important roles in other contexts, such as in the Higgs/Coulomb-branch matching of type-B conformal anomalies and its application on the deconstruction of the 6D $(2,0)$ theory on a torus \cite{Niarchos:2019onf,Niarchos:2020nxk}. In general they can be considered as the next-simple theories after $\cN=4$, and they can represent a bridge between the maximally supersymmetric gauge theory and less constrained $\cN=2$ theories. 
Several results have been obtained lately using the localisation approach for $A_{q-1}$ theories, in particular for Wilson loops and related observables \cite{Mitev:2014yba,Mitev:2015oty,Fiol:2020ojn,Zarembo:2020tpf,Ouyang:2020hwd,GalvPreti2021} and for chiral/antichiral correlators \cite{Pini:2017ouj}, mainly focusing on the $q=2$ case.

In this paper we study the two-point function of a chiral and an anti-chiral multi-trace operator with conformal dimension defined by $n$ inside $A_{q-1}$ SCFTs. In particular we extend the work of \cite{Pini:2017ouj} in many directions.
We exploit the $\cN=2$ localisation technique on the four sphere and we derive a consistent multi-matrix model for the generic $q$ case. Placing the chiral/antichiral operators on the sphere is a non-trivial task due to the conformal anomaly which arises moving from $\mathbb{R}^4$ to $S^4$, which generates a mixing among the operators. Such mixing on the sphere can be disentangled using a normal ordering procedure, which gets more and more involved when increasing the number of nodes $q$ and the conformal dimension $n$. Hence, we implement an algorithm with a high level of efficiency  that keeps track also of the symmetries of the quiver. In this way we are able to generate a big \emph{dataset of results} at a perturbative level for the two point coefficient. This collection of data includes finite and large $N$ perturbative expansions for several multi- and single-trace operators with dimensions between 2 and 8 belonging to different nodes of a quiver with number of nodes between 1 and 6.
The full set of results is available in the attached Mathematica notebook "\texttt{QuiverCorrelators.nb}", while in the main text we are going to highlight specific features about them.

We distinguish many classes of observables based on the nodes of the quiver in which the operators belong (inside the same node of the quiver, or in different nodes with unitary or increasing distance). Besides, we treat the typical observables of orbifold theories, \textit{i.e.} the twisted and untwisted sectors with respect to the orbifold action. All the results are obtained at finite $N$ and for various values of the number of nodes and the conformal dimension. For specific sets of observables we are able to extrapolate general formulas with explicit \emph{parametric dependence on $q$ and $n$}. These results could be a good starting points for exploring new limits of $A_{q-1}$, as already done for other classes of $\cN=2$ theories.

Moreover we consistently write the Lagrangian of $A_{q-1}$ theories following the $\cN=1$ superspace formalism on the 4-dimensional flat space and we compute the two-point function in perturbation theory using the standard Feynman diagrams approach, in order to make explicit checks with the matrix model results. In particular, we produce a full diagrammatic computation for the two-point 
function of operators inserted inside the same node up to a two-loops order and for any values of $n$ and $q$. Moreover, we isolate  some special observables, where the operators are inserted in different nodes of the quiver and the leading perturbative orders arise at very high powers of the couplings. The diagrammatic explanation of this fact is very peculiar, since it \emph{reduces to a combination of simple building blocks}, which nicely capture the leading orders for any values of $n$ and $q$. This pattern is a clear consequence of the high degree of symmetry and also arises in more sophisticated integrability contexts. This is another aspect which is worth to be investigated in the future. 

The manuscript is organised as follows. In section 2 we introduce the gauge theories $A_{q-1}$ defining also the correlators of chiral/antichiral operators. In section 3 we extend the multi-matrix model picture for the necklace $\mathcal{N}=2$ theories focusing on the matching of correlators in the matrix model and in the gauge theory. Moreover we build the full set of normal-ordered orthogonal operators and we discuss about the symmetries of their Gram-Schmidt coefficients and two-point functions. A collection of our results for the correlators are presented in section 3 where we explore different configurations varying the number of quiver nodes, scaling dimensions, positions etc. Finally, we analyse our matrix-model perturbative expansions using the usual Feynman diagrams approach providing some interesting results for specific classes of integrals. Appendices instead are devoted to some technical aspects.

\section{Lagrangian for superconformal quiver theories}
\begin{figure}[!t]
\begin{center}
\includegraphics[scale=0.6]{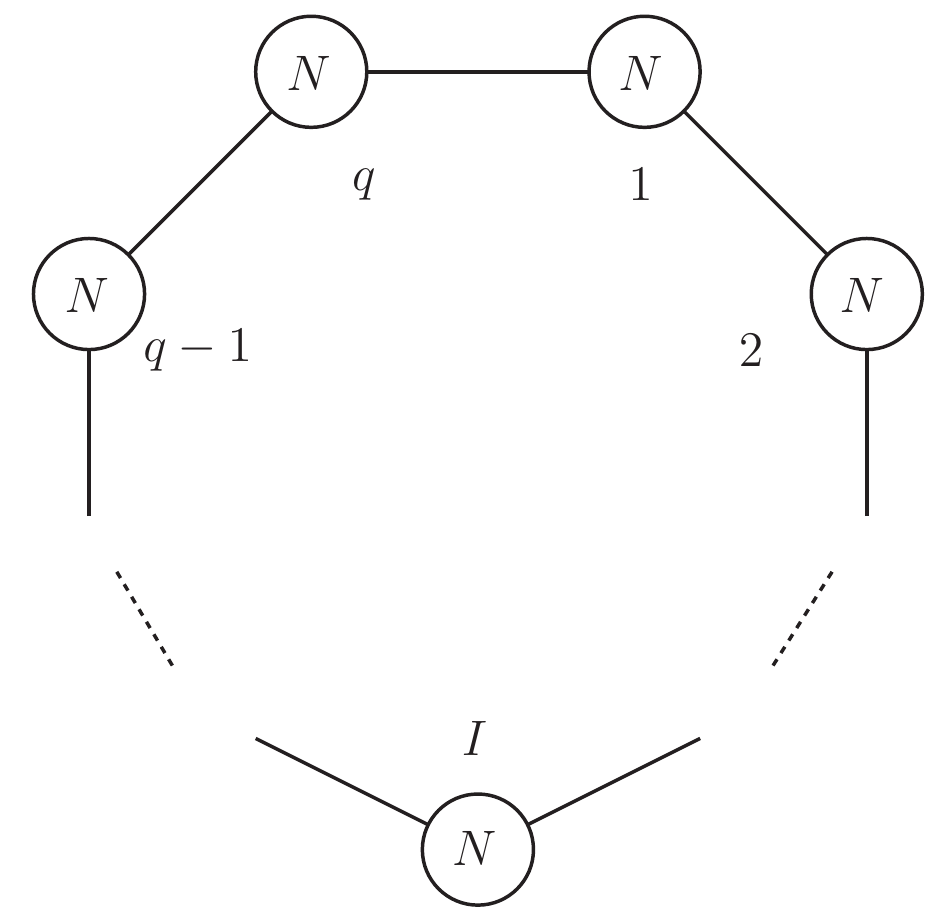}
\caption{The circular quiver with $q$ nodes associated to the theory $A_{q-1}$.}
\label{Fig:quiver}
\end{center}
\end{figure}

We consider $\cN=2$ superconformal quiver theories $A_{q-1}$ which arise as $\mathbb Z_q$ orbifolds of $\cN=4$ theory, and admit a gravity dual with a AdS$_5 \times $ S$^5$ geometry. Such $A_{q-1}$ theories have $q$ nodes, where each node $I$ represents a SU$(N)_I$ vector multiplet with gauge coupling $g_I$, each line connecting two nodes $I$ and $J$ corresponds to a a hypermultiplet in the bifundamental of SU$(N)_I \times $ SU$(N)_J$. This matter content preserves conformal symmetry also at the quantum level. The quiver shape of these theories is displayed in Figure \ref{Fig:quiver}. 

The $A_{q-1}$ quiver theories are also known as the interpolating theory, since it interpolates between $\cN=4$ and $\cN=2$ superconformal QCD (SCQCD) \cite{Gadde:2009dj,Gadde:2010zi,Pomoni:2011jj}. We specify the two interesting limits that we consider throughout the paper:
\begin{itemize}
\item
orbifold point of $\cN=4$, where we take all the couplings equal $g_I=g$;
\item
$\cN=2$ SCQCD with a single gauge group $SU(N)$ and $2N$ flavours, which can be obtained by switching off all the couplings except one ($g_1=g,$ $g_{I\neq 1}=0$).
\end{itemize}
In this sense the $A_{q-1}$ quiver theories can be seen as the $\cN=2$ theories which are closest to $\cN=4$.

Those theories admit a Lagrangian description that can be written in a compact way using the $\cN=1$ Lagrangian formalism.
 We decompose the field content of the theory into $\cN=1$ fields as follows\footnote{This decomposition is organised such that $\begin{pmatrix}
Q \\ \widetilde{Q}^\dagger
\end{pmatrix} $ forms a doublet of SU$(2)_R$.}
\begin{align}\label{N2fields}
I\mathrm{-th}~ \mathrm{Vector}_{(\cN=2)} &= \big(V, \Phi \big)_I ~~~\quad\mathrm{adj~of~SU}(N)_I \notag \\
 \mathrm{Hyper}_{(\cN=2)} &= \big(Q, \widetilde{Q} \big)~~\quad\;\, \big(\square, \bar{\square}\big)  ~\mathrm{of~SU}(N)_I\times \mathrm{SU}(N)_J~.
\end{align}
where $V$ is a $\cN=1$ vector multiplet and $\Phi,~Q,~\tilde{Q}$ are $\cN=1$ chiral multiplets.

We write the Lagrangian for a generic quiver theory $A_{q-1}$, separating the gauge contribution inside each node from the matter contribution which involves the hypermultiplets, such that
\begin{align}\label{Stot}
S_{q-1} = S_{\mathrm{gauge}} + S_{\mathrm{matter}}~.
\end{align}
The action $S_{\mathrm{gauge}}$ is given by multiple copies of a $\cN=2$ pure gauge action
\begin{align}\label{Sgauge1}
S_{\mathrm{gauge}} \!=&\! \sum_{I=1}^q \!\Bigg[ \frac{1}{8g_I^2} \left(\int\!d^4x\,d^2\theta\, \tr(W_I^\a W^I_\a)\!+\!\mathrm{h.c.}\!\right) \!+\!2\!\!\int\!\!d^4x\,d^4\theta\, \tr\!\left( e^{-2g_IV_I}\Phi_I^\dagger e^{2g_IV_I}\Phi_I\right)\!\!\Bigg]~,
\end{align}
where $\tr$ stands for the trace in the fundamental representation, $g_I$ are the gauge couplings and $W^I_\a$ corresponds to the super field strength of V
defined as follows
\begin{align}
W^I_\a=-\frac{1}{4}\bar{D}^2\left(e^{-2g_IV_I}D_\a e^{2g_IV_I}\right)~.
\end{align}
See Appendix \ref{App:Notations} for our notation for the covariant derivatives and spinor indices.
The matter part accounts for all the links between two nodes and it reads
\begin{align}\label{Smatter1}
S_{\mathrm{matter}} =& \sum_{I=1}^q \Bigg( \!\int\!d^4x\,d^4\theta\,\bigg[ \tr \left( Q^\dagger e^{2g_IV_I} Q e^{-2g_{I+1}V_{I+1}}\right) + \tr \left(\widetilde{Q} e^{-2g_IV_I}\widetilde{Q}^\dagger e^{2g_{I+1}V_{I+1}} \right) \bigg] \notag \\
+& \left(\ii \sqrt{2}g_I\!\int\!d^4x\,d^2\theta\,\widetilde{Q} \Phi_I Q +\mathrm{h.c.}\right)+ \left(\ii \sqrt{2}g_{I+1}\!\int\!d^4x\,d^2\theta\,\widetilde{Q} \Phi_{I+1} Q +\mathrm{h.c.}\right) \Bigg)~,
\end{align}
where, since the quiver is circular, the node $I=q+1$ is identified with $I=1$. 
It is useful to introduce also the 't Hooft couplings
\begin{align}\label{tHooft}
\lambda_I = g_I^2 N~.
\end{align}
In the following all the results at both finite and large $N$ will be written in terms of $\lambda_I$.

\subsection{Chiral correlators in gauge theory}

In this framework we study the correlation functions of the following scalar multi-trace local operators 
\begin{align}
	\label{defOn}
		O^{(I)}_{\vec{n}}(x) \equiv 
		\parenth{\frac{g_I^{2}}{2}}^{n/2}\;\tr \varphi_I^{n_1}(x)\, \tr \varphi_I^{n_2}(x) \ldots\,\tr \varphi_I^{n_t}(x)~,
\end{align}
defined on the $I$-th node of the quiver and featured by the vector $\vec n=\{n_1,n_2,...,n_t\}$ with $t$ the number of traces.
The field $\varphi_I$ is the scalar appearing in the vector multiplet. In our conventions it can be extracted as the lowest component of the $\Phi_I$ chiral superfield: 
\begin{equation}
\varphi_I(x) = \Phi_I(x,\theta,\bar \theta)\big|_{\theta = \bar \theta=0}~.
\end{equation}
The operators \eqref{defOn} are known as chiral or 1/2 BPS, {\it{i.e.}} they are annihilated by half of the supercharges.
Their $R$-charge is $n = \sum_i n_i$ and they are automatically normal-ordered because of 
$R$-charge conservation. Since any node of the quiver has gauge group $SU(N)$, one can restrict to $n_i\geq 2$ since $\tr \varphi=0$; thus the independent operators of dimension $n$ are as many as the partitions of $n$ in which the number 1 does not appear.
The analogous anti-chiral operators, constructed with the conjugate field 
$\bar \varphi_{I}(x)$, are denoted by $\bar O^{(I)}_{\vec n}(x)$. 

The general form of two-point functions between chiral and anti-chiral operators is fixed by (super-)conformal symmetry as follows
\begin{equation}
	\label{twopointdef}
		\big\langle O^{(I)}_{\vec n}(x) \,\bar O^{(J)}_{\vec n}(0)\big\rangle_q 
		= \frac{G^{(q,I,J)}_{\vec n}(\lambda_1,...,\lambda_q,N)}{x^{2n}\phantom{\big|}}~.
\end{equation}
where no anomalous dimension is present since the operators are protected by supersymmetry.
Notice that, unlike the $\mathcal{N}=4$ SYM case, in $\mathcal{N}=2$ the coefficient of the two-point correlator $G^{(q,I,J)}_{\vec n}$ is a non-trivial function of the couplings $\{\lambda_1,\lambda_2,...,\lambda_q\}$ and the gauge group rank $N$. This function is captured by taking suitable derivatives of a deformed partition function on the sphere, or equivalently by computing two-point correlators in the associated multi-matrix model \eqref{vevmat}. Indeed, two-point functions on a four-sphere also take the form \eqref{twopointdef}, with $x$ being the chordal distance on $S^4$ and $G^{(q,I,J)}_{\vec n}$ being the same function of flat space. Our goal is to establish a connection between correlators in the multi-matrix model and correlators in the gauge theory. Moreover, since the latter involve chiral and anti-chiral operators that do not have self-contractions, one has also to introduce matrix model operators without self-contractions either, \textit{i.e.}, normal-ordered. 
%

\section{The multi-matrix model}
We review the matrix model approach to $\cN=2$ theories and we apply it to the case of $A_{q-1}$ theories.
Previous computations in these theories on the localised matrix model have been tackled by going to the Cartan of the gauge algebra and solving the remaining integrals with the eigenvalue distribution method. In this paper we follow the full Lie Algebra approach \footnote{See \cite{Beccaria:2020hgy} for a comparison between the two methods.}, which guarantees a nice efficiency for results both at finite and large $N$ and can be implemented in a nice algorithmic way.
\noindent

\subsection{localised partition function}

Let's consider Pestun's localisation results of \cite{Pestun:2007rz}. Then, when the theory described by \eqref{Stot} is placed on a four sphere, the partition function can be reduced to a finite dimensional integral and written as the following multi-matrix model
\begin{equation}
	\label{intda}
		\cZ_{S^4} = \int \prod_{I=1}^q \parenth{da_I ~ e^{- \frac{8\pi^2}{g_{I}^2}\tr a_I^2}} \,  Z_{\mathrm{1-loop}} (a_I)\,\,\big| Z_{\mathrm{inst}}\big|^2~. 
\end{equation}
where ach matrix $a_I$ can be decomposed over a basis of generators $t_a$ of $\mathfrak{su}(N)$
\begin{equation}
	\label{aont}
	a_I = a_I^b \,t_b~,~~~b = 1,\ldots, N^2-1~.
\end{equation}

In this paper we work in the zero-instanton sector, so we can neglect the instanton contributions putting 
$Z_{\mathrm{inst}}=1$. Besides, we will mainly explore the large $N$ limit of $A_{q-1}$ theories and instantons turn out to be exponentially suppressed.
 
The 1-loop partition function $Z_{\mathrm{1-loop}}$ contains interaction terms. Those can be written in terms of the eigenvalues of $a_I$ (where $i=1,\dots N$) as follows
\begin{equation}
	\label{Z1loop}
	\big|Z_{\mathrm{1-loop}}\big|^2\,=\,\dfrac{\prod_{I} \prod_{i<j} H^2(a_i^I-a_j^I)}
	{\prod_{I} \prod_{i,j}^N H(a_i^I-a_j^{I+1})}~,
\end{equation}
where the function $H$ is given in terms of Barnes G-functions, and equivalently expanded as
\begin{equation}\label{logH}
\log H(x) = -\sum_{n=2}^\infty \frac{(-1)^n}{n} \zeta_{2n-1} x^{2n}~,
\end{equation}
with $\zeta_{2n-1}$ the Riemann zeta function $\zeta({2n-1})$.
The main idea is to treat the 1-loop determinant as an interaction action
\begin{equation}
	\label{Z1ltoS}
	\big|Z_{\mathrm{1-loop}}\big|^2\, \equiv \prod_{I=1}^q \rme^{-S_{\mathrm{int}}(a_I)}~,
\end{equation}
to be expanded perturbatively. Recall that the product over the number of nodes in \eqref{Z1ltoS} is such that the node $q+1$ is identified with the node $1$. From \eqref{Z1loop} and \eqref{logH} we build $S_{\mathrm{int}}(a_I)$ in terms of combinations of traces $\tr a_I^n$:
\begin{equation}\label{Sint}
S_{\mathrm{int}}(a_I) =  \sum_{m=2}^\infty \sum_{k=0}^{2m} \frac{(-1)^{m+k}}{m} {\zeta}_{2m-1} \binom{2m}{k} \big( \tr a_I^{2m-k} -\tr a_{I+1}^{2m-k}\big) \big( \tr a_{I}^k- \tr a_{I+1}^k \big)~.
\end{equation}
In this way $\log \big|Z_{\mathrm{1-loop}}\big|^2$ can be treated as a proper interacting term of the $S^4$ matrix model.
Therefore the matrix model that we are going to consider in the following takes the shape:
\begin{equation}
	\label{ZS4def}
		\cZ = 
		\int \prod_{I=1}^q da_I~\rme^{-\frac{8\pi^2}{g_I^2}\mathrm{tr}\, a_I^2 - \cS_{\mathrm{int}}(a_I)}~,
\end{equation}
where the interacting action is given by \eqref{Sint}.

\subsection{Multi-matrix model techniques}\label{sec:mmtech}
We can cope with the multi-matrix model \eqref{ZS4def} by simply extending the techniques introduced in \cite{Billo:2017glv,Billo:2018oog,Billo:2019fbi,Billo:2019job,Beccaria:2020azj,Beccaria:2020hgy}. Firrst of all we rescale all the matrices as follows
\begin{equation}\label{rescaling}
a_I~\rightarrow~ \left(\frac{g_{I}^2}{8\pi^2}\right)^{1/2} a_I~,
\end{equation}
in order to obtain a multi-matrix model with normalised Gaussian factors, while the interactive action can be expanded perturbatively in $g_{I}$.
The crucial point is that, thanks to the structure of the partition function \eqref{ZS4def}, we can expand the interactive action \eqref{Sint} and order by order in perturbation theory we only deal with \emph{$q$ multiple copies of the free Gaussian matrix model} describing the $\mathcal{N}=4$ theory.

As an example we show the first perturbative order (\textit{i.e.} the ${\zeta}_3$ term) of the $A_1$ theory. The partition function \eqref{ZS4def} for $q=2$ after the rescaling \eqref{rescaling} becomes
\begin{align}\label{example11}
\cZ_{A_1} \!\!=\!\! \!\int \!\!d a_1 d a_2&\exp \!\!\Bigg[\!\!-\!\mathrm{tr}\, a_1^2\!-\!\mathrm{tr}\, a_2^2-\frac{3{\color{red}\zeta}_3}{64\pi^4} \Big[{\color{blue}g}_1^4(\mathrm{tr}\, a_1^2)^2\!+\!{\color{blue}g}_2^4(\mathrm{tr}\, a_2^2)^2\!\!-\!2{\color{blue}g}_1^2{\color{blue}g}_2^2\mathrm{tr}\, a_1^2\mathrm{tr}\, a_2^2\Big]\!\!+\!...\Bigg]
\end{align}
then we can factorise the integrals over the two nodes, obtaining
\begin{align}\label{example22}
 \vev{\mathbb{1}}_0\vev{\mathbb{1}}_0\!-\!\frac{3{\color{red}\zeta}_3}{64\pi^4}\!\Big[{\color{blue}g}_1^4\vev{(\mathrm{tr}\, a_1^2)^2}_0\vev{\mathbb{1}}_0 +{\color{blue}g}_2^4\vev{\mathbb{1}}_0\vev{(\mathrm{tr}\, a_2^2)^2}_0 \!-2{\color{blue}g}_1^2{\color{blue}g}_2^2\vev{\mathrm{tr}\, a_1^2}_0\vev{\mathrm{tr}\, a_2^2}_0\Big]\!+...
\end{align}
The subscript 0 indicates that the vacuum expectation value is evaluated in the Gaussian matrix model, whose convention to produce explicit results are explained in the following paragraph. 
It is easy to see that for each perturbative order, the partition function \eqref{example11} factorises into $q=2$ products of Gaussian matrix models.

The generalisation to correlation functions is pretty straightforward. Indeed, one can evaluate gauge invariant observables, generically represented by functions $f(a_I^\ell)$, in the multi-matrix model using the following definition
\begin{align}
	\label{vevmat}
		\vev{ f(a_I^\ell) }\, 
		= \,\frac{\displaystyle{ \int \prod_{I=1}^q da_I~\rme^{-\mathrm{tr}\, a_I^2 - \cS_{\mathrm{int}}(a_I,g_I)}\,f( a_I^\ell)}}
		{ \displaystyle{\int \prod_{I=1}^q da_I~\rme^{-\mathrm{tr}\, a_I^2 - \cS_{\mathrm{int}}(a_I,g_I)}} }\,	= \,\prod_{I=1}^q\frac{\big\langle\,
		\rme^{- S(a_I,g_I)}\,f( a_I^\ell)\,\big\rangle_0\phantom{\Big|}}
		{\big\langle\,\rme^{- S(a_I,g_I)}\,\big\rangle_0
		\phantom{\Big|}}~.
\end{align} 
Also in this case the factorised structure of \eqref{example22} is reproduced. Then, having reduced the computation of vevs in the interacting matrix model to vevs in a Gaussian model, we  can now concentrate on the latter. This means that we need to generalise the recursion formulas of \cite{Billo:2017glv,Billo:2018oog} for our multi-matrix case recalling how this method works in the case of a single gauge group.
 
\paragraph{Recursion relations}

We consider a basis of $\mathfrak{su}(N)$ generators $T_b$, with
$b = 1,\ldots, N^2-1$, normalised as
\begin{equation}
\tr \,T_b \,T_c=\frac{1}{2}\,\delta_{bc}~,
\label{normtrace}
\end{equation}
and write each matrix $a$ as $a=a^b\,T_b$. Then, the flat integration measure appearing above becomes
\begin{align}
	\label{defda}
		da = \prod_b \frac{da^b}{\sqrt{2\pi}}~,
\end{align}
where the normalization has been chosen in such a way that 
$\big\langle\mathbb{1}\big\rangle_0=1$ 
and the ``propagator'' for the components of $a$ is simply
\begin{align}
\label{propa}
\vev{ a^b\, a^c}_0 = \delta^{bc}~.
\end{align} 

As displayed in the examples above \eqref{example22} and \eqref{vevmat}, the basic ingredients for the
calculation of the various observables in the $A_{q-1}$ multi-matrix model
are the expectation values of the multi-trace operators $\vev{\tr a_I^{n_1}\tr a_I^{n_2}\dots}_0$ in the Gaussian theory.
It is convenient to introduce the following notation for the Gaussian vevs
 following the notation:
\begin{equation}
	\label{tn}
		t^{(I)}_{n_1,n_2,\dots} = t^{(I)}_{\vec{n}}= \vev{\tr a_I^{n_1}\tr a_I^{n_2}\dots}_0~.
\end{equation}
Starting from the initial conditions
\begin{align}
	\label{tnodd}
		t^{(I)}_n = 0~~~\text{for $n$ odd},\quad\text{and}\quad t^{(I)}_0=N~,
\end{align}
we can evaluate the general expression for $t^{(I)}_{\vec{n}}$ from the following recursion relations \cite{Billo:2017glv}:
\begin{equation}\begin{split}\label{rrecursion}
t^{(I)}_{[n_1,n_2,...,n_t]} =   &\frac{1}{2} \sum_{m=0}^{n_1-2}  \Big( t^{(I)}_{[m,n_1-m-2,n_2,...,n_t]}
		-\frac{1}{N}\,   t^{(I)}_{[n_1-2,n_2,...,n_t]}  \Big) \\ 
		&+ \sum_{k=2}^{t}\frac{n_k}{2} \,\Big(  t^{(I)}_{[n_1+n_k-2,n_2,...,\slashed{n_k},...,n_t]} -\frac{1}{N} \,t^{(I)}_{[n_1-1,n_2,...,n_k-1,...,n_t]} \Big)~,
\end{split}\end{equation}
where ${[n_1,...,\slashed{n_k},...,n_t]}$ represents the vector of indices without the $k$-th one.
These recursive formulas follow  from the fusion/fission identities satisfied by the $\mathfrak{su}(N)$ generators $T_b$, namely
\begin{equation}
\label{fussion}
\begin{aligned}	
		\tr\big(T_b\, A\, T_b\, B\big) & = \frac{1}{2}\,\tr A\, \tr B
    	-\frac{1}{2N}\,\tr \big(A\, B\big)~,\\
		\tr \big(T_b\, A \big) \,\tr\big(T_b \,B\big) & 
		= \frac{1}{2}\,\tr\big(A\,B\big) -\frac{1}{2N}\,\tr A \,\tr B~,
\end{aligned}
\end{equation}
for two arbitrary $N\times N$ matrices $A$ and $B$. 
In fact, using these identities one can recursively relate any correlator $t^{(I)}_{\vec{n}}$ to the combination of correlators obtained after a single Wick contraction with the propagator \eqref{propa}. The superconformal quiver we are considering contains $q$ vector multiplets with gauge group with the same rank $N$. This produces a big simplification for the correlators. Indeed for a given vector $\vec{n}$, $t^{(I)}_{\vec{n}}$ is the same for any $I$. Then in the rest of the manuscript we drop the index $I$.  

It is useful to provide some examples. Using the recursive relation \eqref{rrecursion} together with \eqref{tnodd} we can compute the correlators $t_{\vec{n}}$ for all the possible partitions of $n=8$
\begin{equation}\small\begin{split}
t_{[8]}&=\frac{7(N^2\!-\!1)(2N^6\!-\!8N^4\!+\!15N^2\!-\!15)}{16N^3},\quad\;\;\,\!\!
t_{[6,2]}=\frac{5(N^2\!-\!1)(N^2\!+\!5)(N^4\!-\!3N^2\!+\!3)}{16N^2},\\
t_{[5,3]}&=\frac{15(N^2\!-\!1)(N^2\!-\!2)(N^2\!-\!4)}{16N^2},\qquad\qquad\;\,\,\!\!
t_{[4,4]}=\frac{(N^2\!-\!1)(4N^6\!+\!20N^4\!-\!99N^2\!+\!135)}{16N^2},\\
t_{[4,2,2]}&=\frac{(N^2\!-\!1)(N^2\!+\!3)(N^2\!+\!5)(2N^2\!-\!3)}{16N},\quad\!\!
t_{[3,3,2]}=\frac{3(N^2\!-\!1)(N^2\!-\!4)(N^2\!+\!5)}{16N},\\
&\qquad\qquad\qquad t_{[2,2,2,2]}=\frac{(N^2\!-\!1)(N^2\!+\!1)(N^2\!+\!3)(N^2\!+\!5)}{16}\,.
\end{split}\end{equation}
We can also evaluate all the Gaussian vevs of formula \eqref{example22}, obtaining the finite $N$ expression of the $A_1$ partition function:
\begin{align}
\cZ_{A_1} = 1-\frac{3{\color{red}\zeta}_3}{64\pi^4}\parenth{\frac{N^4-1}{4N^4}({\color{blue}\lambda}_1^2-{\color{blue}\lambda}_2^2)^2+\frac{1}{N^2} {\color{blue}\lambda}_1^2{\color{blue}\lambda}_2^2}+\dots~,
\end{align}
where the 't Hooft coupling is defined in  \eqref{tHooft}. 

\subsection{Chiral correlators in the multi-matrix model}

Let's define an operator in the multi-matrix model on the sphere as follows
\begin{align}
	\label{defOS4}
		\mathcal{O}^{(I)}_{\vec n} \equiv \parenth{\frac{g_I^2}{8\pi^2}}^{n/2} \tr a_{I}^{n_1}(x)\, \tr a_{I}^{n_2}(x) \ldots \tr a_{I}^{n_t}~,
\end{align}
similarly to the gauge theory case \eqref{defOn} with the proper normalisation factor. 
Our goal is to establish a connection between correlators of this operators and \eqref{twopointdef}. the main ingredient is that $\mathcal{O}^{(I)}_{\vec n}$ has to be normal-ordered.
In order to subtract all the self-contractions of a given operator, one has to make it orthogonal to all the lower-dimensional operators. Indeed, the crucial subtlety is that, due to the conformal anomaly, in $S^4$ operators of different dimensions can mix. 
Furthermore, operators belonging to different nodes of the quiver can also mix. Such intricate mixing system must be disentangled through a Gram-Schmidt procedure as described in \cite{Gerchkovitz:2016gxx}. 

\subsubsection{Normal-ordered operators}

Let be $n$ the scaling dimension of the operator $\mathcal{O}^{(I)}_{\vec n}$, then its corresponding normal-ordering is given by a linear combination of itself and the operators in the basis $\{\mathcal{O}^{(J)}_{\vec p}\}$ with dimensions $\{p\}=\{n-2,n-4,...\}$
\begin{equation}\label{basis}
:\mathcal{O}^{(I)}_{\vec n}:=\mathcal{O}^{(I)}_{\vec n}+\sum_{\substack{\vec p=\text{partitions} \\ \text{of dim. $\{p\}$}}}
\;\sum_{\substack{J=\text{nodes of} \\ \text{quiver $A_{q-1}$}}}
\alpha^{(I,J)}_{\vec n,\vec p}\;\mathcal{O}^{(J)}_{\vec p}~,
\end{equation}
where the coefficient $\alpha$ are functions of the couplings $\lambda_1,...,\lambda_q$ and $N$. In \eqref{basis} we explicitly write the sums on repeated indices for clearance. The scaling dimensions in the set $\{p\}$ differ by 2 since we trade two matrices with a self-contraction. Then, the lower dimensional operator in the set $\{\mathcal{O}^{(J)}_{\vec p}\}$ depends on the parity of the dimension of the operator $\mathcal{O}^{(I)}_{\vec n}$: if $n$ is even the lowest possible dimension is zero corresponding to the identity operator $\mathbb{1}$\footnote{When in \eqref{basis} the operator $\mathcal{O}^{(J)}_{\vec p}$ is equal to the identity $\mathbb{1}$, our convention is to fix $J=I$ since the identity appears only once in the basis.}, if $n$ is odd the lowest possible dimension is three corresponding to the operator $\mathcal{O}^{(I)}_{[3]}$. Let's see some easy examples
\begin{equation}\begin{split}\label{mixexample}
&:\mathcal{O}^{(I)}_{[2]}:\rightarrow\{\mathbb{1}\}~,
\qquad\qquad\qquad\quad\;\;\;:\mathcal{O}^{(I)}_{[3]}:\rightarrow\{\varnothing\}~,\\
&:\mathcal{O}^{(I)}_{[4]}:\rightarrow\{\mathcal{O}^{(1)}_{[2]},...,\mathcal{O}^{(q)}_{[2]},\mathbb{1}\}~,\qquad
:\mathcal{O}^{(I)}_{[2,2]}:\rightarrow\{\mathcal{O}^{(1)}_{[2]},...,\mathcal{O}^{(q)}_{[2]},\mathbb{1}\}~,\\
&:\mathcal{O}^{(I)}_{[5]}:\rightarrow\{\mathcal{O}^{(1)}_{[3]},...,\mathcal{O}^{(q)}_{[3]}\}~,\qquad\quad
:\mathcal{O}^{(I)}_{[3,2]}:\rightarrow\{\mathcal{O}^{(1)}_{[3]},...,\mathcal{O}^{(q)}_{[3]}\}~,\\
&:\mathcal{O}^{(I)}_{[6]}:\rightarrow\{\mathcal{O}^{(1)}_{[4]},...,\mathcal{O}^{(q)}_{[4]},\mathcal{O}^{(1)}_{[2,2]},...,\mathcal{O}^{(q)}_{[2,2]},\mathcal{O}^{(1)}_{[2]},...,\mathcal{O}^{(q)}_{[2]},\mathbb{1}\}~.
\end{split}\end{equation}
The elements in the base does not depend on the chosen partition but only on the operator scaling dimension $n$ and the length of the quiver $q$. Their number is growing fast with $n$ and $q$ as represented in table \ref{tab:tabella}.
\begin{table}[!t]
\begin{center}
 \begin{tabular}{||c c c||} 
 \hline
 dimension $n$ & number of partitions & elements in the base \\ [0.5ex] 
 \hline\hline
 2 & 1 & 1  \\ 
 \hline
 3 & 1 & 0 \\
 \hline
 4 & 2 & $q$+1  \\
 \hline
 5 & 2 & $q$  \\
  \hline
 6 & 4 & 3$q$+1  \\
  \hline
 7 & 4 & 3$q$ \\
  \hline
 8 & 7 & 7$q$+1 \\
  \hline
 9 & 8 & 7$q$ \\
 \hline
 10 & 12 & 14$q$+1  \\ 
  \hline
 11 & 14 & 15$q$  \\ 
 \hline
\end{tabular}
\end{center}
\caption{Number of partitions and number of elements in the orthogonal Gram-Schmidt basis for a given dimension $n$}
\label{tab:tabella}
\end{table}

The mixing coefficients appearing in \eqref{basis} can be determined through the Gram-Schmidt procedure. 
Let's consider $\{\mathcal{O}^{(J)}_{\vec p}\}$ and $\{\mathcal{O}^{(K)}_{\vec s}\}$ two copies of the basis associated to the operator $\mathcal{O}^{(I)}_{\vec n}$. Then we can define the following matrix
\begin{equation}\label{matrixM}
M_{\vec s,\vec p}^{(K,J)}=\langle \mathcal{O}^{(K)}_{\vec s} \;\mathcal{O}^{(J)}_{\vec p} \rangle_q\,,
\end{equation}
as the matrix of all the mixed correlators of the elements of the base with $K,J=1,2,...,q$ and $\vec s$ and $\vec p$ al the possible partitions of the dimensions $\{s\}$ and $\{p\}$. According to the table above, if for instance $n=6$, then $M$ is $3q+1\times 3q+1$ matrix.
Hence one can rewrite the coefficients in \eqref{basis} in terms of the correlators \eqref{matrixM} as follows
\begin{equation}\label{coeffGS}
\alpha^{(I,J)}_{\vec n,\vec p}=-\sum_{\substack{\text{nodes $K$} \\ \text{partitions $\vec s$ of $\{s\}$}}}\langle \mathcal{O}^{(I)}_{\vec n} \;\mathcal{O}^{(K)}_{\vec s} \rangle_q\;\left(M_{\vec s,\vec p}^{(K,J)}\right)^{-1}~,
\end{equation}
where $M^{-1}$ is the inverse of the matrix \eqref{matrixM}. 
It's easy to check that the normal-ordered operator \eqref{basis} with coefficients defined by \eqref{coeffGS} is orthogonal by construction to all operators $\{\mathcal{O}^{(J)}_{\vec p}\}$ of lower dimension for any $J$. In particular its one-point function vanishes
\begin{equation}
\langle \;:\mathcal{O}^{(I)}_{\vec n}: \;\rangle_q=0~,
\end{equation}
since the sums in \eqref{basis} precisely subtract all the self-contractions.
Finally we can relate the two-point function of normal ordered operators in the matrix-model with the correlator in the gauge theory \eqref{twopointdef} as follows
\begin{equation}
	\label{Gnormord}
		G^{(q,I,J)}_{\vec n}(\lambda_1,...,\lambda_q,N)=\big\langle :\mathcal{O}^{(I)}_{\vec n}: \;:\mathcal{O}^{(J)}_{\vec n}:\big\rangle_q ~.
\end{equation}

\paragraph{Examples}$\\$

In the following we present some examples of normal-ordered operators on quivers with various number of nodes. Since the finite $N$ expansions of the coefficients $\alpha$ in \eqref{basis} start to be cumbersome at very low order, for clearance we explicitly write only the first few orders. Anyway, in order to compute several terms in the expansion of the correlators in section \ref{sec:res} and in the notebook attached to this manuscript, we had to compute those coefficients up to very high powers in the couplings $\lambda_1,...,\lambda_q$. 

The lowest dimension operators are extremely simple since they mix with the identity or with the empty set \eqref{mixexample} then
\begin{equation}\label{O2O3}
:\mathcal{O}^{(I)}_{[2]}:\;\equiv\; :\tr a_I^2:\;=\mathcal{O}^{(I)}_{[2]}-\langle\;\mathcal{O}^{(I)}_{[2]}\;\rangle_q~,\qquad\qquad
:\mathcal{O}^{(I)}_{[3]}:\;\equiv\; :\tr a_I^3:\;=\mathcal{O}^{(I)}_{[3]}~,
\end{equation}
where using \eqref{basis} and \eqref{coeffGS} it's easy to see that the coefficient of the identity operator $\alpha_{[2],[0]}^{I,I}$ is the one-point function of the operator itself.
The examples proposed below are of two normal ordered operators with a richer structure. Indeed we consider first the double-trace operator of dimension $n=5$ in quivers with number of nodes $q=1,2,3$. Then the single-trace operator of dimension $n=6$ in quivers with number of nodes $q=1,2$. With $\dots$ we represents higher terms in powers of all the couplings.

\paragraph{Double-trace operator of dimension $n=5$:}

Let's consider for simplicity the operator in the first vector multiplet.
According to \eqref{mixexample}, $\mathcal{O}^{(1)}_{[3,2]}$ mixes with $q$ copies of the single-trace operator of dimension 3, one for any vector multiple of the quiver.

\subparagraph{SCQCD:}
The normal ordered operators is given by
\begin{equation}\begin{split}
:\mathcal{O}^{(1)}_{[3,2]}:\;\equiv\; :\tr a_1^3\;\tr a_1^2:\;=&
\mathcal{O}^{(1)}_{[3,2]}
+\alpha^{(1,1)}_{[3,2],[3]}\;\mathcal{O}^{(1)}_{[3]}~,
\end{split}\end{equation}
with the following coefficient
\begin{equation}\small
\alpha^{(1,1)}_{[3,2],[3]}=
-\frac{N^2\!+\!5}{2}\!+\!\frac{3 \left(N^4\!+\!12 N^2\!+\!35\right) {\color{red}\zeta}_3{\color{blue}\lambda}_1^2}{2 (8\pi^2)^2N^2}\!-\!\frac{15\left(2 N^6\!+\!43 N^4\!+\!60 N^2\!-\!105\right) {\color{red}\zeta}_5{\color{blue}\lambda}_1^3}{4 (8\pi^2)^3N^4}\!+\dots
\end{equation}

\subparagraph{Quiver $A_1$:}
The normal ordered operators is given by
\begin{equation}\begin{split}\label{A132}
:\mathcal{O}^{(1)}_{[3,2]}:\;\equiv\; :\tr a_1^3\;\tr a_1^2:\;=&
\mathcal{O}^{(1)}_{[3,2]}
+\alpha^{(1,1)}_{[3,2],[3]}\;\mathcal{O}^{(1)}_{[3]}
+\alpha^{(1,2)}_{[3,2],[3]}\;\mathcal{O}^{(2)}_{[3]}~,
\end{split}\end{equation}
with the following coefficients
\small\begin{align}
&\alpha^{(1,1)}_{[3,2],[3]}=
-\frac{N^2+5}{2}+\frac{3 \left(N^2+5\right){\color{red}\zeta}_3{\color{blue}\lambda}_1 \left(\left(N^2+7\right) {\color{blue}\lambda}_1-\left(N^2-1\right) {\color{blue}\lambda}_2\right)}{2 (8\pi^2)^2N^2}+\dots\\
&\alpha^{(1,2)}_{[3,2],[3]}=\frac{105 \left(N^6-7 N^4+14 N^2-8\right) {\color{red}\zeta}_7{\color{blue}\lambda}_1^\frac{5}{2} {\color{blue}\lambda}_2^\frac{3}{2}}{4 (8\pi^2)^4N^6}-9\left(N^4\!-\!5 N^2\!+\!4\right)\times \nonumber\\
&\qquad\;\times\frac{{\color{blue}\lambda}_1^\frac{5}{2} {\color{blue}\lambda}_2^\frac{3}{2} \!\left(20\! \left(N^2\!+\!5\right) \!N^2 {\color{red}\zeta}_3{\color{red}\zeta}_5{\color{blue}\lambda}_2\!+\!7 {\color{red}\zeta}_9\!\left(15 \left(N^2\!-\!2\right)^2 \!{\color{blue}\lambda}_2\!+\!4 \left(3 N^4\!-\!10 N^2\!+\!15\right) {\color{blue}\lambda}_1\right)\right)}{4 (8\pi^2)^5N^8}\!+\!\dots
\end{align}\normalsize

\subparagraph{Quiver $A_2$:}
The normal ordered operators is given by
\begin{equation}\begin{split}
:\mathcal{O}^{(1)}_{[3,2]}:\;\equiv\; :\tr a_1^3\;\tr a_1^2:\;=&
\mathcal{O}^{(1)}_{[3,2]}
+\alpha^{(1,1)}_{[3,2],[3]}\;\mathcal{O}^{(1)}_{[3]}
+\alpha^{(1,2)}_{[3,2],[3]}\;\mathcal{O}^{(2)}_{[3]}
+\alpha^{(1,3)}_{[3,2],[3]}\;\mathcal{O}^{(3)}_{[3]}~,
\end{split}\end{equation}
with the following coefficients
\small\begin{align}
&\alpha^{(1,1)}_{[3,2],[3]}=
-\frac{N^2+5}{2}+\frac{3 \left(N^2+5\right){\color{red}\zeta}_3 {\color{blue}\lambda}_1 \left(2 \left(N^2+7\right) {\color{blue}\lambda}_1-\left(N^2-1\right) \left({\color{blue}\lambda}_2+{\color{blue}\lambda}_3\right)\right)}{4 (8\pi^2)^2N^2}+\dots\\
&\alpha^{(1,2)}_{[3,2],[3]}=
\frac{105 \left(N^6-7 N^4+14 N^2-8\right) {\color{red}\zeta}_7 {\color{blue}\lambda}_1^\frac{5}{2} {\color{blue}\lambda}_2^\frac{3}{2}}{8 (8\pi^2)^4N^6}-9\left(N^4\!-\!5 N^2\!+\!4\right) \times\nonumber\\
&\qquad\;\times\!\frac{{\color{blue}\lambda}_1^\frac{5}{2} {\color{blue}\lambda}_2^\frac{3}{2} \!\left(10 \left(N^2\!+\!5\right) \!N^2 {\color{red}\zeta}_3 {\color{red}\zeta}_5 {\color{blue}\lambda}_2\!+\!7 {\color{red}\zeta}_9 \!\left(15 \left(N^2\!-\!2\right)^2\! {\color{blue}\lambda}_2\!+\!4 \left(3 N^4\!-\!10 N^2\!+\!15\right)\! {\color{blue}\lambda}_1\right)\right)}{8 (8\pi^2)^5N^8}\!+\!\dots\\
&\alpha^{(1,3)}_{[3,2],[3]}=
\frac{105 \left(N^6-7 N^4+14 N^2-8\right) {\color{red}\zeta}_7 {\color{blue}\lambda}_1^\frac{5}{2} {\color{blue}\lambda}_3^\frac{3}{2}}{8 (8\pi^2)^4N^6}-9\left(N^4\!-\!5 N^2\!+\!4\right) \times\nonumber\\
&\qquad\;\times\!\frac{{\color{blue}\lambda}_1^\frac{5}{2} {\color{blue}\lambda}_3^\frac{3}{2} \!\left(10 \left(N^2\!+\!5\right) \!N^2 {\color{red}\zeta}_3 {\color{red}\zeta}_5 {\color{blue}\lambda}_3\!+\!7 {\color{red}\zeta}_9 \!\left(15 \left(N^2\!-\!2\right)^2\! {\color{blue}\lambda}_3\!+\!4 \left(3 N^4\!-\!10 N^2\!+\!15\right)\! {\color{blue}\lambda}_1\right)\right)}{8 (8\pi^2)^5N^8}\!+\!\dots
\end{align}\normalsize

\noindent
In the last example it's easy to see that 
\begin{equation}\label{example1}
\alpha^{(1,3)}_{[3,2],[3]}=\alpha^{(1,2)}_{[3,2],[3]}\biggl|_{{\lambda}_2\leftrightarrow {\lambda}_3}~.
\end{equation}
This is consequence of a general symmetry of the coefficients $\alpha$ and the correlators itself in quivers. More details about it are presented in section \ref{sec:symmetry}.

\paragraph{Single-trace operator of dimension $n=6$:}

Let's consider also in this case the operator in the first vector multiplet.
According to \eqref{mixexample}, $\mathcal{O}^{(1)}_{[6]}$ mixes with $q$ copies of the single- and double-trace operators of dimension 4 and 2, one for any vector multiple of the quiver. Moreover, since the operator has even dimension, it mixes also with the identity operator.

\subparagraph{SCQCD:}
The normal ordered operators is given by
\begin{equation}\begin{split}
:\mathcal{O}^{(1)}_{[6]}:\;\equiv\; :\tr a_1^6:\;=&
\mathcal{O}^{(1)}_{[6]}
+\alpha^{(1,1)}_{[6],[4]}\;\mathcal{O}^{(1)}_{[4]}
+\alpha^{(1,1)}_{[6],[2,2]}\;\mathcal{O}^{(1)}_{[2,2]}
+\alpha^{(1,1)}_{[6],[2]}\;\mathcal{O}^{(1)}_{[2]}
+\alpha^{(1,1)}_{[6],[0]}~,
\end{split}\end{equation}
with the following coefficients
\begin{equation}\small
\alpha^{(1,1)}_{[6],[4]}=\frac{15-6N^2}{2 N}+\frac{9 \left(2 N^4\!+\!13 N^2\!-\!45\right){\color{red}\zeta}_3{\color{blue}\lambda}_1^2}{2 (8\pi^2)^2N^3}-\frac{15\left(16 N^6\!+\!90 N^4\!-\!81 N^2\!+\!495\right) {\color{red}\zeta}_5{\color{blue}\lambda}_1^3}{4 (8\pi^2)^3 N^5}+\dots
\end{equation}
\small\begin{align}
\alpha^{(1,1)}_{[6],[2,2]}&=-\frac{3}{2}+\frac{9 \left(N^2+9\right) {\color{red}\zeta}_3 {\color{blue}\lambda}_1^2}{2 (8\pi^2)^2N^2}-\frac{15\left(18 N^6+95 N^4-183 N^2-60\right) {\color{red}\zeta}_5 {\color{blue}\lambda}_1^3}{4 (8\pi^2)^3 N^6}+\dots\\
\alpha^{(1,1)}_{[6],[2]}&=\frac{15 \left(N^4-3 N^2+3\right)}{4 N^2}-\frac{45\left(N^6+5 N^4-21 N^2+24\right) {\color{red}\zeta}_3 {\color{blue}\lambda}_1^2}{2  (8\pi^2)^2N^4}\nonumber\\
&\qquad\qquad\qquad\qquad\quad+\frac{45 \left(16 N^8+70 N^6-161 N^4+300 N^2-445\right) {\color{red}\zeta}_5 {\color{blue}\lambda}_1^3}{4  (8\pi^2)^3N^6}+\dots\\
\alpha^{(1,1)}_{[6],[0]}&=-\frac{5 \left(N^6-4 N^4+6 N^2-3\right)}{8 N^2}+\frac{45 (N^2-1) \left(N^2+7\right) \left(N^4-3 N^2+3\right){\color{red}\zeta}_3 {\color{blue}\lambda}_1^2}{8 (8\pi^2)^2N^4}\nonumber\\
&\qquad\qquad\;-\frac{15(N^2-1)\left(46 N^8+177 N^6-434 N^4+810 N^2-1165\right) {\color{red}\zeta}_5 {\color{blue}\lambda}_1^3}{16 (8\pi^2)^3N^6}+\dots
\end{align}\normalsize

\subparagraph{Quiver $A_1$:}
The normal ordered operators is given by
\begin{equation}\begin{split}
:\mathcal{O}^{(1)}_{[6]}:\;\equiv\; :\tr a_1^6:\;=&
\mathcal{O}^{(1)}_{[6]}
+\alpha^{(1,1)}_{[6],[4]}\;\mathcal{O}^{(1)}_{[4]}
+\alpha^{(1,2)}_{[6],[4]}\;\mathcal{O}^{(2)}_{[4]}
+\alpha^{(1,1)}_{[6],[2,2]}\;\mathcal{O}^{(1)}_{[2,2]}\\
&+\alpha^{(1,2)}_{[6],[2,2]}\;\mathcal{O}^{(2)}_{[2,2]}
+\alpha^{(1,1)}_{[6],[2]}\;\mathcal{O}^{(1)}_{[2]}
+\alpha^{(1,2)}_{[6],[2]}\;\mathcal{O}^{(2)}_{[2]}
+\alpha^{(1,1)}_{[6],[0]}~,
\end{split}\end{equation}
with the following coefficients
\small\begin{align}
\alpha^{(1,1)}_{[6],[4]}&\!=\frac{15-6N^2}{2 N}+\frac{9 \left(2 N^2-5\right){\color{red}\zeta}_3 {\color{blue}\lambda}_1
 \left(\left(N^2+9\right) {\color{blue}\lambda}_1
-\left(N^2-1\right) {\color{blue}\lambda}_2
\right)}{2(8\pi^2)^2 N^3}+\dots\\
\alpha^{(1,2)}_{[6],[4]}&\!=\frac{63 \left(N^2-1\right) \left(N^8+6 N^6-60 N^4+600\right){\color{red}\zeta}_9 {\color{blue}\lambda}_1^3 {\color{blue}\lambda}_2^2}{8 (8\pi^2)^5N^9}+\dots\\
\alpha^{(1,1)}_{[6],[2,2]}&\!=-\frac{3}{2}+\frac{9 {\color{red}\zeta}_3 {\color{blue}\lambda}_1 \left(\left(N^2+9\right) {\color{blue}\lambda}_1-\left(N^2-1\right) {\color{blue}\lambda}_2\right)}{2(8\pi^2)^2 N^2}+\dots\\
\alpha^{(1,1)}_{[6],[2,2]}&\!=\frac{45 \left(N^2-1\right) \left(2 N^6-13 N^4+60 N^2-90\right)  {\color{red}\zeta}_3 {\color{red}\zeta}_5 {\color{blue}\lambda}_1^3 {\color{blue}\lambda}_2^2}{2(8\pi^2)^5N^8}+\dots\\
\alpha^{(1,1)}_{[6],[2]}&\!=\frac{15 \left(N^4\!-3 N^2\!+3\right)}{4 N^2}-\frac{45\left(N^4-3 N^2+3\right) {\color{red}\zeta}_3 {\color{blue}\lambda}_1 \left(\left(N^2+8\right) {\color{blue}\lambda}_1\!-\left(N^2-1\right) {\color{blue}\lambda}_2\right)}{2(8\pi^2)^2N^4}\!+\dots\\
\alpha^{(1,2)}_{[6],[2]}&\!=-\frac{21\left(N^2-1\right) \left(N^8+6 N^6-60 N^4+600\right) {\color{red}\zeta}_7{\color{blue}\lambda}_1^3{\color{blue}\lambda}_2}{16 (8\pi^2)^4N^8}
+\dots\\
\alpha^{(1,1)}_{[6],[0]}&\!=\!-\frac{5\! \left(\!N^6\!\!-\!4 N^4\!+\!6 N^2\!-\!3\right)\!}{8 N^2}\!+\!\frac{45\! \left(\!N^6\!\!-\!4 N^4\!+\!6 N^2\!-\!3\right)\! {\color{red}\zeta}_3{\color{blue}\lambda}_1 \!\left(\!\left(\!N^2\!+\!7\right)\!{\color{blue}\lambda}_1\!\!-\!\left(\!N^2\!-\!1\right)\!{\color{blue}\lambda}_2\!\right)\!}{8(8\pi^2)^2 N^4}\!+\!...
\end{align}\normalsize

\noindent
In all the previous examples we can define a pattern about the orders in which a coefficient starts to contribute. Indeed if we consider $\alpha^{(I,J)}_{\vec n,\vec p}$ with $I=J$ the expansion starts from $O({\lambda}^0)$, but if $J$ differs from $I$ it starts at higher order in ${\lambda}$ and also in transcendentality.
Moreover, the coefficients of the identity operator show another interesting feature. Indeed, similarly to the simplest case in \eqref{O2O3} for dimension $n=2$, they are simply given by the one-point functions of all the operators involved in the normal ordering with their corresponding coefficients
\begin{equation}
\alpha^{(I,I)}_{\vec n,[0]}=-\,\langle\,\mathcal{O}^{(I)}_{\vec n}\,\rangle_q-\sum_{\substack{\vec p=\text{partitions} \\ \text{of dim. $\{p\}\setminus 0$}}}
\;\sum_{\substack{J=\text{nodes of} \\ \text{quiver $A_{q-1}$}}}
\alpha^{(I,J)}_{\vec n,\vec p}\;\langle\,\mathcal{O}^{(J)}_{\vec p}\,\rangle_q~,
\end{equation}
where the set $\{\mathcal{O}^{(J)}_{\vec p}\}$ does not include the identity.

\subsubsection{Symmetries of Gram-Schmidt coefficients and correlators}\label{sec:symmetry}

The circular quiver presents some interesting symmetries for the Gram-Schmidt coefficients and the two-point functions. 
They are generated by the rotations and reflections of regular polygons with number of vertices equal to the number of vector multiplets $q$ in the quiver.
Those symmetries form a discrete finite group: the Dihedral group $D_q$.

Let's consider the Gram-Schmidt coefficients  $\alpha^{(I,J)}_{\vec n,\vec p}$ in a quiver with $q$ nodes. For a given pair of vectors $\vec n$ and $\vec p$, we can classify the coefficients by the "distance" between the node $I$ and $J$ defined as follows
\begin{equation}\label{dist}
d(q,I,J)\equiv ||J-I-q/2|-q/2|~.
\end{equation}  
For instance if $I=1$, the nodes $J=2$ and $J=q$ have distance $d=1$, the nodes $J=3$ and $J=q-1$ have distance $d=2$ and so on.
The coefficients associated to nodes at the same distance are the same with a suitable change of the 't Hooft couplings as follows
\begin{equation}\label{sym1}
\alpha^{(I,I+q-k)}_{\vec n,\vec p}=\alpha^{(I,I+k)}_{\vec n,\vec p}\biggl|_{\substack{{\lambda}_{I+1}\leftrightarrow {\lambda}_{I+q-1}\\
{\lambda}_{I+2}\leftrightarrow {\lambda}_{I+q-2}\\ \dots}}~,
\end{equation}
\begin{equation}\label{sym2}
\alpha^{(I+k,I)}_{\vec n,\vec p}=\alpha^{(I,I+k)}_{\vec n,\vec p}\biggl|_{\substack{{\lambda}_{I}\leftrightarrow {\lambda}_{I+k}\\
{\lambda}_{I-1}\leftrightarrow {\lambda}_{I+k-1}\\ \dots}}~,
\end{equation}
\begin{equation}\label{sym3}
\alpha^{(I+q-k,I)}_{\vec n,\vec p}=\alpha^{(I,I+k)}_{\vec n,\vec p}\biggl|_{\substack{{\lambda}_{I}\rightarrow {\lambda}_{I+q-k}\\
{\lambda}_{I+1}\rightarrow {\lambda}_{I+q-k+1}\\ \dots}}~,
\end{equation}
with $k=1,2,...,q$ and where we consider the circular symmetry of the quiver such that $q+n=n$.
Formulas \eqref{sym1} and \eqref{sym2} are associated to reflections respect to some axes cutting the quiver in half, where formula \eqref{sym3} is associated to a clock-wise rotation of $q-k$ steps. Indeed, reflections produce transformations in the couplings with double arrows and rotations with single arrows. Notice that the symmetry observed in the example in the previous section \ref{example1} is given by \eqref{sym1} with $q=3$, $I=1$ and $k=1$.
The last case to consider is when $J=I$. Indeed we have
\begin{equation}\label{sym4}
\alpha^{(I+k,I+k)}_{\vec n,\vec p}=\alpha^{(I,I)}_{\vec n,\vec p}\biggl|_{\substack{{\lambda}_{I}\leftrightarrow {\lambda}_{I+k}\\
{\lambda}_{I+1}\leftrightarrow {\lambda}_{I+k-1}\\ \dots}}~,
\end{equation}
Notice that at the orbifold point, when all the couplings are equal, all the equalities above are always true without any coupling exchange.

\begin{figure}[!t]
    \begin{minipage}[t]{.48\textwidth}
        \centering
        \includegraphics[width=\textwidth]{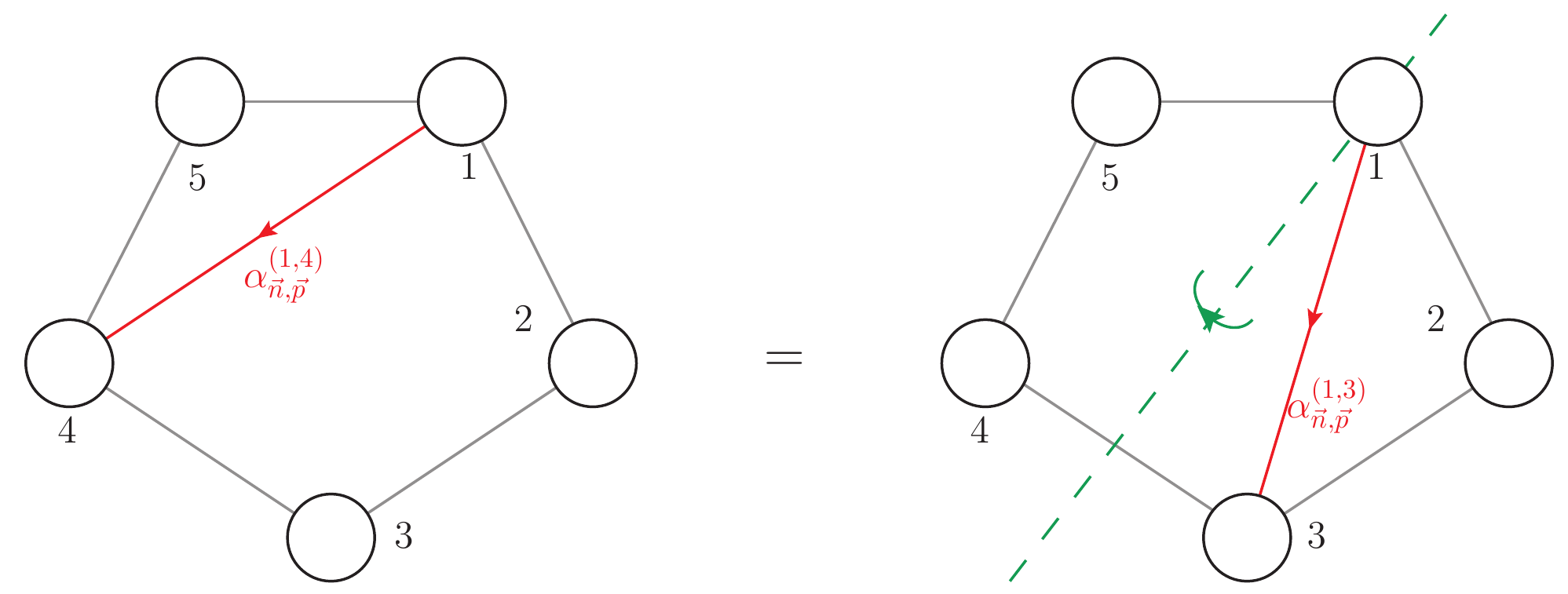}
        \subcaption{Graphical representation of \eqref{sym1}}\label{fig:1}
    \end{minipage}
    \hfill
    \begin{minipage}[t]{.48\textwidth}
        \centering
        \includegraphics[width=\textwidth]{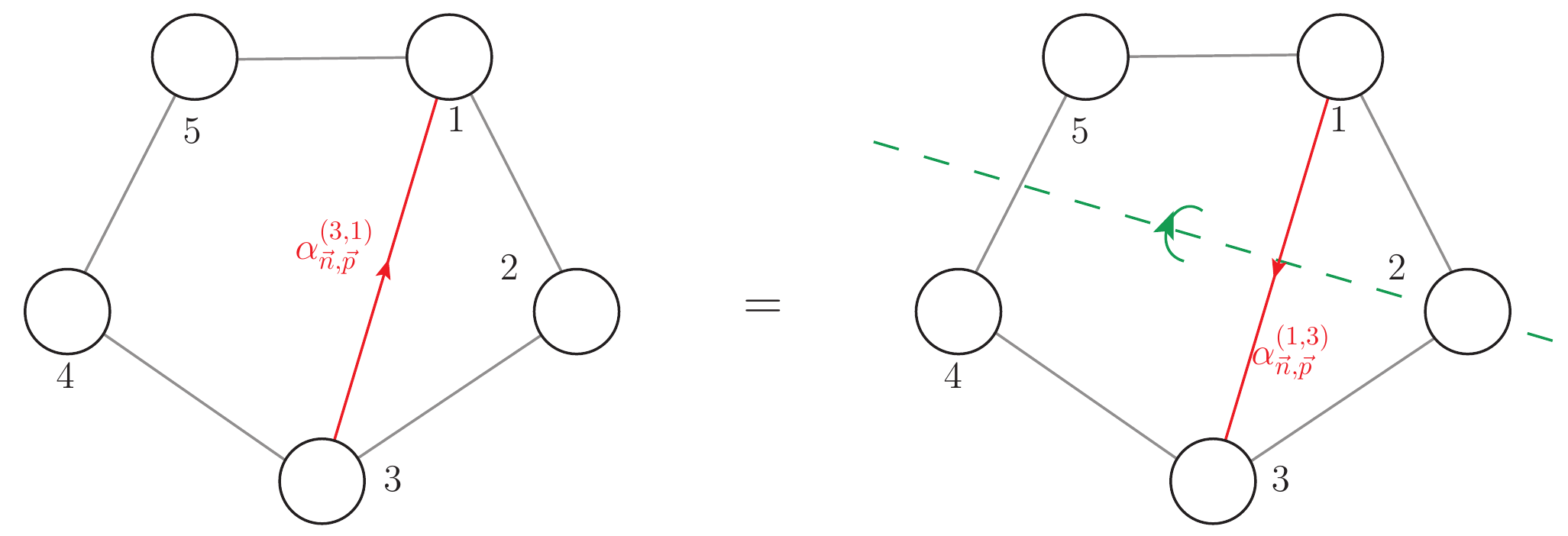}
        \subcaption{Graphical representation of \eqref{sym2}}\label{fig:2}
    \end{minipage}  
     \hfill
    \begin{minipage}[t]{.48\textwidth}
        \centering
        \includegraphics[width=\textwidth]{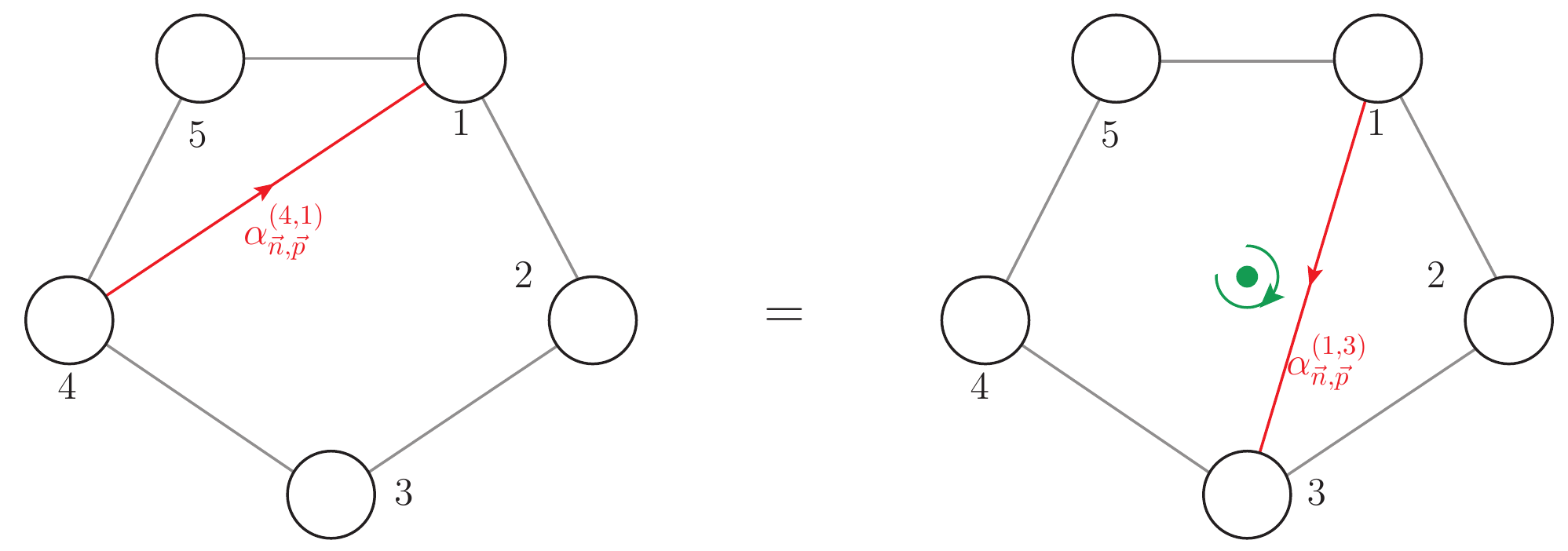}
        \subcaption{Graphical representation of \eqref{sym3}}\label{fig:3}
    \end{minipage}  
 \hfill
    \begin{minipage}[t]{.48\textwidth}
        \centering
        \includegraphics[width=\textwidth]{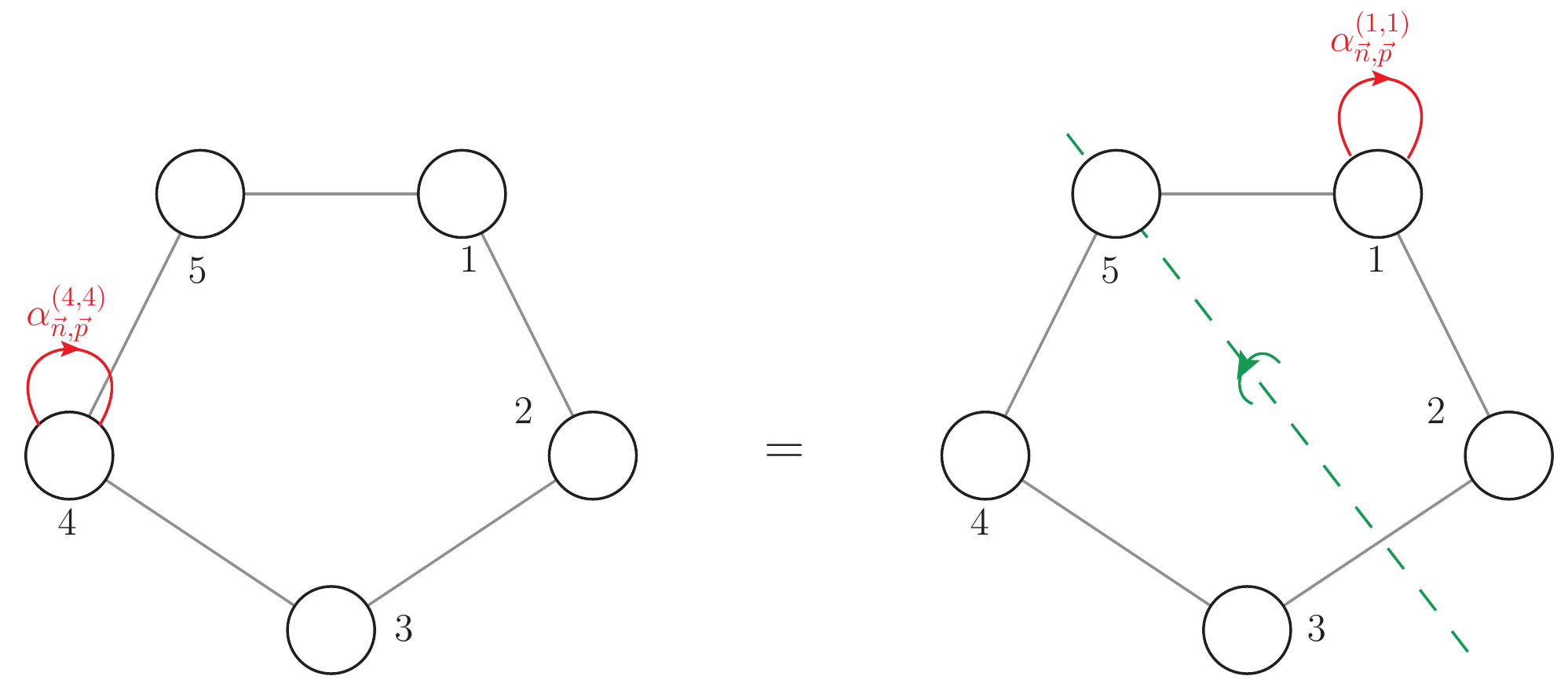}
        \subcaption{Graphical representation of \eqref{sym4}}\label{fig:4}
    \end{minipage}  
    \caption{A graphical representation for the symmetries between Gram-Schmidt coefficients $\alpha$. We consider a quiver with $q=5$ nodes. The coefficients $\alpha^{(I,J)}_{\vec n,\vec p}$ are represented by red arrows going from $I$ to $J$. The symmetries are encoded to the reflections and rotations around the axis in green.}
        \label{fig:simmetrie}
\end{figure}
In order to clarify the previous expressions, it is useful to produce an example.
Let's consider the theory $A_4$ on a quiver with $q=5$ nodes. Then it is possible to write the following equalities  
\begin{equation}\begin{split}
\alpha^{(1,4)}_{\vec n,\vec p}=\alpha^{(1,3)}_{\vec n,\vec p}\biggl|_{\substack{{\lambda}_{2}\leftrightarrow {\lambda}_{5}\\
{\lambda}_{3}\leftrightarrow {\lambda}_{4}}}\qquad\qquad\qquad\quad\!
\alpha^{(3,1)}_{\vec n,\vec p}=\alpha^{(1,3)}_{\vec n,\vec p}\biggl|_{\substack{{\lambda}_{1}\leftrightarrow {\lambda}_{3}\\
{\lambda}_{5}\leftrightarrow {\lambda}_{4}}}\\
\alpha^{(4,1)}_{\vec n,\vec p}=\alpha^{(1,3)}_{\vec n,\vec p}\biggl|_{\substack{{\lambda}_{1}\rightarrow {\lambda}_{4},\;{\lambda}_{4}\rightarrow {\lambda}_{2}\\
{\lambda}_{2}\rightarrow {\lambda}_{5},\;{\lambda}_{5}\rightarrow {\lambda}_{3}\\
{\lambda}_{3}\rightarrow {\lambda}_{1}}}\qquad\qquad
\alpha^{(4,4)}_{\vec n,\vec p}=\alpha^{(1,1)}_{\vec n,\vec p}\biggl|_{\substack{{\lambda}_{1}\leftrightarrow {\lambda}_{4}\\
{\lambda}_{2}\leftrightarrow {\lambda}_{3}}}~,
\end{split}\end{equation}
where we consider coefficients at a distance of 2 and zero nodes. 
Those relations can also be interpreted graphically as in figure \ref{fig:simmetrie}. Indeed, if we represent the coefficients $\alpha^{(I,J)}_{\vec n,\vec p}$ with arrows going from the node $I$ to the node $J$ of the quiver, performing the symmetries \eqref{sym1}, \eqref{sym2}, \eqref{sym3} and \eqref{sym4} means to rotate or reflect the figure in order to have the two arrows to completely overlap. This is equivalent to identify the axis of symmetry with respect to which to perform the reflection or rotation. Then it is easy to pick out the exchanges of the couplings corresponding to the symmetry of the labels of the  nodes.

Finally, we can extend the previous symmetries for the coefficients $\alpha$ to the full correlators.
Since the same considerations of before hold, we can write   
\begin{equation}\begin{split}\label{symG}
G^{(q,I,I+q-k)}_{\vec n}=G^{(q,I,I+k)}_{\vec n}\biggl|_{\substack{{\lambda}_{I+1}\leftrightarrow {\lambda}_{I+q-1}\\
{\lambda}_{I+2}\leftrightarrow {\lambda}_{I+q-2}\\ \dots}}\qquad\quad
G^{(q,I+k,I)}_{\vec n}=G^{(q,I,I+k)}_{\vec n}\biggl|_{\substack{{\lambda}_{I}\leftrightarrow {\lambda}_{I+k}\\
{\lambda}_{I-1}\leftrightarrow {\lambda}_{I+k-1}\\ \dots}}\\
G^{(q,I+q-k,I)}_{\vec n}=G^{(q,I,I+k)}_{\vec n}\biggl|_{\substack{{\lambda}_{I}\rightarrow {\lambda}_{I+q-k}\\
{\lambda}_{I+1}\rightarrow {\lambda}_{I+q-k+1}\\ \dots}}
\qquad
G^{(q,I+k,I+k)}_{\vec n}=G^{(q,I,I)}_{\vec n}\biggl|_{\substack{{\lambda}_{I}\leftrightarrow {\lambda}_{I+k}\\
{\lambda}_{I+1}\leftrightarrow {\lambda}_{I+k-1}\\ \dots}}~.
\end{split}\end{equation}
and at the orbifold point, correlators of two operators at the same distance are all the same for any node of the quiver
\begin{equation}\label{Gorb}
G^{(q,I,J)}_{\vec n}=G^{(q,1,d)}_{\vec n}
\end{equation}
with $d$ given by \eqref{dist}.

\section{Results for correlation functions}\label{sec:res}

In this section we present part of the main results for the chiral correlators \eqref{twopointdef}. We use the matrix-model representation for the function $G^{(q,I,J)}_{\vec n}$ as in \eqref{Gnormord} where the normal ordered operators are defined by \eqref{basis}. For example, if we consider a quiver with $q=2$ nodes and the normal ordered double-trace operator with dimension $n=5$ defined in \eqref{A132} we have 
\begin{align}
&G^{(2,1,2)}_{[3,2]}(\lambda_1,\lambda_2,N)\equiv\big\langle :\mathcal{O}^{(1)}_{[3,2]}: \;:\mathcal{O}^{(2)}_{[3,2]}:\big\rangle_2 
=\langle\mathcal{O}^{(1)}_{[3,2]}\mathcal{O}^{(2)}_{[3,2]}\rangle_2
+\alpha^{(1,1)}_{[3,2],[3]}\alpha^{(2,1)}_{[3,2],[3]}\langle\mathcal{O}^{(1)}_{[3]}\mathcal{O}^{(1)}_{[3]}\rangle_2\nonumber\\
&+\left(\alpha^{(1,2)}_{[3,2],[3]}\alpha^{(2,1)}_{[3,2],[3]}+\alpha^{(1,1)}_{[3,2],[3]}\alpha^{(2,2)}_{[3,2],[3]}\right)\langle\mathcal{O}^{(1)}_{[3]}\mathcal{O}^{(2)}_{[3]}\rangle_2
+\alpha^{(1,2)}_{[3,2],[3]}\alpha^{(2,2)}_{[3,2],[3]}\langle\mathcal{O}^{(2)}_{[3]}\mathcal{O}^{(2)}_{[3]}\rangle_2\\
&+\!\alpha^{(1,1)}_{[3,2],[3]}\langle\mathcal{O}^{(1)}_{[3]}\mathcal{O}^{(2)}_{[3,2]}\rangle_2
\!+\!\alpha^{(1,2)}_{[3,2],[3]}\langle\mathcal{O}^{(2)}_{[3]}\mathcal{O}^{(2)}_{[3,2]}\rangle_2
\!+\!\alpha^{(2,2)}_{[3,2],[3]}\langle\mathcal{O}^{(1)}_{[3,2]}\mathcal{O}^{(2)}_{[3]}\rangle_2
\!+\!\alpha^{(2,1)}_{[3,2],[3]}\langle\mathcal{O}^{(1)}_{[3]}\mathcal{O}^{(1)}_{[3,2]}\rangle_2\nonumber~,
\end{align}
where we consider the first operator in the node $I=1$ and the second in the node $J=2$ of the quiver.
The last step consists in the substitution of the Gram-Schmidt coefficients with their definition \eqref{coeffGS} in order to have the function $G$ written only in terms of two-point functions on the sphere matrix model of non-normal-ordered operator. Then using the perturbative techniques of section \eqref{sec:mmtech} together with the recursion relations \eqref{rrecursion}, we obtain the expansion for the $G$-function.

Since this procedure has a very well defined algorithmic structure, it is possible to implement it in a Mathematica package \cite{Preti2021maybe}. In principle with the package one is able to compute all the relevant quantities for this computation as: the functions $t_{n_1,n_2,...}$, the Gram-Schmidt coefficients $\alpha^{(I,J)}_{\vec n,\vec p}$ and obviously the correlators $G^{(I,J)}_{\vec n}$ for any choice of parameters. However, some steps of the algorithm are extremely expensive from a computational point of view. In particular, the computation of the Gram-Schmidt coefficients involve the inversion of a big matrix with a very long expansion on the couplings $\lambda_1,...,\lambda_q$ and $N$ for any element. This is obviously a bottleneck for the package. However, even with an average laptop, it was possible for us to generate a very large amount of data for correlators varying the dimension of the operators, the length of the quiver and the operator positions in it. We attach to this manuscript a notebook with a collection of finite and large $N$ results for single- and multi-trace operators with dimension between 2 and 8, on quivers with number of nodes from 1 to 6 up to very high order in the couplings. In the following, considering that those perturbative expansions are extremely cumbersome, we write only few selected results.

In this section we organise our results depending on the distance \eqref{dist} between the two operators on the quiver. In particular we consider operators belonging to the same node, to adjacent nodes and to nodes at higher distance. Depending on what is more convenient to show, we present results at finite or large $N$ and for some specific value of the parameters $q$, $I$, $J$ and $\vec n$. 

\subsection{Operators on the same node}

Let’s consider correlators of two normal-ordered operators in the same node of a quiver of length $q$. 
The smallest quiver we can study is the one with $q=1$ that corresponds to the SCQCD theory. 

\subsubsection{SCQCD}\label{subsec:QCD}

\begin{figure}[!t]
\begin{center}
\includegraphics[scale=0.7]{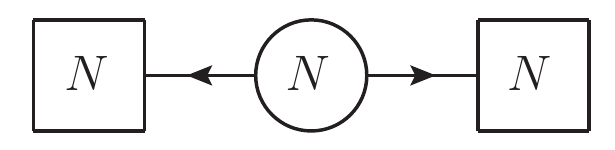}
\end{center}
\caption{Quiver representation of SCQCD. The square nodes now represent the flavour symmetry.}
\label{Fig:SCQCDquiver}
\end{figure}

The simplest theory we can study is  the SCQCD theory. It can be obtain from the $A_{q-1}$ theory by truncating the quiver to a single node, which defines a unique $SU(N)$ gauge theory with coupling $\lambda_1$. The residual interaction with the ungauged neighbouring nodes gives rise to the $2N$ hypermultiplets, which now transform in the fundamental representation of the unique gauge group, see Figure \ref{Fig:SCQCDquiver}. 

In the following we write the first few orders at finite $N$ of the two-point function of single- and multi-trace operators of dimensions from 2 to 6. The complete expansions can be found in the notebook "\texttt{QuiverCorrelators.nb}".

\noindent\underline{Operator of dimension $n=2$}
\small\begin{align}\label{scqcd2}
&G_{[2]}^{(1,1,1)}=\frac{\left(N^2-1\right)  {\color{blue}\lambda}_1^2}{2 (8\pi^2)^2N^2}-\frac{9\left(N^4-1\right) {\color{red}\zeta}_3 {\color{blue}\lambda}_1^4}{2 (8\pi^2)^4N^4}+\frac{15 \left(2 N^6-N^4-2 N^2+1\right) {\color{red}\zeta}_5 {\color{blue}\lambda}_1^5}{(8\pi^2)^5N^6}\nonumber\\
&+\frac{5 \left(N^2-1\right)  {\color{blue}\lambda}_1^6 \left(72 N^2 \left(N^4+3 N^2+2\right) {\color{red}\zeta}_3^2-35 \left(8 N^6+4 N^4-3 N^2+3\right) {\color{red}\zeta}_7\right)}{8(8\pi^2)^6 N^8}\\
&-\frac{45 \left(N^2\!-\!1\right) {\color{blue}\lambda}_1^7 \left(60 N^2 \left(2 N^6\!+\!7 N^4\!+\!2 N^2\!-\!3\right) {\color{red}\zeta}_3 {\color{red}\zeta}_5\!-7 \left(26 N^8\!+\!28 N^6\!-\!3 N^4\!+\!6 N^2\!-\!9\right) {\color{red}\zeta}_9\right)}{8 (8\pi^2)^7N^{10}}+O\!\left({\color{blue}\lambda}_1 ^{8}\right)\nonumber
\end{align}\normalsize

\noindent\underline{Operator of dimension $n=3$}
\small\begin{align}
&G_{[3]}^{(1,1,1)}\!=\!\!\frac{3\! \left(N^4\!-\!5 N^2\!+\!4\right) \!{\color{blue}\lambda}_1^3}{8 (8\pi^2)^3N^4}\!-\!\frac{\!27\! \left(N^6\!-\!2 N^4\!-\!11 N^2\!+\!12\right) \!{\color{red}\zeta}_3 {\color{blue}\lambda}_1^5}{8(8\pi^2)^5 N^6}\!+\!\frac{\!15 (N^2\!-\!4)\!\left(N^2\!-\!1\right)^2 \!\left(22 N^2\!+\!53\right)\! {\color{red}\zeta}_5 {\color{blue}\lambda}_1^6}{16 (8\pi^2)^6N^8}\nonumber\\
&+\frac{9 \left(N^4\!-\!5 N^2\!+\!4\right) {\color{blue}\lambda}_1^7 \left(36 N^2 \left(3 N^4\!+\!20 N^2\!+\!37\right) {\color{red}\zeta}_3^2-35 \left(11 N^6\!+\!12 N^4\!-\!16 N^2\!+\!29\right) {\color{red}\zeta}_7\right)}{32 (8\pi^2)^7N^{10}}\\
&-\!\frac{\!27\!\left(\!N^2\!\!-\!4\right) \!\left(\!N^2\!\!-\!1\right)^2\! {\color{blue}\lambda}_1^8 \!\left(20 N^2 \!\left(50 N^4\!\!+\!373 N^2\!\!+\!633\right) \!{\color{red}\zeta}_3 {\color{red}\zeta}_5\!-\!7 \!\left(192 N^6\!\!+\!552 N^4\!\!+\!515 N^2\!\!+\!645\right) \!{\color{red}\zeta}_9\right)\!\!}{64 (8\pi^2)^8N^{12}}\!+\!O\!\!\left({\color{blue}\lambda}_1^{9}\right)\nonumber
\end{align}\normalsize

\noindent\underline{Operators of dimension $n=4$}
\small\begin{align}
&G_{[4]}^{(1,1,1)}\!=\frac{\left(N^2\!-\!1\right) \left(N^4\!-\!6 N^2\!+\!18\right) {\color{blue}\lambda}_1^4}{4 (8\pi^2)^4N^6}\!-\!\frac{3\left(N^8\!+\!N^6\!-\!20 N^4\!+\!99 N^2\!-\!81\right) {\color{red}\zeta}_3 {\color{blue}\lambda}_1^6}{(8\pi^2)^6N^8}\nonumber\\
&\qquad\qquad\qquad\qquad\qquad\qquad+\frac{5 \left(N^2\!-\!1\right) \left(8 N^8\!+\!19 N^6\!-\!9 N^4\!+\!432 N^2\!-\!648\right) {\color{red}\zeta}_5 {\color{blue}\lambda}_1^7}{2 (8\pi^2)^7N^{10}}+O\left({\color{blue}\lambda}_1 ^{8}\right)\\
&G_{[2,2]}^{(1,1,1)}\!=\!\frac{\left(N^4\!-\!1\right)  {\color{blue}\lambda}_1^4}{2 (8\pi^2)^4N^4}\!-\!\frac{9\left(N^2\!-\!1\right) \left(N^4\!+\!4 N^2\!+\!3\right) {\color{red}\zeta}_3 {\color{blue}\lambda}_1^6}{(8\pi^2)^6N^6}+\frac{30 \left(N^4\!-\!1\right) \left(N^2\!+\!6\right) \left(2 N^2\!-\!1\right) {\color{red}\zeta}_5 {\color{blue}\lambda}_1^7}{(8\pi^2)^7N^8}\nonumber\\
&+\!\frac{\!\left(\!N^2\!\!-\!1\right) \! {\color{blue}\lambda}_1^8 [18 N^2\! \!\left(N^2\!+\!1\right) \!\!\left(29 N^4\!\!+\!204 N^2\!+\!355\right) \!{\color{red}\zeta}_3^2\!-\!\!175\! \left(N^2\!+\!10\right) \!\!\left(8 N^6\!\!+\!4 N^4\!-\!3 N^2\!+\!3\right) \!{\color{red}\zeta}_7]}{4 (8\pi^2)^8N^{10}}\!+\!O\!\left({\color{blue}\lambda}_1 ^{9}\right)
\end{align}\normalsize

\noindent\underline{Operators of dimension $n=5$}
\small\begin{align}
G_{[5]}^{(1,1,1)}\!=&
\frac{5 \left(N^4+24\right) \left(N^4-5 N^2+4\right)  {\color{blue}\lambda}_1^5}{32 (8\pi^2)^5N^8}-\frac{75 \left(N^4-5 N^2+4\right) \left(N^6+11 N^4+144\right) {\color{red}\zeta}_3 {\color{blue}\lambda}_1^7}{32 (8\pi^2)^7 N^{10}}\nonumber\\
&+\frac{125 \left(N^4-5 N^2+4\right) \left(8 N^8+103 N^6+129 N^4+264 N^2-1464\right){\color{red}\zeta}_5 {\color{blue}\lambda}_1^8}{64 (8\pi^2)^8N^{12}}+O\left({\color{blue}\lambda}_1 ^{9}\right)\\
G_{[3,2]}^{(1,1,1)}\!=&
\frac{3 \left(N^6-21 N^2+20\right) {\color{blue}\lambda}_1^5}{16(8\pi^2)^5 N^6}-\frac{27  \left(N^2+5\right)^2 \left(N^4-5 N^2+4\right) {\color{red}\zeta}_3 {\color{blue}\lambda}_1^7}{8 (8\pi^2)^7N^8}\nonumber\\
&\qquad\qquad\qquad\;\;\,+\frac{15 \left(N^2-4\right) \left(N^2-1\right)^2 \left(46 N^4+703 N^2+1525\right) {\color{red}\zeta}_5 {\color{blue}\lambda}_1^8}{32 (8\pi^2)^8N^{10}}+O\left({\color{blue}\lambda}_1 ^{9}\right)
\end{align}\normalsize

\noindent\underline{Operators of dimension $n=6$}
\small\begin{align}
G_{[6]}^{(1,1,1)}\!=&
\frac{3 \left(N^2\!\!-\!1\right)\! \left(N^8\!\!+\!6 N^6\!\!-\!60 N^4\!\!+\!600\right){\color{blue}\lambda}_1^6}{32  (8\pi^2)^6N^{10}}\nonumber\\
&\qquad\qquad-\frac{27 \left(N^2\!\!-\!1\right)\! \left(N^{10}\!\!+\!20 N^8\!\!-\!76 N^6\!\!-\!50 N^4\!\!-\!300 N^2\!\!+\!4500\right) \!{\color{red}\zeta}_3 {\color{blue}\lambda}_1^8}{16  (8\pi^2)^8N^{12}}\!+\!O\!\left({\color{blue}\lambda}_1^{9}\right)\\
G_{[4,2]}^{(1,1,1)}\!=&
\frac{\left(N^2-1\right) \left(N^6+17 N^4-72 N^2+162\right) {\color{blue}\lambda}_1^6}{8 (8\pi^2)^6N^8}\nonumber\\
&\qquad\qquad-\frac{3 \left(N^2-1\right) \left(7 N^8+210 N^6+227 N^4-2286 N^2+7290\right) {\color{red}\zeta}_3{\color{blue}\lambda}_1^8}{8 (8\pi^2)^8N^{10}}+O\left({\color{blue}\lambda}_1^{9}\right)\\
G_{[3,3]}^{(1,1,1)}\!=&
\frac{9\! \left(\!N^2\!\!-\!4\right)^2\!\! \left(N^2\!\!-\!1\right)\! \!\left(N^2\!\!+\!8\right) \! {\color{blue}\lambda}_1^6}{32(8\pi^2)^6 N^8}\!-\!\frac{\!81\!\left(N^4\!\!-\!5 N^2\!\!+\!4\right)\!\! \left(N^6\!\!+\!10 N^4\!\!-\!2 N^2\!\!-\!240\right) \!{\color{red}\zeta}_3{\color{blue}\lambda}_1^8\!}{16 (8\pi^2)^8N^{10}}\!+\!O\!\!\left(\!{\color{blue}\lambda}_1^{9}\right)\\
G_{[2,2,2]}^{(1,1,1)}\!=&
\frac{3 \left(N^2-1\right) \left(N^4+4 N^2+3\right)  {\color{blue}\lambda}_1^6}{4 (8\pi^2)^6N^6}-\frac{81(N^4-1)\left(N^2+3\right) \left(N^2+5\right) {\color{red}\zeta}_3 {\color{blue}\lambda}_1^8}{4 (8\pi^2)^8N^8}+O\left({\color{blue}\lambda}_1^{9}\right)~.
\end{align}\normalsize

\noindent
The number of two-point function for a fixed dimension is given by the number of partitions presented in table \ref{tab:tabella}. 
From the previous expansions, it is easy to derive a general pattern in term of the coupling and transcendentality. Indeed we have
\begin{equation}\label{patt1}
G_{\vec n}^{(1,1,1)}=c_1{\color{blue}\lambda}_1^n+ 0\times {\color{blue}\lambda}_1^{n+1}+
c_3 {\color{red}\zeta}_3{\color{blue}\lambda}_1^{n+2}+
c_4 {\color{red}\zeta}_5{\color{blue}\lambda}_1^{n+3}+
(c_5 {\color{red}\zeta}_3^2+c_5 {\color{red}\zeta}_7){\color{blue}\lambda}_1^{n+4}+\dots
\end{equation}  
where the coefficients $c_k$ are functions of $N$. Notice that the 1-loop order is proportional to the 1-loop coefficient of the Beta function, hence $O(\lambda_1^{n+1})$ always vanishes in conformal theories. 
In general, the expansion \eqref{patt1} includes an infinite series of corrections to the $\mathcal{N}=4$ correlator with terms proportional to products of Riemann zeta functions. We can unambiguously isolate any term that it could be compared with a similar term that should appear in a direct Feynman diagram calculation.
It's interesting to notice that the large $N$ limit does not suppress any coefficient in the expansion \eqref{patt1}, in other words the transcendentality is conserved by the large $N$ limit. Indeed we have
\small\begin{align}
G_{[2]}^{(1,1,1)}\!&=\frac{2{\color{blue}\lambda}_1^2}{(4\pi)^4}\biggl[
1-\frac{9}{4} {\color{red}\zeta}_3 \frac{{\color{blue}\lambda}_1^2}{(2\pi)^4}+\frac{15}{2} {\color{red}\zeta}_5 \frac{{\color{blue}\lambda}_1^3}{(2\pi)^6}+\frac{5}{8}  \left(9 {\color{red}\zeta}_3^2-35 {\color{red}\zeta}_7\right) \frac{{\color{blue}\lambda}_1^4}{(2\pi)^8}\nonumber\\
&+\frac{45}{64}  \left(91 {\color{red}\zeta}_9-60 {\color{red}\zeta}_3 {\color{red}\zeta}_5\right) \frac{{\color{blue}\lambda}_1^5}{(2\pi)^{10}}-\frac{21}{512} \left(360 {\color{red}\zeta}_3^3-3360 {\color{red}\zeta}_7 {\color{red}\zeta}_3-1900 {\color{red}\zeta}_5^2+4697 {\color{red}\zeta}_{11}\right) \frac{{\color{blue}\lambda}_1^6}{(2\pi)^{12}}\nonumber\\
&+\frac{7}{256}  \left(20 \left(324 {\color{red}\zeta}_5 {\color{red}\zeta}_3^2-819 {\color{red}\zeta}_9 {\color{red}\zeta}_3-917 {\color{red}\zeta}_5 {\color{red}\zeta}_7\right)+21879 {\color{red}\zeta}_{13}\right) \frac{{\color{blue}\lambda}_1^7}{(2\pi)^{14}}\biggl]+O\left({\color{blue}\lambda}_1^{10}\right)~,\label{largqcd1}
\end{align}\normalsize
\small\begin{align}
G_{[3]}^{(1,1,1)}\!&=\frac{3{\color{blue}\lambda}_1^3}{(4\pi)^6}\biggl[
1-\frac{9}{4} {\color{red}\zeta}_3 \frac{{\color{blue}\lambda}_1^2}{(2\pi)^4}+\frac{55}{8} {\color{red}\zeta}_5 \frac{{\color{blue}\lambda}_1^3}{(2\pi)^6}+\frac{3}{64}  \left(108 {\color{red}\zeta}_3^2-385 {\color{red}\zeta}_7\right) \frac{{\color{blue}\lambda}_1^4}{(2\pi)^8}\nonumber\\
&+\frac{9}{32}\!  \left(168 {\color{red}\zeta}_9\!-\!125 {\color{red}\zeta}_3 {\color{red}\zeta}_5\right) \!\frac{{\color{blue}\lambda}_1^5}{(2\pi)^{10}}\!-\frac{1}{512}\! \left(6048 {\color{red}\zeta}_3^3\!-53550 {\color{red}\zeta}_7 {\color{red}\zeta}_3\!-31100 {\color{red}\zeta}_5^2\!+65373 {\color{red}\zeta}_{11}\right)\! \frac{{\color{blue}\lambda}_1^6}{(2\pi)^{12}}\nonumber\\
&+\frac{9}{1024}  \left(560 {\color{red}\zeta}_5 \left(27 {\color{red}\zeta}_3^2-73 {\color{red}\zeta}_7\right)-34986 {\color{red}\zeta}_3 {\color{red}\zeta}_9+40755 {\color{red}\zeta}_{13}\right) \frac{{\color{blue}\lambda}_1^7}{(2\pi)^{14}}\biggl]+O\left({\color{blue}\lambda}_1^{11}\right)~,\\
G_{[4]}^{(1,1,1)}\!&=\frac{4{\color{blue}\lambda}_1^4}{(4\pi)^8}\biggl[
1-3  {\color{red}\zeta}_3 \frac{{\color{blue}\lambda}_1^2}{(2\pi)^4}+10  {\color{red}\zeta}_5 \frac{{\color{blue}\lambda}_1^3}{(2\pi)^6}+\frac{7}{64}  \left(72 {\color{red}\zeta}_3^2-265 {\color{red}\zeta}_7\right) \frac{{\color{blue}\lambda}_1^4}{(2\pi)^8}\nonumber\\
&+\frac{3}{32} \left(903 {\color{red}\zeta}_9-620 {\color{red}\zeta}_3 {\color{red}\zeta}_5\right) \frac{{\color{blue}\lambda}_1^5}{(2\pi)^{10}}+\frac{3}{64} \left(-432 {\color{red}\zeta}_3^3+3990 {\color{red}\zeta}_7 {\color{red}\zeta}_3+2250 {\color{red}\zeta}_5^2-5467 {\color{red}\zeta}_{11}\right) \frac{{\color{blue}\lambda}_1^6}{(2\pi)^{12}}\nonumber\\
&+\frac{3}{64}  \left(5100 {\color{red}\zeta}_5 {\color{red}\zeta}_3^2-12873 {\color{red}\zeta}_9 {\color{red}\zeta}_3+55 \left(312 {\color{red}\zeta}_{13}-259 {\color{red}\zeta}_5 {\color{red}\zeta}_7\right)\right) \frac{{\color{blue}\lambda}_1^7}{(2\pi)^{14}}\biggl]+O\left({\color{blue}\lambda}_1^{12}\right)~,\\
G_{[5]}^{(1,1,1)}\!&=\frac{5{\color{blue}\lambda}_1^5}{(4\pi)^{10}}\biggl[
1-\frac{15}{4} {\color{red}\zeta}_3 \frac{{\color{blue}\lambda}_1^2}{(2\pi)^4}+\frac{25}{2}  {\color{red}\zeta}_5 \frac{{\color{blue}\lambda}_1^3}{(2\pi)^6}+\frac{5}{64}  \left(144 {\color{red}\zeta}_3^2-455 {\color{red}\zeta}_7\right) \frac{{\color{blue}\lambda}_1^4}{(2\pi)^8}\nonumber\\
&+\frac{21}{32}\left(153 {\color{red}\zeta}_9\!-\!125 {\color{red}\zeta}_3 {\color{red}\zeta}_5\right)\! \frac{{\color{blue}\lambda}_1^5}{(2\pi)^{10}}-\frac{5}{512}  \left(3240 {\color{red}\zeta}_3^3-26250 {\color{red}\zeta}_7 {\color{red}\zeta}_3-15100 {\color{red}\zeta}_5^2+29799 {\color{red}\zeta}_{11}\right) \!\frac{{\color{blue}\lambda}_1^6}{(2\pi)^{12}}\nonumber\\
&+\frac{5}{1024} \left(220 {\color{red}\zeta}_5 \left(342 {\color{red}\zeta}_3^2-847 {\color{red}\zeta}_7\right)-162414 {\color{red}\zeta}_3 {\color{red}\zeta}_9+178035 {\color{red}\zeta}_{13}\right) \frac{{\color{blue}\lambda}_1^7}{(2\pi)^{14}}\biggl]+O\left({\color{blue}\lambda}_1^{13}\right)~,\\
G_{[6]}^{(1,1,1)}\!&=\frac{6{\color{blue}\lambda}_1^6}{(4\pi)^{12}}\biggl[
1-\frac{9}{2} {\color{red}\zeta}_3 \frac{{\color{blue}\lambda}_1^2}{(2\pi)^4}+15  {\color{red}\zeta}_5 \frac{{\color{blue}\lambda}_1^3}{(2\pi)^6}+\frac{3}{32}\left(162 {\color{red}\zeta}_3^2-455 {\color{red}\zeta}_7\right) \frac{{\color{blue}\lambda}_1^4}{(2\pi)^8}\nonumber\\
&+\frac{27}{64} \left(287 {\color{red}\zeta}_9\!-\!260 {\color{red}\zeta}_3 {\color{red}\zeta}_5\right) \!\frac{{\color{blue}\lambda}_1^5}{(2\pi)^{10}}-\frac{3}{512}  \left(7920 {\color{red}\zeta}_3^3-57960 {\color{red}\zeta}_7 {\color{red}\zeta}_3-33400 {\color{red}\zeta}_5^2+60599 {\color{red}\zeta}_{11}\right) \!\frac{{\color{blue}\lambda}_1^6}{(2\pi)^{12}}\nonumber\\
&+\frac{9}{256} \left(70 {\color{red}\zeta}_5 \left(216 {\color{red}\zeta}_3^2-487 {\color{red}\zeta}_7\right)-29778 {\color{red}\zeta}_3 {\color{red}\zeta}_9+30745 {\color{red}\zeta}_{13}\right) \frac{{\color{blue}\lambda}_1^7}{(2\pi)^{14}}\biggl]+O\left({\color{blue}\lambda}_1^{14}\right)~.\label{largqcd2}
\end{align}\normalsize
In the previous list of results, we did not consider the multi-trace operators. Indeed in the large $N$ limit, correlators of multi-trace operators are factorized in products of correlators of the corresponding single-traces as follows
\begin{equation}\label{factorization}
\big\langle :\mathcal{O}^{(I)}_{[n_1,n_2,...,n_t]}: \;:\mathcal{O}^{(J)}_{[n_1,n_2,...,n_t]}:\big\rangle_q =
\prod_{k=1}^t\big\langle :\mathcal{O}^{(I)}_{[n_k]}: \;:\mathcal{O}^{(J)}_{[n_k]}:\big\rangle_q ~.
\end{equation} 
This statement is totally general and we tested it in several cases at very high order.
The large $N$ expansions above contain and extend the ones found in \cite{Rodriguez-Gomez:2016ijh,Pini:2017ouj} to higher orders and considering all the terms proportional to a product of Riemann zetas.

\subsubsection{Effective couplings for $\mathcal{N}=2$ SCQCD}\label{subsub:QCD}

In \cite{Pomoni:2013poa}, it was observed that in the purely gluonic $SU(2,1|2)$ sub-sector of any planar $\mathcal{N}=2$ SYM theory, observables can be computed replacing the coupling of the $\mathcal{N}=4$ SYM result. In particular, it was shown in \cite{Mitev:2014yba} that the VEV of the supersymmetric circular Wilson loop in SCQCD resembles the  one in $\mathcal{N}=4$ SYM after an effective coupling replacement. This function of the coupling was computed in \cite{Mitev:2014yba,Fraser:2015xha,Mitev:2015oty}. However in \cite{Pini:2017ouj}, it was shown that in the case of correlators of local operators with even dimension in the planar SCQCD, the terms linear on the Riemann zetas of the effective coupling is not universal but it depends on the dimension of the considered operators. With the recursive method presented above and the large $N$ results given in formulas from \eqref{largqcd1} to \eqref{largqcd2}, we are able to compute the effective couplings for even and odd dimensions and also for terms proportional to products of ${\color{red}\zeta}$-functions. Indeed we have
\begin{align}
{\lambda}^{\text{eff}}_{[2]}={\color{blue}\lambda}_1\biggl[&1-\frac{9{\color{red}\zeta}_{3} {\color{blue}\lambda}_1^2}{8(4\pi^2)^2}+\frac{15  {\color{red}\zeta}_{5} {\color{blue}\lambda}_1^3}{4(4\pi^2)^3}+\frac{\left(279 {\color{red}\zeta}_{3}^2-1400 {\color{red}\zeta}_{7}\right) {\color{blue}\lambda}_1^4}{128(4\pi^2)^4}  -\frac{45 \left(48 {\color{red}\zeta}_{3} {\color{red}\zeta}_{5}-91 {\color{red}\zeta}_{9}\right) {\color{blue}\lambda}_1^5}{128(4\pi^2)^5}\nonumber\\
&\qquad-\frac{3 \left(1683 {\color{red}\zeta}_{3}^3-19320 {\color{red}\zeta}_{7} {\color{red}\zeta}_{3}-10900 {\color{red}\zeta}_{5}^2+32879 {\color{red}\zeta}_{11}\right) {\color{blue}\lambda}_1^6}{1024(4\pi^2)^6}
+O\left({\color{blue}\lambda}_1^{7}\right)\biggl]~,\\
{\lambda}^{\text{eff}}_{[3]}={\color{blue}\lambda}_1\biggl[&
1-\frac{3{\color{red}\zeta}_{3} {\color{blue}\lambda}_1^2}{4(4\pi^2)^2} +\frac{55{\color{red}\zeta}_{5} {\color{blue}\lambda}_1^3}{24(4\pi^2)^3}  +\frac{\left(72 {\color{red}\zeta}_{3}^2-385 {\color{red}\zeta}_{7}\right) {\color{blue}\lambda}_1^4}{64(4\pi^2)^4}  +\frac{\left(504 {\color{red}\zeta}_{9}-265 {\color{red}\zeta}_{3} {\color{red}\zeta}_{5}\right) {\color{blue}\lambda}_1^5}{32(4\pi^2)^5} \nonumber\\
&\qquad-\frac{ \left(9720 {\color{red}\zeta}_{3}^3-119070 {\color{red}\zeta}_{7} {\color{red}\zeta}_{3}-69100 {\color{red}\zeta}_{5}^2+196119 {\color{red}\zeta}_{11}\right) {\color{blue}\lambda}_1^6}{4608(4\pi^2)^6}
+O\left({\color{blue}\lambda}_1^{7}\right)\biggl]~,\\
{\lambda}^{\text{eff}}_{[4]}={\color{blue}\lambda}_1\biggl[&
1-\frac{3{\color{red}\zeta}_{3} {\color{blue}\lambda}_1^2}{4(4\pi^2)^2} +\frac{5{\color{red}\zeta}_{5} {\color{blue}\lambda}_1^3}{2(4\pi^2)^3}+\frac{ \left(288 {\color{red}\zeta}_{3}^2-1855 {\color{red}\zeta}_{7}\right) {\color{blue}\lambda}_1^4}{256(4\pi^2)^4} -\frac{3\left(380 {\color{red}\zeta}_{3} {\color{red}\zeta}_{5}-903 {\color{red}\zeta}_{9}\right) {\color{blue}\lambda}_1^5}{128(4\pi^2)^5} \nonumber\\
&\qquad-\frac{3  \left(720 {\color{red}\zeta}_{3}^3-10395 {\color{red}\zeta}_{7} {\color{red}\zeta}_{3}-5800 {\color{red}\zeta}_{5}^2+21868 {\color{red}\zeta}_{11}\right) {\color{blue}\lambda}_1^6}{1024(4\pi^2)^6}
+O\left({\color{blue}\lambda}_1^{7}\right)\biggl]~,\\
{\lambda}^{\text{eff}}_{[5]}={\color{blue}\lambda}_1\biggl[&
1-\frac{3{\color{red}\zeta}_{3} {\color{blue}\lambda}_1^2}{4(4\pi^2)^2} +\frac{5{\color{red}\zeta}_{5} {\color{blue}\lambda}_1^3}{2(4\pi^2)^3}+\frac{\left(72 {\color{red}\zeta}_{3}^2-455 {\color{red}\zeta}_{7}\right) {\color{blue}\lambda}_1^4}{64(4\pi^2)^4}  +\frac{3\left(1071 {\color{red}\zeta}_{9}-475 {\color{red}\zeta}_{3} {\color{red}\zeta}_{5}\right) {\color{blue}\lambda}_1^5}{160(4\pi^2)^5}  \nonumber\\
&\qquad-\frac{3\left(360 {\color{red}\zeta}_{3}^3-5110 {\color{red}\zeta}_{7} {\color{red}\zeta}_{3}-2900 {\color{red}\zeta}_{5}^2+9933 {\color{red}\zeta}_{11}\right) {\color{blue}\lambda}_1^6}{512(4\pi^2)^6}
+O\left({\color{blue}\lambda}_1^{7}\right)\biggl]~,\\
{\lambda}^{\text{eff}}_{[6]}={\color{blue}\lambda}_1\biggl[&
1-\frac{3{\color{red}\zeta}_{3} {\color{blue}\lambda}_1^2}{4(4\pi^2)^2}+\frac{5{\color{red}\zeta}_{5} {\color{blue}\lambda}_1^3}{2(4\pi^2)^3}+\frac{\left(72 {\color{red}\zeta}_{3}^2-455 {\color{red}\zeta}_{7}\right) {\color{blue}\lambda}_1^4}{64(4\pi^2)^4}  -\frac{3\left(380 {\color{red}\zeta}_{3} {\color{red}\zeta}_{5}-861 {\color{red}\zeta}_{9}\right) {\color{blue}\lambda}_1^5}{128(4\pi^2)^5} \nonumber\\
&\qquad-\frac{ \left(2160 {\color{red}\zeta}_{3}^3-30660 {\color{red}\zeta}_{7} {\color{red}\zeta}_{3}-17400 {\color{red}\zeta}_{5}^2+60599 {\color{red}\zeta}_{11}\right) {\color{blue}\lambda}_1^6}{1024(4\pi^2)^6}
+O\left({\color{blue}\lambda}_1^{7}\right)\biggl]~,
\end{align}
where, at any order, the highest transcendentality term is in agreement with the results of \cite{Pini:2017ouj}.

As it was noticed in \cite{Pini:2017ouj} for the linear terms in the ${\zeta}$-functions,  ${\lambda}^{\text{eff}}_{[n+1]}-{\lambda}^{\text{eff}}_{[n]}\sim O({\color{red}\zeta}_{2n-1}{\color{blue}\lambda}_1^n)$. It's interesting to notice that this relation holds also for non-linear terms in the Riemann zetas, \textit{i.e.} in the difference only the highest transcendentality terms appear. Then we can conclude that
\begin{equation}\begin{split}
\big\langle {O}^{(1)}_{[n]} \;\bar{O}^{(1)}_{[n]}\big\rangle_1({\color{blue}\lambda}_1)=&
\big\langle {O}^{(1)}_{[n]} \;\bar{O}^{(1)}_{[n]}\big\rangle_{\mathcal{N}=4}({\lambda}_{\text{eff}})[1+O({\color{red}\zeta}_{2n-1}{\color{blue}\lambda}_1^n)]\\
=&\frac{n{\lambda}_{\text{eff}}^n}{(4\pi)^{2n}}[1+O({\color{red}\zeta}_{2n-1}{\color{blue}\lambda}_1^n)]~,
\end{split}\end{equation}
with the effective coupling given by
\begin{equation}\begin{split}
{\lambda}_{\text{eff}}={\color{blue}\lambda}_1\biggl[&1-\frac{3{\color{red}\zeta}_3{\color{blue}\lambda}_1^2}{4(4\pi^2)^2}+\frac{5 {\color{red}\zeta}_5{\color{blue}\lambda}_1^3}{2(4\pi^2)^3} +\frac{\left(72 {\color{red}\zeta}_3^2-455 {\color{red}\zeta}_7\right){\color{blue}\lambda}_1^4}{64(4\pi^2)^4}  -\frac{3\left(380 {\color{red}\zeta}_3 {\color{red}\zeta}_5-861 {\color{red}\zeta}_9\right){\color{blue}\lambda}_1^5}{128(4\pi^2)^5}\\
&-\frac{3 \left(360 {\color{red}\zeta}_3^3-5110 {\color{red}\zeta}_7 {\color{red}\zeta}_3-2900 {\color{red}\zeta}_5^2+10087 {\color{red}\zeta}_{11}\right){\color{blue}\lambda}_1^6}{512(4\pi^2)^6}\\
&+\frac{\left(13860 {\color{red}\zeta}_5 {\color{red}\zeta}_3^2-50589 {\color{red}\zeta}_9 {\color{red}\zeta}_3-56770 {\color{red}\zeta}_5 {\color{red}\zeta}_7+91806 {\color{red}\zeta}_{13}\right){\color{blue}\lambda}_1^7}{512(4\pi^2)^7}  +O\left({\color{blue}\lambda}_1^{8}\right)\biggl]~,
\end{split}\end{equation}
where to compute ${\lambda}_{\text{eff}}$ we used also the results for the correlators of single-trace operators of dimensions 7 and 8.

\subsubsection{Results for $A_{q-1}$ theories with $q\geq 2$}\label{subsec:SameNode}

\begin{figure}[!t]
\begin{center}
\includegraphics[scale=0.6]{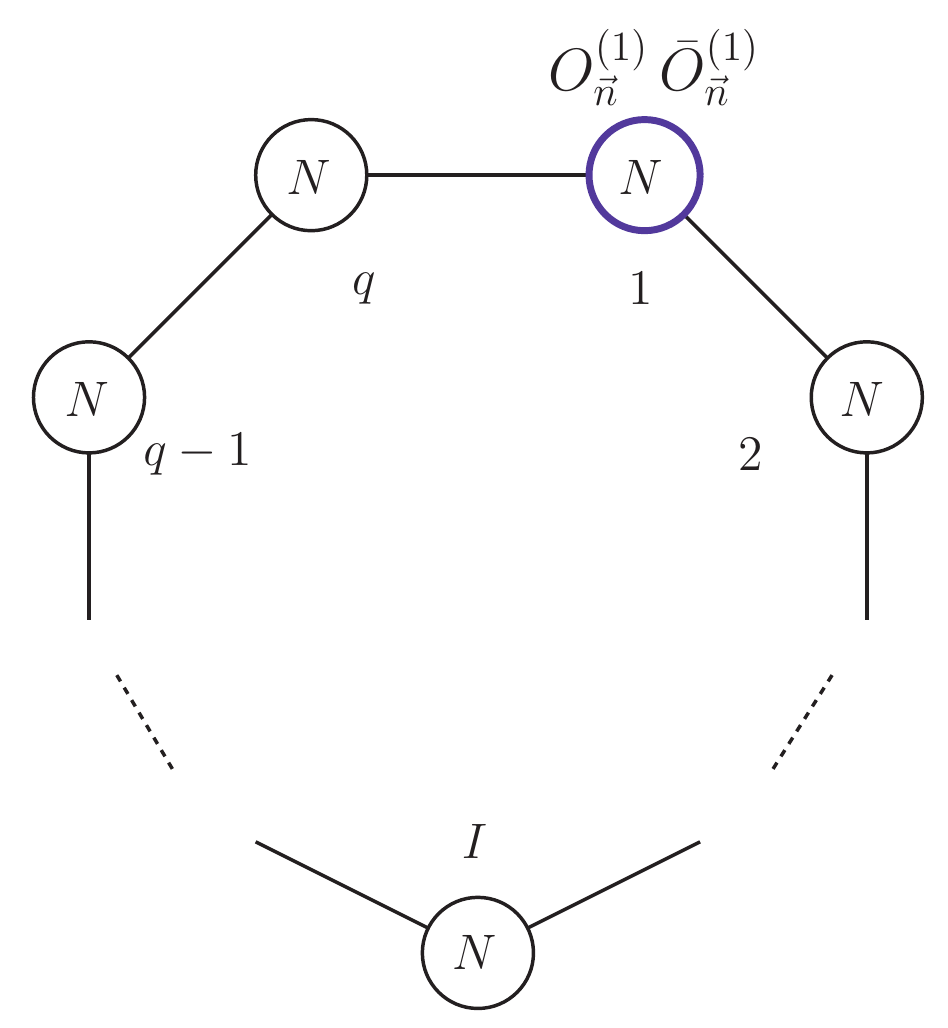}
\end{center}
\caption{The circular quiver with $q$ nodes associated to the theory $A_{q-1}$. The operators $O_{\vec n}^{(1)}$ and $\bar O_{\vec n}^{(1)}$ belong to the same node, highlighted in blue.}
\label{Fig:quiverSameNode}
\end{figure}
We now move to general $A_{q-1}$ theories, starting from the case where the two normal-ordered operators in the same node of a quiver of length $q$.
Taking into account of the symmetries of correlators presented in \eqref{symG}, we can restrict our computation in the case in which both operators are in the node $I=1$ without loss of generality, see Figure \ref{Fig:quiverSameNode}.
 
With the same techniques explained above, it is possible to compute the correlators at finite $N$ for several scaling dimensions and number of nodes $q$. Since those expressions are extremely long, it is useful to give an example in the large $N$ limit 
\small\begin{align}
&G_{[2]}^{(5,1,1)}\!=\frac{2{\color{blue}\lambda}_1^2}{(4\pi)^4}\biggl[
1\!-\!\frac{3{\color{red}\zeta}_{3} {\color{blue}\lambda}_1 \left(3 {\color{blue}\lambda}_1\!-\!{\color{blue}\lambda}_2\!-\!{\color{blue}\lambda}_5\right)}{4(2\pi)^4} \!+\!\frac{5{\color{red}\zeta}_{5} {\color{blue}\lambda}_1 \left(12 {\color{blue}\lambda}_1^2\!-\!3 \left({\color{blue}\lambda}_2\!+\!{\color{blue}\lambda}_5\right) {\color{blue}\lambda}_1\!-\!{\color{blue}\lambda}_2^2\!-\!{\color{blue}\lambda}_5^2\right)}{8(2\pi)^6}  \biggl]\!+O\!\left({\color{blue}\lambda}_1^{6}\right)\,,\label{4.26}\\
&G_{[3]}^{(5,1,1)}\!\!=\!\!\frac{3{\color{blue}\lambda}_1^3}{(4\pi)^6}\biggl[
1\!-\!\frac{9{\color{red}\zeta}_{3} {\color{blue}\lambda}_1 \!\left(2 {\color{blue}\lambda}_1\!-\!{\color{blue}\lambda}_2\!-\!{\color{blue}\lambda}_5\right)}{8(2\pi)^4}\!+\!\frac{\!5 {\color{red}\zeta}_{5} {\color{blue}\lambda}_1 \!\left(22 {\color{blue}\lambda}_1^2\!-\!9 \left({\color{blue}\lambda}_2\!+\!{\color{blue}\lambda}_5\right) \!{\color{blue}\lambda}_1\!-\!3 \left({\color{blue}\lambda}_2^2\!+\!{\color{blue}\lambda}_5^2\right)\!\right)\!}{16(2\pi)^6} \biggl]\!+O\!\left({\color{blue}\lambda}_1^{7}\right)\,\!,\\
&G_{[4]}^{(5,1,1)}\!=\frac{4{\color{blue}\lambda}_1^4}{(4\pi)^8}\biggl[
1\!-\!\frac{3{\color{red}\zeta}_{3} {\color{blue}\lambda}_1 \left(2 {\color{blue}\lambda}_1\!-\!{\color{blue}\lambda}_2\!-\!{\color{blue}\lambda}_5\right)}{2(2\pi)^4}\!+\!\frac{5{\color{red}\zeta}_{5} {\color{blue}\lambda}_1 \left(8 {\color{blue}\lambda}_1^2\!-\!3 \left({\color{blue}\lambda}_2\!+\!{\color{blue}\lambda}_5\right) {\color{blue}\lambda}_1\!-\!{\color{blue}\lambda}_2^2\!-\!{\color{blue}\lambda}_5^2\right)}{4(2\pi)^6}  \biggl]+O\!\left({\color{blue}\lambda}_1^{8}\right)\,,\\
&G_{[5]}^{(5,1,1)}\!=\!\frac{5{\color{blue}\lambda}_1^5}{(4\pi)^{10}}\biggl[
1\!-\!\frac{15{\color{red}\zeta}_{3} {\color{blue}\lambda}_1 \left(2 {\color{blue}\lambda}_1\!-\!{\color{blue}\lambda}_2\!-\!{\color{blue}\lambda}_5\right)\!}{8(2\pi)^4} \!+\!\frac{\!25{\color{red}\zeta}_{5} {\color{blue}\lambda}_1 \left(8 {\color{blue}\lambda}_1^2\!-\!3 \left({\color{blue}\lambda}_2\!+\!{\color{blue}\lambda}_5\right) {\color{blue}\lambda}_1\!-\!{\color{blue}\lambda}_2^2\!-\!{\color{blue}\lambda}_5^2\right)\!}{16(2\pi)^6}  \biggl]\!+O\!\left({\color{blue}\lambda}_1^{9}\right)\,\!,\\
&G_{[6]}^{(5,1,1)}\!=\!\frac{6{\color{blue}\lambda}_1^6}{(4\pi)^{12}}\biggl[
1\!-\!\frac{9{\color{red}\zeta}_{3} {\color{blue}\lambda}_1 \left(2 {\color{blue}\lambda}_1\!-\!{\color{blue}\lambda}_2\!-\!{\color{blue}\lambda}_5\right)\!}{4(2\pi)^4} \!+\!\frac{\!15 {\color{red}\zeta}_{5} {\color{blue}\lambda}_1 \left(8 {\color{blue}\lambda}_1^2\!-\!3 \left({\color{blue}\lambda}_2\!+\!{\color{blue}\lambda}_5\right) {\color{blue}\lambda}_1\!-\!{\color{blue}\lambda}_2^2\!-\!{\color{blue}\lambda}_5^2\right)\!}{8(2\pi)^6} \biggl]\!+O\!\left({\color{blue}\lambda}_1^{10}\right)\,\!,
\end{align}\normalsize
where $q=5$ and the operators are both in the node 1. 
Analysing these results together with the one in the notebook attached to this manuscript, it is possible to elaborate a general pattern for the expansions in the $A_{q-1}$ theories.  
\begin{equation}\small\begin{split}
G_{\vec n}^{(2,1,1)}&=G_{\vec n}^{(1,1,1)}+O({\lambda}^{n+2})~,\\
G_{\vec n}^{(3,1,1)}&=G_{\vec n}^{(1,1,1)}+G_{\vec n}^{(2,1,1)}\biggl|_{{\lambda}_2\rightarrow\frac{{\lambda}_2+{\lambda}_3}{2}}
+O({\lambda}^{n+3})~,\\
G_{\vec n}^{(4,1,1)}&=G_{\vec n}^{(1,1,1)}+G_{\vec n}^{(2,1,1)}\biggl|_{{\lambda}_2\rightarrow\frac{{\lambda}_2+{\lambda}_4}{2}}
+G_{\vec n}^{(3,1,1)}\biggl|_{{\lambda}_3\rightarrow{\lambda}_4}
+O({\lambda}^{n+4})~,
\end{split}\end{equation}
where $O({\lambda}^{\ell})$ stand for any combinations of the couplings with power $\ell$.
Finally, at the orbifold point when all the couplings are equal, this pattern simplify even further 
\begin{equation}\small\begin{split}
G_{\vec n}^{(q,1,1)}&=\sum_{k=1}^{q-1}\,G_{\vec n}^{(k,1,1)}+O({\lambda}_1^{n+q})~.
\end{split}\end{equation}

\subsection{Operators on adjacent nodes}\label{subsec:MinDist}

\begin{figure}[!t]
\begin{center}
\includegraphics[scale=0.6]{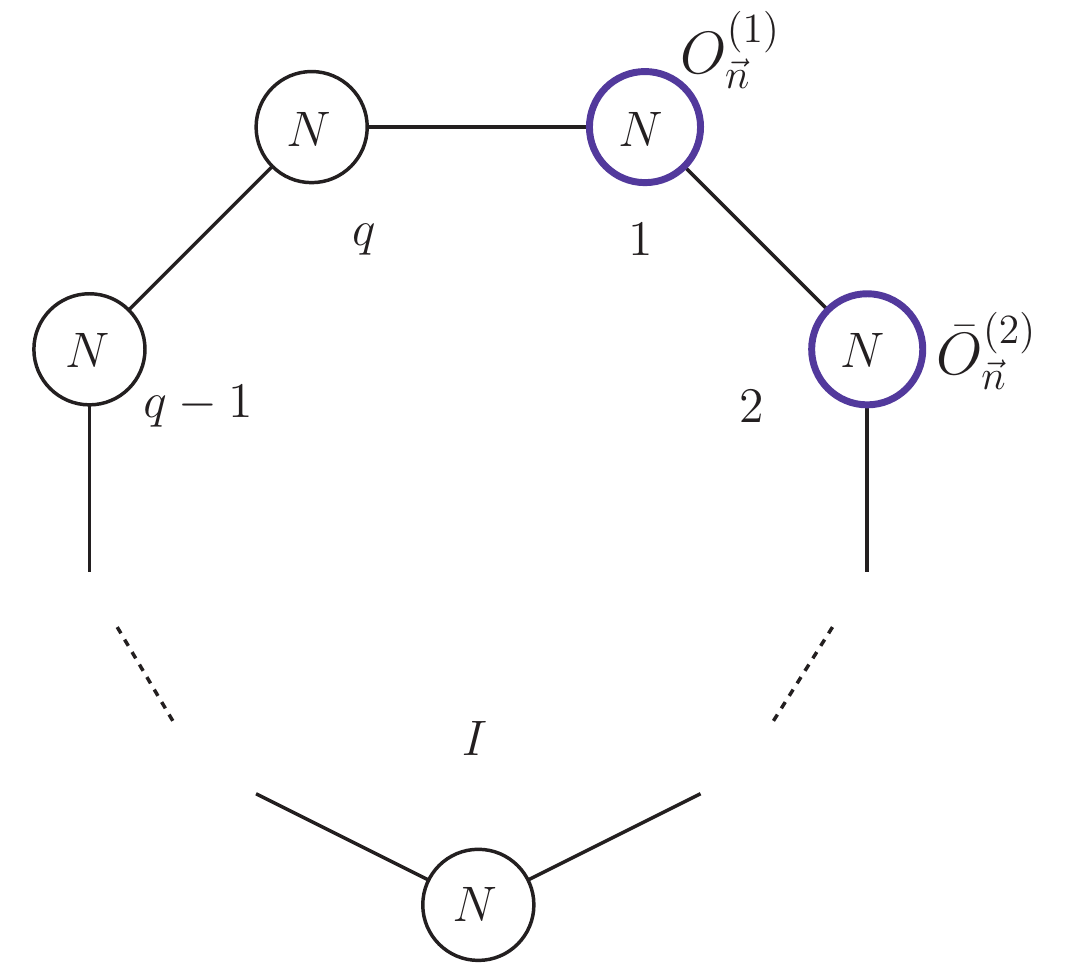}
\end{center}
\caption{The circular quiver with $q$ nodes associated to the theory $A_{q-1}$. We highlight the nodes at a distance $d=1$ in which the operators $O_{\vec n}^{(1)}$ and $\bar O_{\vec n}^{(2)}$ belongs to.}
\label{Fig:quiverMinDistance}
\end{figure}

We consider the case of chiral/antichiral operators in two neighbouring nodes of the quiver. Since the quiver has a circular invariance, we consider operators in the nodes $I=2$ and $J=2$ for simplicity without any loss of generality (see Figure \ref{Fig:quiverMinDistance}). 
Correlators of operators associated to different nodes have no longer the $\cN=4$ result as a leading order. Indeed, the operators belong to different vector multiplets and therefore they need to be corrected using hypers in a pure $\cN=2$ effect. This fact has an important consequence: the leading order of each correlator depends on the dimension of the operator, following the pattern
\begin{equation}\label{patd1}
G_{\vec n}^{(q,1,2)}\big|_{\mathrm{LO}} \propto ({\color{blue}\lambda}_1{\color{blue}\lambda}_2)^n~.
\end{equation} 
This fact is totally general and holds also at finite $N$. Indeed, if we consider the first few operators of dimensions $n=2,3,4$, the leading order of their correlators are
\small\begin{align}
&G_{[2]}^{(q,1,2)}\big|_{\mathrm{LO}} \!=\frac{3 \left(N^2-1\right)^2 {\color{red}\zeta}_3 {\color{blue}\lambda}_1^2 {\color{blue}\lambda}_2^2}{4 (8\pi^2)^4 N^4 }~,\\
&G_{[3]}^{(q,1,2)}\big|_{\mathrm{LO}} \!=\frac{15 \left(N^2-1\right)^2\left(N^2-4\right)^2 {\color{red}\zeta}_5 {\color{blue}\lambda}_1^3 {\color{blue}\lambda}_2^3}{16(8\pi^2)^6 N^8 }~,\\
&G_{[4]}^{(q,1,2)}\big|_{\mathrm{LO}} \!\!=\!\frac{ \!\left(N^2\!\!-\!1\right)^2 \!\!{\color{blue}\lambda}_1^4 {\color{blue}\lambda}_2^4}{32(8\pi^2)^8 N^{12}}\!\left[35{\color{red}\zeta}_7(N^8\!\!-\!12N^6\!\!+\!72N^4\!-\!216N^2\!+\!324) \!+\!144{\color{red}\zeta}_3^2(N^6\!\!-\!3N^4\!+\!3N^2)  \right] \\
&G_{[2,2]}^{(q,1,2)}\big|_{\mathrm{LO}} \!=\frac{ \left(N^2-1\right)^2 {\color{blue}\lambda}_1^4 {\color{blue}\lambda}_2^4}{32(8\pi^2)^8 N^{10}}\left[140{\color{red}\zeta}_7(4N^4-12N^2+9) +36{\color{red}\zeta}_3^2(N^6-2N^4+N^2)  \right]~.
\end{align}\normalsize
From the previous examples, it's interesting to notice that also the transcendentality grows with $n$ and the power of the couplings. In particular, already at leading order, different combinations of Riemann zetas can appear. Unlike the case of operators in the same node studied in the previous section, terms with different transcendentality at the same perturbative order can multiply polynomials with different degrees in $N$. As a consequence, the large $N$ limit selects only terms at a specific transcendentality
\begin{align}
G_{[4]}^{(q,1,2)}\!\xrightarrow[N\to\infty]{}\frac{35 }{32(8\pi^2)^8}{\color{red}\zeta}_7 {\color{blue}\lambda}_1^4 {\color{blue}\lambda}_2^4~, \hspace{1.5cm}G_{[2,2]}^{(q,1,2)}\!\xrightarrow[N\to\infty]{}\frac{9 }{8(8\pi^2)^8}{\color{red}\zeta}_3^2 {\color{blue}\lambda}_1^4 {\color{blue}\lambda}_2^4 ~.
\end{align}
The fact that the selected term in $G_{[2,2]}^{(q,1,2)}$ is the one proportional to ${\zeta}_3^2$ is expected because of the factorisation property of multi-trace correlators at large $N$ as in \eqref{factorization}. Instead, the correlator of single-traces pick the highest transcendentality. 

In the following we concentrate on large $N$ results for lack or space, therefore we consider only correlators of single-trace operators. Let's consider the theory $A_3$ for $q=4$, then up to NNLO we have
\begin{equation}\label{4.39}\small\begin{split}
&G_{[2]}^{(4,1,2)}\!=\frac{3}{4(8\pi^2)^4}  {\color{red}\zeta}_3 {\color{blue}\lambda}_1^2 {\color{blue}\lambda}_2^2-\frac{5 }{2(8\pi^2)^5}{\color{red}\zeta}_5 {\color{blue}\lambda}_1^2 {\color{blue}\lambda}_2^2 \left({\color{blue}\lambda}_1+{\color{blue}\lambda}_2\right) \\&+\frac{{\color{blue}\lambda}_1^2 {\color{blue}\lambda}_2^2}{16(8\pi^2)^{6}} \left(35 {\color{red}\zeta}_7 \left(3 {\color{blue}\lambda}_1^2+8 {\color{blue}\lambda}_2 {\color{blue}\lambda}_1+3 {\color{blue}\lambda}_2^2\right)-36 {\color{red}\zeta}_3^2 \left(3({\color{blue}\lambda}_1^2+{\color{blue}\lambda}_2^2)-2{\color{blue}\lambda}_1{\color{blue}\lambda}_2-{\color{blue}\lambda}_1{\color{blue}\lambda}_4-{\color{blue}\lambda}_2{\color{blue}\lambda}_3\right)\right)+\dots 
\end{split}\end{equation}
\begin{equation}\small\begin{split}
&G_{[3]}^{(4,1,2)}\!=\frac{15 }{16 (8\pi^2)^{6}}{\color{red}\zeta}_5 {\color{blue}\lambda}_1^3 {\color{blue}\lambda}_2^3-\frac{315 }{64(8\pi^2)^{7}}{\color{red}\zeta}_7 {\color{blue}\lambda}_2^3 {\color{blue}\lambda}_1^3 \left({\color{blue}\lambda}_1+{\color{blue}\lambda}_2\right) \\&+\frac{27{\color{blue}\lambda}_1^3 {\color{blue}\lambda}_2^3 }{64(8\pi^2)^{8}} \left(21 {\color{red}\zeta}_9 \left(2 {\color{blue}\lambda}_1\!+\!{\color{blue}\lambda}_2\right) \left({\color{blue}\lambda}_2\!+\!2 {\color{blue}\lambda}_2\right)-10 {\color{red}\zeta}_3 {\color{red}\zeta}_5 \left(2({\color{blue}\lambda}_1^2+{\color{blue}\lambda}_2^2-{\color{blue}\lambda}_1{\color{blue}\lambda}_2)-{\color{blue}\lambda}_1{\color{blue}\lambda}_4-{\color{blue}\lambda}_2{\color{blue}\lambda}_3\right)\right)+\dots
\end{split}\end{equation}
\begin{equation}\small\begin{split}
&G_{[4]}^{(4,1,2)}\!=\frac{35 }{32(8\pi^2)^{8}}{\color{red}\zeta}_7 {\color{blue}\lambda}_1^4 {\color{blue}\lambda}_2^4-\frac{63}{8(8\pi^2)^{9}}{\color{red}\zeta}_9 {\color{blue}\lambda}_1^4 {\color{blue}\lambda}_2^4 \left({\color{blue}\lambda}_1+{\color{blue}\lambda}_2\right) \\&+\frac{21{\color{blue}\lambda}_1^4 {\color{blue}\lambda}_2^4 }{32(8\pi^2)^{10}} \left(11 {\color{red}\zeta}_{11} \left(5 {\color{blue}\lambda}_1^2\!+\!12 {\color{blue}\lambda}_2 {\color{blue}\lambda}_1\!+5 {\color{blue}\lambda}_2^2\right)-10 {\color{red}\zeta}_3 {\color{red}\zeta}_7 \left(2({\color{blue}\lambda}_1^2\!+{\color{blue}\lambda}_2^2\!-\!{\color{blue}\lambda}_1{\color{blue}\lambda}_2)\!-\!{\color{blue}\lambda}_1{\color{blue}\lambda}_4\!-\!{\color{blue}\lambda}_2{\color{blue}\lambda}_3\right)\right)+\dots
\end{split}\end{equation}
\begin{equation}\small\begin{split}
&G_{[5]}^{(4,1,2)}\!=\frac{315 }{256(8\pi^2)^{10}}{\color{red}\zeta}_9 {\color{blue}\lambda}_1^5 {\color{blue}\lambda}_2^5-\frac{5775 }{512(8\pi^2)^{11}}{\color{red}\zeta}_{11} {\color{blue}\lambda}_1^5 {\color{blue}\lambda}_2^5 \left({\color{blue}\lambda}_1+{\color{blue}\lambda}_2\right)\\&+\frac{75{\color{blue}\lambda}_1^5 {\color{blue}\lambda}_2^5}{512(8\pi^2)^{12}}  \left(143 {\color{red}\zeta}_{13} \left(3 {\color{blue}\lambda}_1^2\!+\!7 {\color{blue}\lambda}_2 {\color{blue}\lambda}_1\!+\!3 {\color{blue}\lambda}_2^2\right)-63 {\color{red}\zeta}_3 {\color{red}\zeta}_9 \left(2({\color{blue}\lambda}_1^2\!+\!{\color{blue}\lambda}_2^2\!-\!{\color{blue}\lambda}_1{\color{blue}\lambda}_2)\!-\!{\color{blue}\lambda}_1{\color{blue}\lambda}_4\!-\!{\color{blue}\lambda}_2{\color{blue}\lambda}_3\right)\right)+\dots
\end{split}\end{equation}
\begin{equation}\label{4.43}\small\begin{split}
&G_{[6]}^{(4,1,2)}\!=\frac{693}{512(8\pi^2)^{12}} {\color{red}\zeta}_{11} {\color{blue}\lambda}_1^6 {\color{blue}\lambda}_2^6-\frac{3861  }{256(8\pi^2)^{13}}{\color{red}\zeta}_{13} {\color{blue}\lambda}_1^6 {\color{blue}\lambda}_2^6 \left({\color{blue}\lambda}_1+{\color{blue}\lambda}_2\right) \\&+\frac{891 {\color{blue}\lambda}_1^6 {\color{blue}\lambda}_2^6}{4096(8\pi^2)^{14}} \left(65 {\color{red}\zeta}_{15} \left(7 {\color{blue}\lambda}_1^2\!+\!16 {\color{blue}\lambda}_2 {\color{blue}\lambda}_1\!+\!7 {\color{blue}\lambda}_2^2\right)\!-\!56 {\color{red}\zeta}_3 {\color{red}\zeta}_{11} \left(2({\color{blue}\lambda}_1^2\!+\!{\color{blue}\lambda}_2^2\!-\!{\color{blue}\lambda}_1{\color{blue}\lambda}_2)\!-\!{\color{blue}\lambda}_1{\color{blue}\lambda}_4\!-\!{\color{blue}\lambda}_2{\color{blue}\lambda}_3\right)\right)\!+...
\end{split}\end{equation}
In section \ref{sec:feyadjnodes} we provide a  general diagrammatic interpretation for LO and NLO terms of these correlation functions.
For finite $N$ expansions, cases with different $q$ and higher order terms all the informations are in the attached notebook "\texttt{QuiverCorrelators.nb}".

\subsection{Operators on nodes at a distance $d\geq 2$ }\label{subsec:maxdist}

\begin{figure}[!t]
\begin{center}
\includegraphics[scale=0.6]{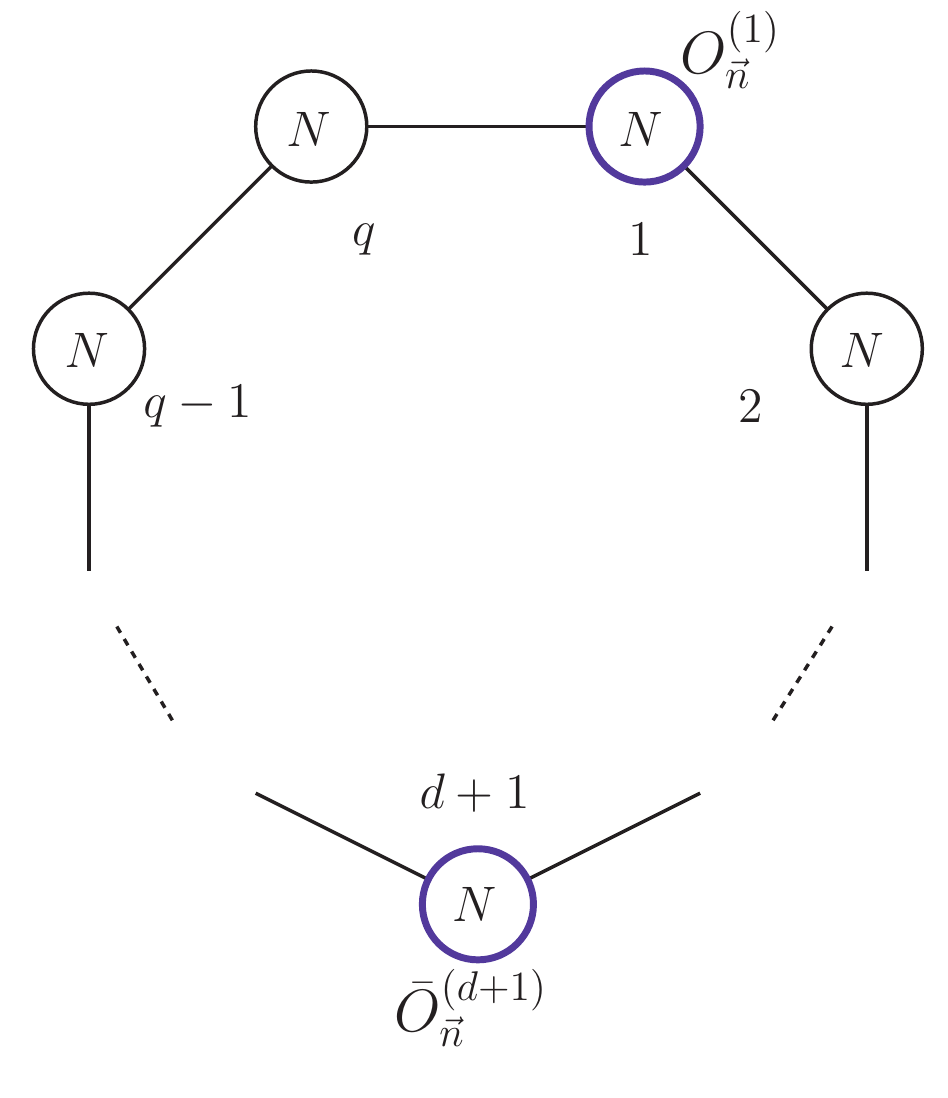}
\end{center}
\caption{Insertion of operators $O_{\vec n}^{(1)}$ and $\bar O_{\vec n}^{(d+1)}$ with distance $d$.}
\label{Fig:quiverMaxDistance}
\end{figure}
Let's consider the definition of distance $d$  between two nodes on the quiver given in \eqref{dist}. In this section we consider the case of chiral/antichiral operators in two nodes of the quiver at a distance $d\geq 2$. For simplicity, we can consider the first operator on the node $I=1$ and the second on the node $J=d+1$ (see figure \ref{Fig:quiverMaxDistance}). As the previous case, correlators of operators belonging to different vector multiplets are given by pure $\mathcal{N}=2$ effects. The generalization of the pattern \eqref{patd1} for any $d$ is highly not trivial: indeed it depends on the choice of $q$ and $d$. 
In the special case in which $q$ is even and $d=q/2$ we have more than one possible combinations of the couplings: starting from the node $I=1$ we can connect the operator inserted in node $J=q/2+1$ following both clockwise and anti-clockwise paths. For instance if $q=4$, $d=2$ and $n=4$, the possible combinations at leading order are ${\lambda}_1^4{\lambda}_2^4{\lambda}_3^4$,  ${\lambda}_1^4{\lambda}_3^4{\lambda}_4^4$  and  ${\lambda}_1^4{\lambda}_2^2{\lambda}_3^4{\lambda}_4^2$. For all the other cases
\begin{equation}
G_{\vec n}^{(q,1,d+1)}\big|_{\mathrm{LO}} \propto ({\color{blue}\lambda}_1{\color{blue}\lambda}_2...{\color{blue}\lambda}_{d+1})^n~.
\end{equation} 
Let's consider for instance the leading order of correlators of operators of dimension $n=2,3$
\begin{equation}\begin{split}
G_{[2]}^{(q,1,d+1)}\big|_{\mathrm{LO}} &=\frac{3^d(N^2-1)^{d+1}}{2^{d+1}N^{2(d+1)}}{\color{red}\zeta}_3^d({\color{blue}\lambda}_1{\color{blue}\lambda}_2...{\color{blue}\lambda}_{d+1})^2~,\\
G_{[3]}^{(q,1,d+1)}\big|_{\mathrm{LO}} &=\frac{3 \times 5^d(N^4-5N^2+4)^{d+1}}{2^{d+3}N^{4(d+1)}}{\color{red}\zeta}_5^d({\color{blue}\lambda}_1{\color{blue}\lambda}_2...{\color{blue}\lambda}_{d+1})^3~.
\end{split}\end{equation}
As in the previous section for $d=1$, correlators for $n\geq 4$ contains more than one combinations of Riemann zeta functions. The large $N$ limit selects for the single-trace operators the highest transcendentality and for the multi-trace operators the lowest one.   

In the following we concentrate on large $N$ results for lack of space, therefore we consider only correlators of single-trace operators. We show some results in the theory $A_3$ for $d=2$ up to NLO
\begin{equation}\label{4.46}\small
G_{[2]}^{(4,1,3)}\!=\frac{9\,{\color{red}\zeta}_3^2}{8(8\pi^2)^6}   {\color{blue}\lambda}_1^2 {\color{blue}\lambda}_3^2({\color{blue}\lambda}_2^2+{\color{blue}\lambda}_4^2)\!-\!\frac{15\,{\color{red}\zeta}_3{\color{red}\zeta}_5 }{4(8\pi^2)^7} {\color{blue}\lambda}_1^2 {\color{blue}\lambda}_3^2  \left(2({\color{blue}\lambda}_2^3+{\color{blue}\lambda}_4^3)+{\color{blue}\lambda}_3({\color{blue}\lambda}_2^2+{\color{blue}\lambda}_4^2)+{\color{blue}\lambda}_1({\color{blue}\lambda}_2^2+{\color{blue}\lambda}_4^2)\right)\!+...
\end{equation}
\begin{equation}\small
G_{[3]}^{(4,1,3)}\!=\frac{75\,{\color{red}\zeta}_5^2}{32(8\pi^2)^9}   {\color{blue}\lambda}_1^3 {\color{blue}\lambda}_3^3({\color{blue}\lambda}_2^3\!+\!{\color{blue}\lambda}_4^3)\!-\!\frac{1574\,{\color{red}\zeta}_5{\color{red}\zeta}_7 }{128(8\pi^2)^{10}} {\color{blue}\lambda}_1^3 {\color{blue}\lambda}_3^3  \left(2({\color{blue}\lambda}_2^4\!+\!{\color{blue}\lambda}_4^4)\!+\!{\color{blue}\lambda}_3({\color{blue}\lambda}_2^3\!+\!{\color{blue}\lambda}_4^3)\!+\!{\color{blue}\lambda}_1({\color{blue}\lambda}_2^3\!+\!{\color{blue}\lambda}_4^3)\right)\!+...
\end{equation}
\begin{equation}\small
G_{[4]}^{(4,1,3)}\!=\frac{1225\,{\color{red}\zeta}_7^2}{256(8\pi^2)^{12}}   {\color{blue}\lambda}_1^4 {\color{blue}\lambda}_3^4({\color{blue}\lambda}_2^4\!+\!{\color{blue}\lambda}_4^4)\!-\!\frac{2205\,{\color{red}\zeta}_7{\color{red}\zeta}_9 }{64(8\pi^2)^{13}} {\color{blue}\lambda}_1^4 {\color{blue}\lambda}_3^4  \!\left(2({\color{blue}\lambda}_2^5\!+\!{\color{blue}\lambda}_4^5)\!+\!{\color{blue}\lambda}_3({\color{blue}\lambda}_2^4\!+\!{\color{blue}\lambda}_4^4)\!+\!{\color{blue}\lambda}_1({\color{blue}\lambda}_2^4\!+\!{\color{blue}\lambda}_4^4)\!\right)\!+...
\end{equation}
We present also few expansions for $q=6$ and $d=3$ as follows
\begin{equation}\small\begin{split}
G_{[2]}^{(6,1,4)}\!&=\frac{27{\color{red}\zeta}_3^3 }{16(8\pi^2)^{8}}  {\color{blue}\lambda}_1^2 {\color{blue}\lambda}_4^2 \left({\color{blue}\lambda}_2^2 {\color{blue}\lambda}_3^2+{\color{blue}\lambda}_5^2 {\color{blue}\lambda}_6^2\right)\\
&\!\!-\frac{45 {\color{red}\zeta}_3^2 {\color{red}\zeta}_5}{8(8\pi^2)^{9}}  {\color{blue}\lambda}_1^2 {\color{blue}\lambda}_4^2 \left(2 {\color{blue}\lambda}_5^2 {\color{blue}\lambda}_6^3+{\color{blue}\lambda}_5^2 \left({\color{blue}\lambda}_1+{\color{blue}\lambda}_4+2 {\color{blue}\lambda}_5\right) {\color{blue}\lambda}_6^2+{\color{blue}\lambda}_2^2 {\color{blue}\lambda}_3^2 \left({\color{blue}\lambda}_1+2 \left({\color{blue}\lambda}_2+{\color{blue}\lambda}_3\right)+{\color{blue}\lambda}_4\right)\right)+...
\end{split}\end{equation}
\begin{equation}\small\begin{split}\label{4.50}
G_{[3]}^{(6,1,4)}\!&=\frac{375 {\color{red}\zeta}_5^3}{64(8\pi^2)^{12}}  {\color{blue}\lambda}_1^3 {\color{blue}\lambda}_4^3 \left({\color{blue}\lambda}_2^3 {\color{blue}\lambda}_3^3+{\color{blue}\lambda}_5^3 {\color{blue}\lambda}_6^3\right)\\
&\!\!-\frac{7875{\color{red}\zeta}_5^2 {\color{red}\zeta}_7}{256(8\pi^2)^{13}} {\color{blue}\lambda}_1^3 {\color{blue}\lambda}_4^3 \left(2 {\color{blue}\lambda}_5^3 {\color{blue}\lambda}_6^4\!+{\color{blue}\lambda}_5^3 \left({\color{blue}\lambda}_1\!+{\color{blue}\lambda}_4\!+2 {\color{blue}\lambda}_5\right) {\color{blue}\lambda}_6^3\!+{\color{blue}\lambda}_2^3 {\color{blue}\lambda}_3^3 \left({\color{blue}\lambda}_1\!+2 \left({\color{blue}\lambda}_2\!+{\color{blue}\lambda}_3\right)\!+{\color{blue}\lambda}_4\right)\right)+...
\end{split}\end{equation}
We can foresee some nice patterns also in this case, we will discuss them with the help of the Feynman diagram analysis in section \ref{sec:feynd}.

\subsection{The twisted and untwisted operators}
We provide additional results for a class of observables originally introduced in \cite{Gukov:1998kk,Lee:1998bxa}, corresponding to twisted and untwisted sectors of the $A_{q-1}$ quiver theories. We define the single untwisted operator and the $q$ twisted operators as:
\begin{align}\label{ut}
U_{\vec n} = \frac{1}{\sqrt{q}} \sum_{I=1}^q O_{\vec n}^{(I)}~, \hspace{1cm} T^{(I)}_{\vec n} = \frac{1}{\sqrt{2}} \parenth{O_{\vec n}^{(I)}-O_{\vec n}^{(I+1)}}~.
\end{align}
These operators have well defined transformation properties under the $\mathbb{Z}_q$ orbifold action, and therefore they have corresponding holographic dual states. Under gauge groups exchanges, the operator $U_{\vec n}$ is even while the operators $T^{(I)}_{\vec n} $ are odd. Their correlation function have been approached for the $q=2$ case and in the large $N$ limit in \cite{Pini:2017ouj}. Using our dataset of results for chiral correlators we can extend this analysis and find new features for these observables.
For simplicity reasons, in this manuscript we provide expansions only at the orbifold point, \textit{i.e.} all the couplings are the same $\lambda_1=...=\lambda_q=\lambda$.

At the orbifold point, the twisted and untwisted sectors are orthogonal. For $q=2$ this can be easily checked since the correlator $\langle U_{\vec n}\,T^{(1)}_{\vec n}\rangle$ is always proportional to the difference of the couplings $\lambda_1$ and $\lambda_2$.
Let's consider first the untwisted sector computing the correlation function of two untwisted operators. Following the definition \eqref{ut}, it is clear that we can decompose it in terms of the correlators $G$ \eqref{Gnormord} computed with the multi-matrix model. For instance, considering $q=3$ we have
\begin{equation}
\big\langle U_{\vec n} \,\bar U_{\vec n} \rangle_3=\frac{1}{3\,x^{2n}}\biggl[
G_{\vec n}^{(3,1,1)}+G_{\vec n}^{(3,2,2)}+G_{\vec n}^{(3,3,3)}+2\left(G_{\vec n}^{(3,1,2)}+G_{\vec n}^{(3,1,3)}+G_{\vec n}^{(3,2,3)}\right)
\biggl]
\end{equation}
Using the symmetry \eqref{Gorb} at the orbifold point, we can reduce the computation to simply
\begin{equation}
\big\langle U_{\vec n} \,\bar U_{\vec n} \rangle_3=\frac{G_{\vec n}^{(3,1,1)}+2G_{\vec n}^{(3,1,2)}}{x^{2n}}
\end{equation}
and then
\begin{align}
\big\langle U_{[2]} \,\bar U_{[2]} \rangle_3&=
\frac{\left(N^2-1\right) {\color{blue}\lambda}^2}{2(8\pi^2)^2 N^2}-\frac{9 \left(N^2-1\right) {\color{red}\zeta}_3{\color{blue}\lambda}^4}{(8\pi^2)^4N^4}+\frac{30 \left(3 N^4-5 N^2+2\right) {\color{red}\zeta}_5{\color{blue}\lambda}^5}{(8\pi^2)^5 N^6}+...\\
\big\langle U_{[3]} \,\bar U_{[3]} \rangle_3&\!=\!
\frac{3 \!\left(\!N^4\!\!-\!5 N^2\!\!+\!4\right)\! {\color{blue}\lambda}^3\!}{8(8\pi^2)^3N^4}\!-\!\frac{27\! \left(\!N^4\!\!-\!5 N^2\!\!+\!4\right)\! {\color{red}\zeta}_3{\color{blue}\lambda}^5\!}{2(8\pi^2)^5N^6}\!+\!\frac{405\! \left(\!N^2\!\!-\!4\right)\!\! \left(\!N^2\!\!-\!1\right)^{\!2}\! {\color{red}\zeta}_5{\color{blue}\lambda}^6\!}{4(8\pi^2)^6 N^8}\!+\!...\\
\big\langle U_{[4]} \,\bar U_{[4]} \rangle_3&=
\frac{\left(N^2-1\right) \left(N^4-6 N^2+18\right) {\color{blue}\lambda}^4}{4(8\pi^2)^4N^6}-\frac{9 \left(N^2-1\right) \left(3 N^4-14 N^2+33\right) {\color{red}\zeta}_3{\color{blue}\lambda}^6}{(8\pi^2)^6N^8}\nonumber\\
&\qquad\qquad\qquad+\frac{45 \left(N^2-1\right) \left(5 N^6-22 N^4+72 N^2-66\right) {\color{red}\zeta}_5{\color{blue}\lambda}^7}{(8\pi^2)^7N^{10}}+...\\
\big\langle U_{[5]} \,\bar U_{[5]} \rangle_3&=
\frac{5 \left(N^4+24\right)\! \left(N^4-5 N^2+4\right) {\color{blue}\lambda}^5}{32(8\pi^2)^5N^8}\!-\!\frac{225\left(N^4-5 N^2+4\right)\! \left(N^4-2 N^2+14\right) {\color{red}\zeta}_3{\color{blue}\lambda}^7}{8(8\pi^2)^7N^{10}}\nonumber\\
&\qquad\qquad+\frac{375 \left(N^4-5 N^2+4\right) \left(9 N^6-5 N^4+116 N^2-200\right) {\color{red}\zeta}_5{\color{blue}\lambda}^8}{16(8\pi^2)^8 N^{12}}+...
\end{align}
Performing the large $N$ limit, the expansions above reduces simply to the leading terms. Those terms are identical to the $\mathcal{N}=4$ SYM counterpart for operators of dimension $n$. This is true in general, indeed
\begin{equation}
\big\langle U_{\vec n} \,\bar U_{\vec n} \rangle_q=\frac{n{\lambda}^n}{(4\pi)^{2n}}
\end{equation}
as also noticed in \cite{Rodriguez-Gomez:2016ijh,Pini:2017ouj}. We tested this formula for several dimensions of the operators and for $q=1,2,...,6$.

Let's focus now on the twisted sector. This sector is much more complicated because of the mixing of the operators $T^{(I)}_{\vec n}$.
Similarly to the untwisted case, we can express any correlator in terms of the functions $G$ computed with the multi-matrix model.
At the orbifold point, correlators of operators with the same index $I$ are all the same and for instance for $q=4$ we have
\begin{align}
\big\langle T_{[2]} \bar T_{[2]} \rangle_4&\!\!=
\frac{1}{2(8\pi^2)^2}  {\color{blue}\lambda}^2\!-\!\frac{9}{4(8\pi^2)^4}  {\color{red}\zeta}_{3} {\color{blue}\lambda}^4\!+\frac{15}{(8\pi^2)^5} {\color{red}\zeta}_{5} {\color{blue}\lambda}^5\!+\frac{15}{8(8\pi^2)^6} \left(6 {\color{red}\zeta}_{3}^2\!-\!49 {\color{red}\zeta}_{7}\right) {\color{blue}\lambda}^6\!+...\\
\big\langle T_{[3]} \bar T_{[3]} \rangle_4&\!\!=
\frac{3}{8(8\pi^2)^3}  {\color{blue}\lambda}^3-\frac{45}{16(8\pi^2)^6} {\color{red}\zeta}_{5} {\color{blue}\lambda}^6+\frac{945}{32(8\pi^2)^7} {\color{red}\zeta}_{7} {\color{blue}\lambda}^7-\frac{15309}{64(8\pi^2)^8} {\color{red}\zeta}_{9} {\color{blue}\lambda}^8+...\\
\big\langle T_{[4]} \bar T_{[4]} \rangle_4&\!\!=
\frac{1}{4(8\pi^2)^4}  {\color{blue}\lambda}^4-\frac{105}{32(8\pi^2)^8} {\color{red}\zeta}_{7} {\color{blue}\lambda}^8+\frac{189}{4(8\pi^2)^9} {\color{red}\zeta}_{9} {\color{blue}\lambda}^9-\frac{7623}{16(8\pi^2)^{10}} {\color{red}\zeta}_{11} {\color{blue}\lambda}^{10}+...\\
\big\langle T_{[5]} \bar T_{[5]} \rangle_4&\!\!=\!\!
\frac{5}{32(8\pi^2)^5} {\color{blue}\lambda}^5\!\!-\!\frac{945}{256(8\pi^2)^{10}}{\color{red}\zeta}_{9} {\color{blue}\lambda}^{10}\!\!+\!\frac{17325}{\!256(8\pi^2)^{11}} {\color{red}\zeta}_{11} {\color{blue}\lambda}^{11}\!\!-\!\frac{418275}{512(8\pi^2)^{12}} {\color{red}\zeta}_{13} {\color{blue}\lambda}^{12}\!\!+\!\!...
\end{align}
We checked similar correlators at very high orders also for higher dimensions and different $q$ and we deduced a pattern similar to the one proposed in \cite{Pini:2017ouj}. 
\begin{equation}
\big\langle T_{\vec n} \bar T_{\vec n} \rangle_q=\frac{n{\lambda}^n}{(4\pi)^{2n}}+O({\lambda}^{2n})
\end{equation}
In \cite{Niarchos:2020nxk,Niarchos:2019onf} it was noticed that the gap between the leading order identical to the one of $\mathcal{N}=4$ SYM and the NLO could be filled by instanton contributions.
Finally we can briefly analyze correlators of $T_{\vec n}^{(I)} $ and $\bar T_{\vec n}^{(J)} $ in the case in which $J=I+k$. If $k=1$ the correlator is the same for the case of $J=1$ multiplied by $-1/2$. For $k\geq2$ the leading order starts at very high order. For instance for $k=2$ we have $\big\langle T_{\vec n}^{(I)} \bar T_{\vec n}^{(I+2)} \rangle_q=O({\lambda}^{2n})$.

\section{Feynman diagrams analysis}

In this section we evaluate the chiral/antichiral correlators using the $\cN=1$ superspace formalism. The motivation for this section is twofold. The direct perturbative computation allows to explain and visualize the results obtained from the matrix model approach.
Besides, the comparison with the matrix model helps us to push the computation of some specific Feynman integrals up to very high orders in perturbation theory and derive general formulas for any values of $n$ and $q$. 

The starting point for a perturbative analysis of the two-point function \eqref{twopointdef} is represented by the diagram of Figure \ref{Fig:tree}. Such diagram must be fully connected at each perturbative order by using the Feynman rules of Appendix \ref{App:Feynman}. 
\begin{figure}[!t]
\begin{center}
\includegraphics[scale=0.7]{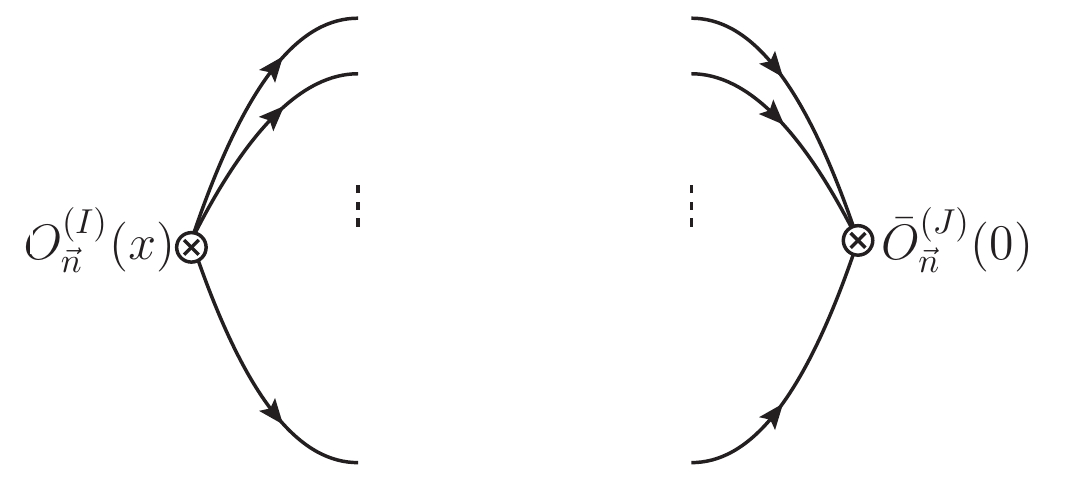}
\end{center}
\caption{The diagram representing the insertion of a chiral operator $O_{\vec n}^{(I)}$ from the $I$ -th node in $x$ 
and of the corresponding anti-chiral operator $\bar{O}_{\vec n}^{(J)}$ from the $J$ -th node in the origin. The umber of propagators arising from the insertion points is $n$.}
\label{Fig:tree}
\end{figure}

\subsection{Review of SCQCD}\label{subsec:Feynman1}
The perturbative analysis of SCQCD has been deeply analysed in \cite{Billo:2017glv}, and has a full diagrammatic explanation up to a two-loops order. We review some aspects of that computation as a warm up.

The two point function \eqref{twopointdef} can be seen as the insertion of a chiral $O_{\vec{n}}$ and an antichiral operator $\bar O_{\vec{n}}$ inside the four dimensional flat spacetime, see Figure \ref{Fig:tree} where $I=J$ (so we can drop the $(I),(J)$ indices). 

The tree level is obtained by connecting all the lines with $n$ tree level scalar propagators, which can be obtained from the $\theta=\bar\theta =0$ component of the propagator \eqref{propagators}:
\begin{equation}\label{scalarprop}
\vev{\varphi^a(x) \varphi^b(0)} = \frac{\delta_{ab}}{4\pi^2x^2}~.
\end{equation}
The spacetime factor reproduces the conformal factor of \eqref{twopointdef}, from all the possible colour contractions we obtain the full $\cN=4$ result, where observables \eqref{twopointdef} do not receive any perturbative correction. Such result only depends on the normalization and the trace structure of the chiral operators \eqref{defOn}.
We show an explicit example for $O_{2}(x)= g^2/2 \tr T^aT^b \varphi_a\varphi_b$. Using the propagator \eqref{scalarprop} and the conventions for the fundamental traces \eqref{normtrace}, we immediately find:
\begin{equation}
G_{[2]}^{(1,1,1)}\big|_{\cN=4} = \frac{(N^2-1) {\color{blue}\lambda}^2}{2N^2 (8\pi^2)^2}~,
\end{equation}
which is the leading order of the matrix model result \eqref{scqcd2}.
In this way all the LO results of subsection \ref{subsec:QCD} are reproduced.

The two-point function \eqref{twopointdef} in SCQCD has an infinite tower of perturbative corrections, which are simplified by the presence of conformal symmetry. This means that the spacetime factor $(4\pi^2x^2)^{-1}$ is preserved at the quantum level, and all the nonvanishing diagrams provide only finite contributions. Moreover, there is \emph{no 1-loop correction}, due to the vanishing of the 1-loop coefficient of the Beta function (see for example \cite{Billo:2017glv} for an explicit proof). 

The first nontrivial $\cN=2$ perturbative correction arises at the two-loops order.
The only two subdiagrams appearing at the two-loops order are the two-loops correction of the scalar propagator and a box diagram. These diagrams have been computed in \cite{Billo:2017glv} and their contribution can be displayed as a product of the color factor with the spacetime integral:
\begin{align}\label{subdiag}
\parbox[c]{.2\textwidth}{\includegraphics[width = .2\textwidth]{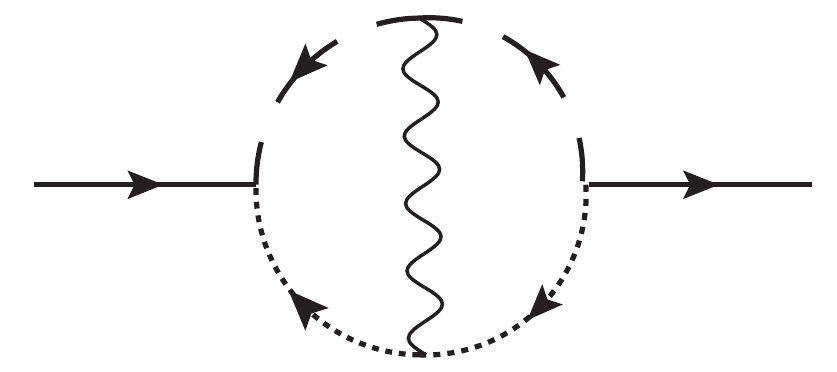}} = C_2(g,N)~ W_2(x_{12}) ~, \hspace{1cm}
\parbox[c]{.2\textwidth}{\includegraphics[width = .2\textwidth]{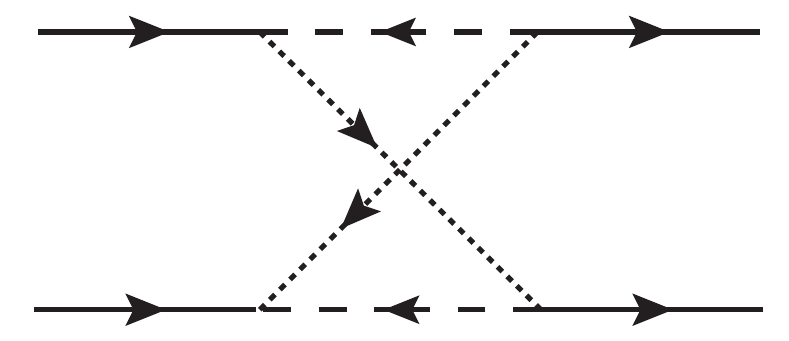}} = C_4(g,N)~ W_4(x_{12})~. 
\end{align}

\noindent The color factors $C_2$ and $C_4$ are computed in Appendix \ref{App:2}, while the spacetime integrals $W_2$ and $W_4$ have been computed in Appendix C of \cite{Billo:2017glv} and read \footnote{We repeat the computation of $W_4$ in Appendix \ref{App:superspace}, since an extension of this integral will be useful in the following.}
\begin{align}\label{spacetimeint}
W_2(x_{12}) = \frac{-3{\color{red}\zeta}_3}{(16\pi^2)^2}\frac{1}{4\pi^2x_{12}^2}~, \hspace{2cm}
W_4(x_{12}) = \frac{6{\color{red}\zeta}_3}{(16\pi^2)^2}\frac{1}{(4\pi^2x_{12}^2)^2}~.
\end{align}

Therefore the two-loops order at finite $N$ of the two-point correlator \eqref{twopointdef} is given by the insertion (with the proper multiplicity factor) of the two subdiagrams inside Figure \ref{Fig:tree}. We provide an example for the $n=2$ case, whose correlator is simply given by the combination:
\begin{align}\label{5.5}
 2\times \parbox[c]{.2\textwidth}{\includegraphics[width = .2\textwidth]{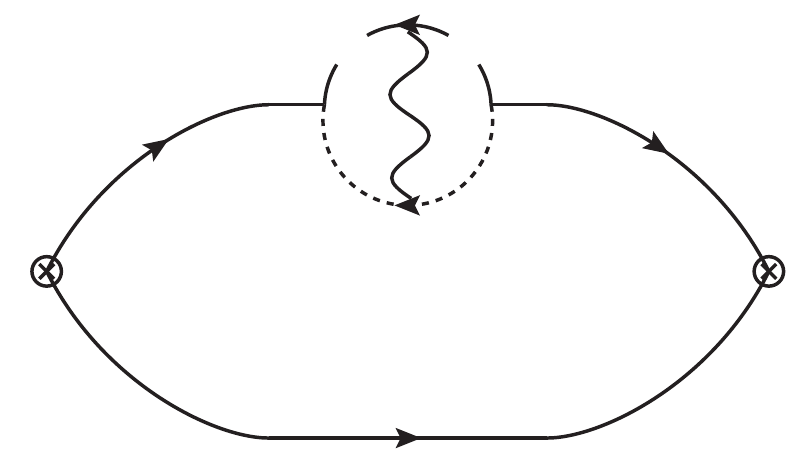}}+\parbox[c]{.2\textwidth}{\includegraphics[width = .2\textwidth]{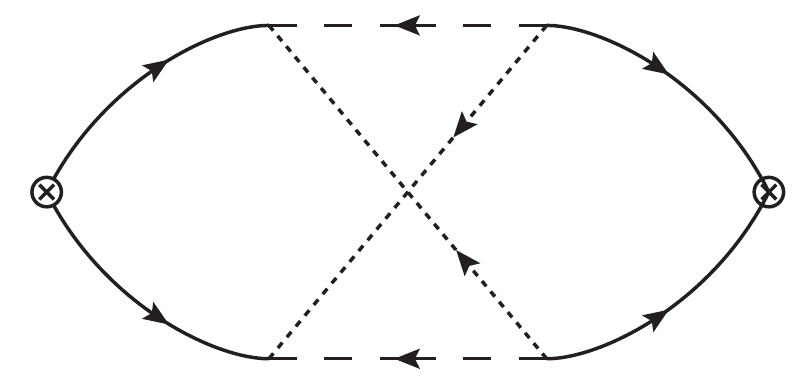}}~,
\end{align}
and such combination returns (see equations \eqref{B2} and \eqref{B3} for the computation of the color factors):
\begin{align}
G_{[2]}^{(1,1,1)}\big|_{{\color{red}\zeta}_3} =- \frac{3{\color{red}\zeta}_3{\color{blue}\lambda}^4}{N^4(8\pi^2)^4}\parenth{(N^4-1)+\frac{N^4-1}{2}} = - \frac{9(N^4-1){\color{red}\zeta}_3{\color{blue}\lambda}^4}{2N^4(8\pi^2)^4}~,
\end{align}
which captures the two-loops term of \eqref{scqcd2} as expected.
 The same reasoning holds for any dimension $n$ of the operators, so that we perfectly recover the ${\zeta}_3$ terms of all the results of subsection \ref{subsec:QCD}.

Just a brief remark for the large $N$ case at two-loops: it turns out that the contribution of the box diagram is planar only when inserted inside $G_{[2]}$. This fact explains the difference in the ${\zeta}_3$ terms of the effective couplings between ${\lambda}^{\text{eff}}_{[2]}$ and ${\lambda}^{\text{eff}}_{[n>2]}$, see the discussion inside Subsubsection \ref{subsub:QCD}.

\subsection{$A_{q-1}$ theories: operators in the same node}\label{subsec:Feynman2}
We extend the diagrammatic analysis to $A_{q-1}$ theories and we start by considering the case of operators in the same node (we choose node 1 as in subsection \ref{subsec:SameNode}).
The starting point is the same of SCQCD, so the insertion of a chiral and an antichiral operator in the same node, see Figure \ref{Fig:tree} with $I=J=1$. The tree level is the same as $\cN=4$ theory. 

The presence of the other nodes comes out in perturbative corrections. At the two loops order we must include the contribution of the two neighboring nodes $J=2$ and $J=q$. In particular the full ${\zeta}_3$ term is captured by the following subdiagrams, where we use labels to specify the node where each adjoint field comes from:
\begin{align}
\parbox[c]{.2\textwidth}{\includegraphics[width = .2\textwidth]{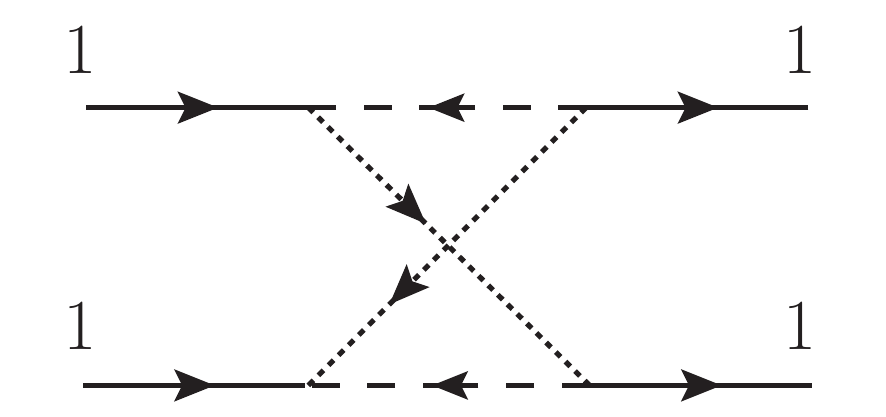}}~,~~~~\parbox[c]{.2\textwidth}{\includegraphics[width = .2\textwidth]{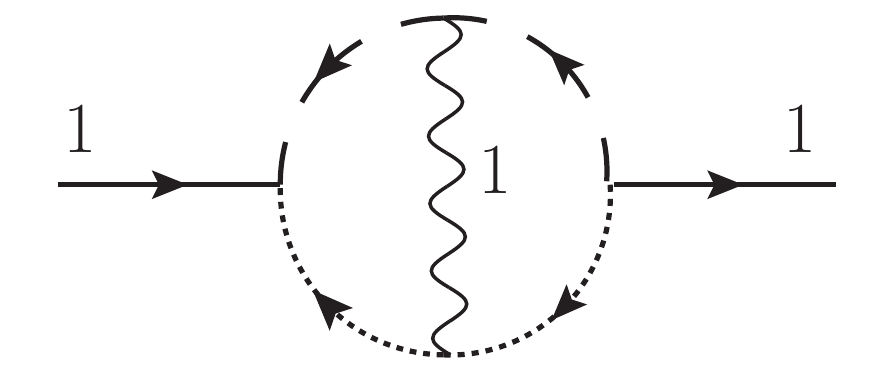}}~,~~~~\parbox[c]{.2\textwidth}{\includegraphics[width = .2\textwidth]{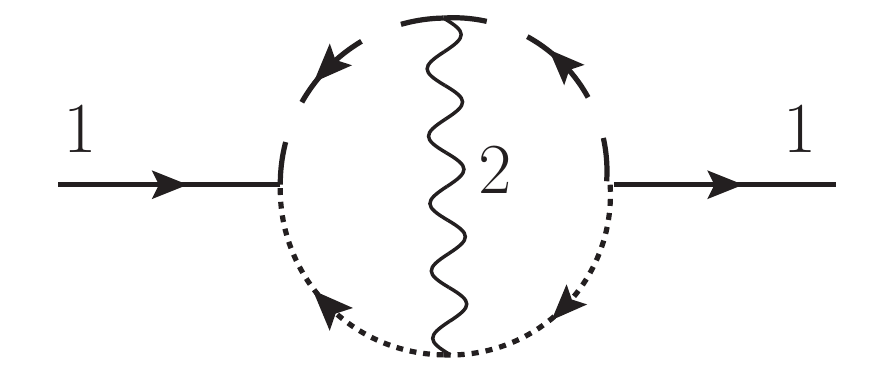}}~,~~~~\parbox[c]{.2\textwidth}{\includegraphics[width = .2\textwidth]{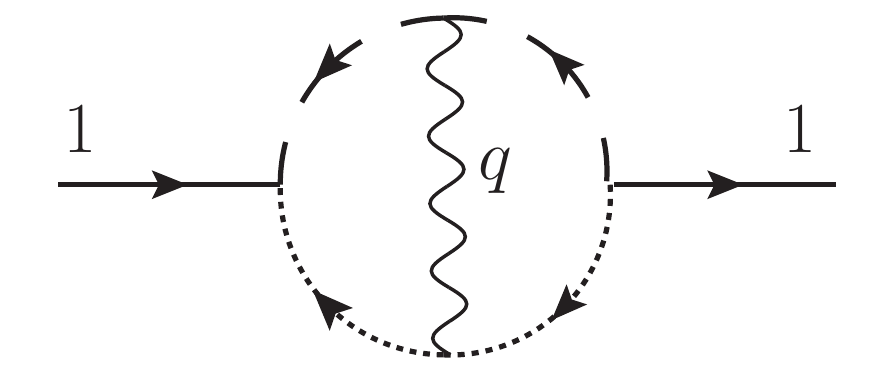}}~.
\end{align}
The structure of these subdiagrams is the same as \eqref{subdiag}, so the space-time integrals are equivalent to \eqref{spacetimeint}. The only difference  arises in the colour factor of the two-loops correction of the propagator $C_2(g_I,N)$ which is now a function of all the couplings $g_I$ and is computed in Appendix \ref{App:2}, see equation \eqref{C5}. Therefore also for $A_{q-1}$ theories it is possible to compute the correlators of operators in the same node up to a two-loops order at finite $N$. We give again an explicit example for the $n=2$ case. The subdiagrams are combined in the same way as the SCQCD result \eqref{5.5}  and therefore we get:
\begin{align}
G_{[2]}^{(q,1,1)}\big|_{{\color{red}\zeta}_3} = - \frac{3(N^2-1){\color{red}\zeta}_3{\color{blue}\lambda}_1^3}{2N^4(8\pi^2)^4}\parenth{3(N^2+1){\color{blue}\lambda}_1-(N^2-1) ({\color{blue}\lambda}_2+{\color{blue}\lambda}_q)}~,
\end{align}
which reproduces the matrix model findings \eqref{4.26} when we take the large $N$ limit.

A final brief observation for these observables: at the orbifold point $g_I=g~, \forall I$ and in the large $N$ limit, the contribution of the two-loops correction to the propagator vanishes (see explicitly in \eqref{C5}). This is a nice examples at the diagrammatic level of the cancellations occurring at the orbifold point of $A_{q-1}$ quivers. It would be interesting to explore this limit with a systematic analysis at higher orders.

\subsection{$A_{q-1}$ theories: operators at $d=1$}\label{sec:feyadjnodes}

In this case we do not have any $\cN=4$ result, since it is impossible to close the diagram with tree-level propagators (the two operators come from different vector multiplets). In order to have perturbative corrections we need to connect the two sides of Figure \ref{Fig:tree} inserting the bifundamental hypermultiplets. 
Therefore it is reasonable to expect that the first non-trivial contribution at the perturbative level of each two-point function $G_{\vec n}^{(q,1,2)}$ explicitly depends on the conformal dimension $n$.

We can give a precise diagrammatic explanation at large N. For each correlator $G_{\vec n}^{(q,1,2)}$ the leading order in perturbation theory at large $N$ is captured by a single diagram.

\begin{figure}[!t]
    \begin{minipage}[t]{.48\textwidth}
        \centering
        \includegraphics[width=\textwidth]{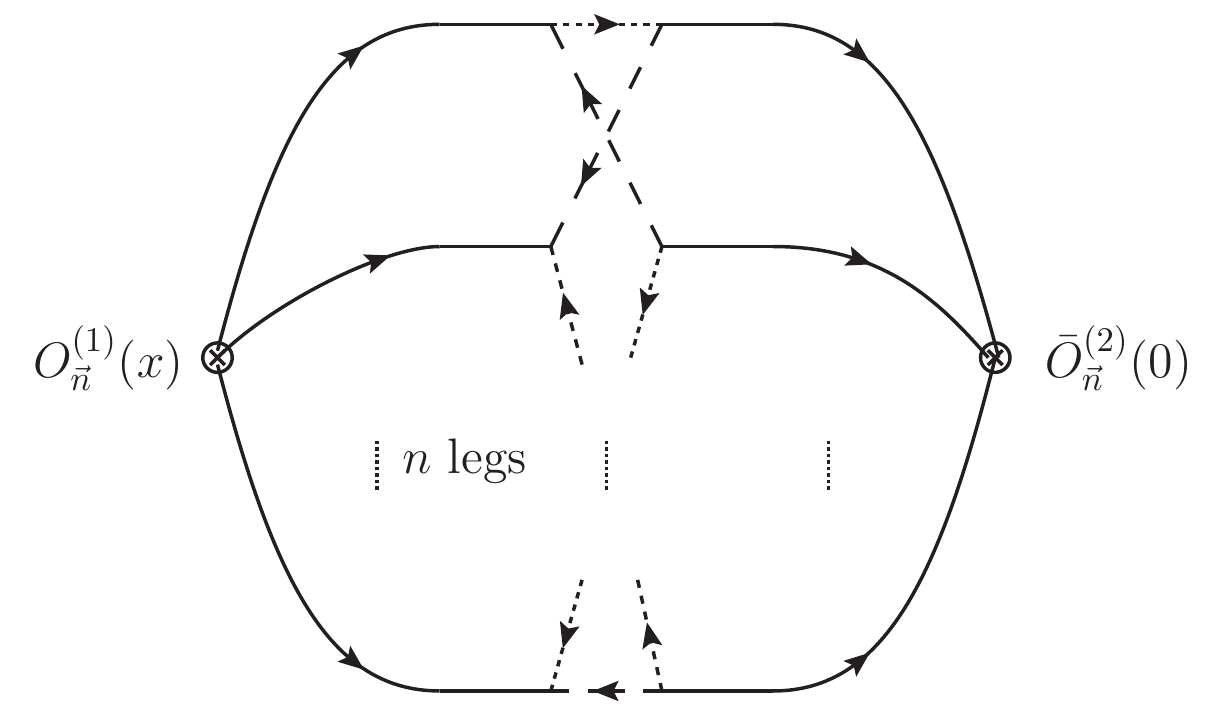}
        \subcaption{LO order}\label{Fig:Q2LO}
    \end{minipage}
    \hfill
    \begin{minipage}[t]{.48\textwidth}
        \centering
        \includegraphics[width=\textwidth]{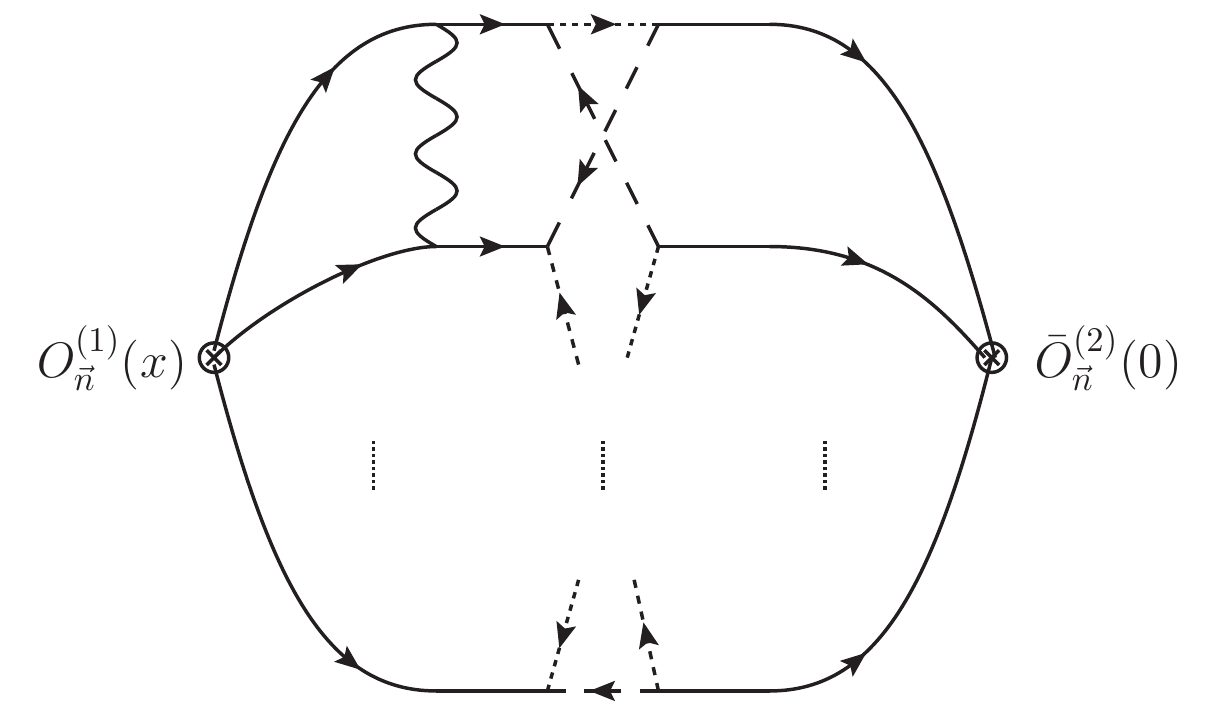}
        \subcaption{NLO order}\label{Fig:Q2NLO}
    \end{minipage}  

    \caption{Leading and Next-to-Leading Orders for $G_{\vec n}^{(q,1,2)}$. The NLO includes all the ways of correcting the legs of both nodes 1 and 2 with the vector multiplet}
\end{figure}

This diagram, given by a hypermultiplet loop with $n$ adjoint legs from the node 1, and $n$ from the node 2, produces an exact result for the space-time integral, and can be computed exactly:
\begin{align}
\label{Davydichev}
\parbox[c]{.15\textwidth}{\includegraphics[width = .15\textwidth]{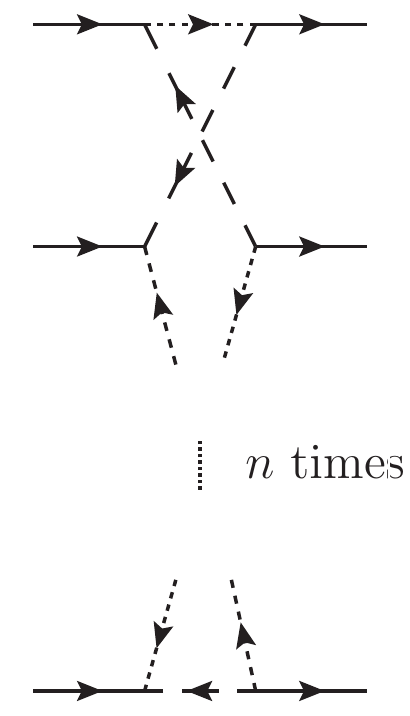}} \quad
=~ \Big(\frac{-1}{16\pi^2}\Big)^{n} \,\,\binom{2n}{n} \,\frac{{\color{red}\zeta}_{2n-1}}{n} \,\times \,
\Big(\frac{1}{4\pi^2 x^2}\Big)^n \equiv L(n)\frac{1}{x^{2n}}~.
\end{align}

 The result \eqref{Davydichev} is fully derived in Appendix \ref{App:superspace} (see also \cite{Billo:2017glv} and \cite{Beccaria:2020hgy}).
 
We insert this subdiagram inside Figure \ref{Fig:tree} with the proper multiplicity factor $n^2$ to keep it planar, obtaining the final diagram represented in Figure \ref{Fig:Q2LO}. Besides, the colour factor returns the correct leading power in $N$. The full result at leading order for generic $n$ and $q$ reads:
 \begin{equation}\label{LOdiag}
G_{\vec n}^{(q,1,2)}\big|_{\mathrm{LO}} = n^2 g_1^{2n} g_2^{2n} N^{2n} L(n) = n\frac{{\color{blue}\lambda}_1^n{\color{blue}\lambda}_2^n}
{(16\pi^2)^{2n}} \,\binom{2n}{n} \,{\color{red}\zeta}_{2n-1}~.
\end{equation}

We can also infer the diagrammatic explanation for the Next-to-Leading-Order with a similar reasoning. From the matrix model achievements we see that a unique set of Feynman diagrams can contribute at NLO, having the following structure:
\begin{align}
\label{DavydichevNLO}
\parbox[c]{.15\textwidth}{\includegraphics[width = .15\textwidth]{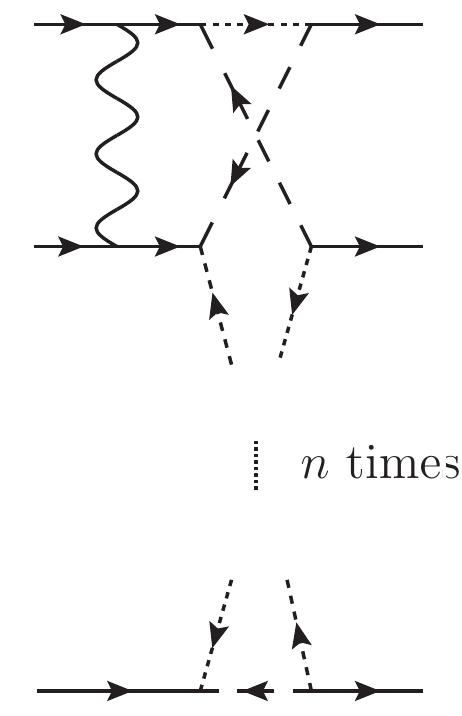}} \quad
=~ \Big(\frac{-1}{16\pi^2}\Big)^{n+1} \,n\,\binom{2n+2}{n+1} \,\frac{{\color{red}\zeta}_{2n+1}}{n} \,\times \,
\Big(\frac{1}{4\pi^2 x^2}\Big)^n \equiv N(n)\frac{1}{x^{2n}}~.
\end{align}
Notice the additional $n$ factor which counts the way of inserting the vector $V$ among the legs. Inserting the subdiagram \eqref{DavydichevNLO} inside Figure \ref{Fig:tree} we obtain the diagram of Figure \ref{Fig:Q2NLO}. Again there is the $n^2$ multiplicity factor, while the coupling dependence includes the $V$ insertion among the legs of both sides, node 1 and node 2. The final result at NLO reads:
 \begin{equation}\label{NLOdiag}
G_{\vec n}^{(q,1,2)}\big|_{\mathrm{NLO}} = n^2 g_1^{2n}g_2^{2n} (g_1^2+g_2^2)N^{n+1} N(n) = n^2\frac{{\color{blue}\lambda}_1^n{\color{blue}\lambda}_2^n ({\color{blue}\lambda}_1+{\color{blue}\lambda}_2)}{(16\pi^2)^{2n+1}}\,\binom{2n+2}{n+1} \,{\color{red}\zeta}_{2n+1}~.
\end{equation}
The sum of \eqref{LOdiag} with \eqref{NLOdiag} perfectly reproduces the matrix model results of \eqref{4.39}-\eqref{4.43}, and extends them to the general $n$ case.

\subsection{$A_{q-1}$ theories: operators  at $d \geq 2$}\label{sec:feynd}

\begin{figure}[!t]
    \begin{minipage}[t]{.48\textwidth}
        \centering
        \includegraphics[width=\textwidth]{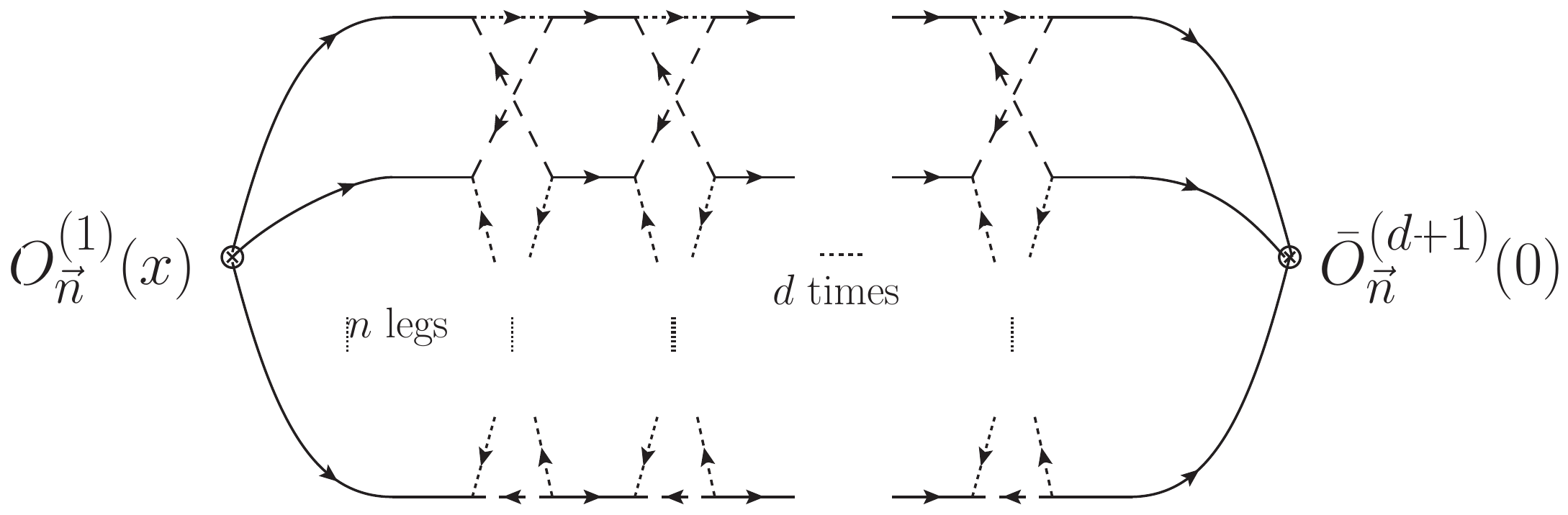}
        \subcaption{LO order}\label{Fig:QdLO}
    \end{minipage}
    \hfill
    \begin{minipage}[t]{.48\textwidth}
        \centering
        \includegraphics[width=\textwidth]{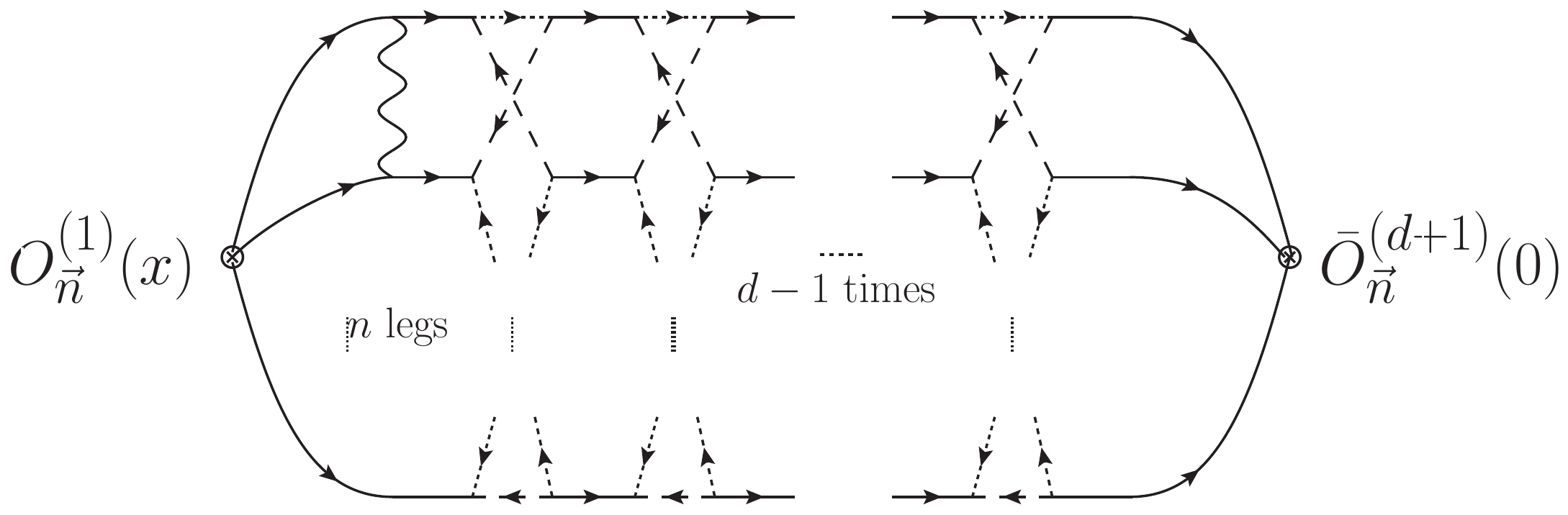}
        \subcaption{NLO order}\label{Fig:QdNLO}
    \end{minipage}  

    \caption{Leading and Next-to-Leading Orders for $G_{\vec n}^{(q,1,d+1)}$. The NLO includes all the ways of correcting all the adjoint legs with the vector multiplet}
\end{figure}

As seen in section \ref{subsec:maxdist}, when we increase the distance between the nodes where the operators are inserted, the leading order of each correlator arises at higher perturbative orders. At the diagrammatic level the general subdiagrams shown in \eqref{Davydichev} and \eqref{DavydichevNLO} represent some building blocks. First of all the leading order requires the insertion of multiple copies of the diagram \eqref{Davydichev}.
\begin{align}
\label{MultiDavydichev}
\parbox[c]{.35\textwidth}{\includegraphics[width = .35\textwidth]{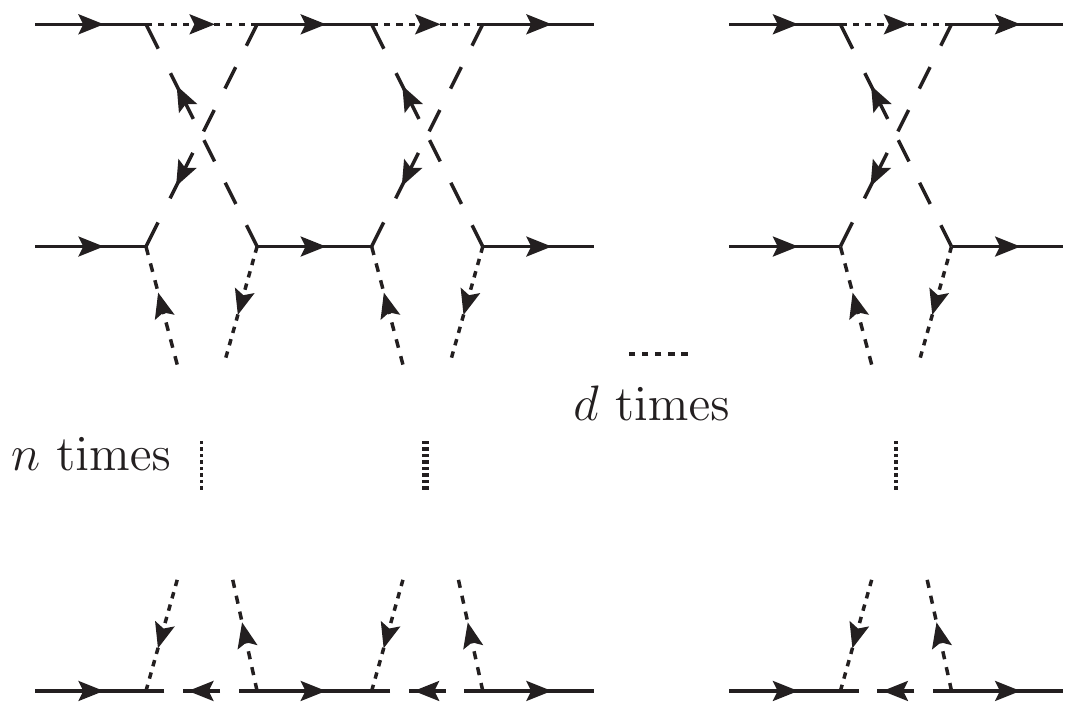}} \!= n^{d-1}   \left[\parenth{\frac{-1}{16\pi^2}}^{n}\binom{2n}{n}\frac{{\color{red}\zeta}_{2n-1}}{n}\right]^d\Big(\frac{1}{4\pi^2 x^2}\Big)^n \equiv L_d(n)\frac{1}{x^{2n}},
\end{align}
which  precisely explains the leading order of these correlators at a diagrammatic level:
\begin{equation}\label{LOdiagMULT}
G_{\vec n}^{(q,1,d+1)}\big|_{\mathrm{LO}} = n^2 {\color{blue}\lambda}_1^n \dots {\color{blue}\lambda}_{d+1}^n L_d(n) = n\frac{{\color{blue}\lambda}_1^n \dots {\color{blue}\lambda}_{d+1}^n}{(16\pi^2)^{n(d+1)}}\,\left[\binom{2n}{n} \,{\color{red}\zeta}_{2n-1}\right]^d~.
\end{equation}
The case of maximal distance (\textit{i.e.} $d=q/2$ for even $q$) is slightly more involved in the coupling dependence, since we must include the possibility of connecting the nodes $I_1$ with $J=q/2+1$ in both clockwise and anti-clockwise directions. With this remark, we see that \eqref{LOdiagMULT} perfectly reproduces the matrix model computations at leading order displayed in \eqref{4.46}-\eqref{4.50} for small values of $q$ and $n$.  

Another interesting comment about the pattern of equation \eqref{MultiDavydichev} is that this kind of Feynman diagram has the same structure of the fermionic wheel diagrams appearing in the integrable fishnet theories studied in \cite{Caetano:2016ydc,Kazakov:2018gcy,Pittelli:2019ceq,Levkovich-Maslyuk:2020rlp}. In particular the fermionic fishnet theory deriving from the double scaling limit of the superconformal $\mathcal{N}=2$ quivers admit the same diagrammatics for the computation of some specific 4-point functions \eqref{Pittelli:2019ceq,PittelliPreti2021}. In that case, the Yukawa vertices are purely vector multiplets contributions while in our present case, they are coming only from hypermultiplets contributions. Interestingly, performing  the $\gamma$-deformation of $\mathcal{N}=2$ but considering a different scaling limit, it is possible to select those hypers contributions. It would be interesting explore a little bit more this direction.

Moving to the Next-to-Leading-Order, it is hard to present a closed result like \eqref{LOdiag}, \eqref{NLOdiag} and \eqref{LOdiagMULT}, but we can provide a general idea.
The NLO can be obtained as a combination of the building blocks of the previous section, in particular a single diagram \eqref{DavydichevNLO} and $d-1$ of the form \eqref{Davydichev}.
\\[0.4cm]
\noindent\begin{minipage}{0.5\linewidth}
\includegraphics[width = .8\linewidth]{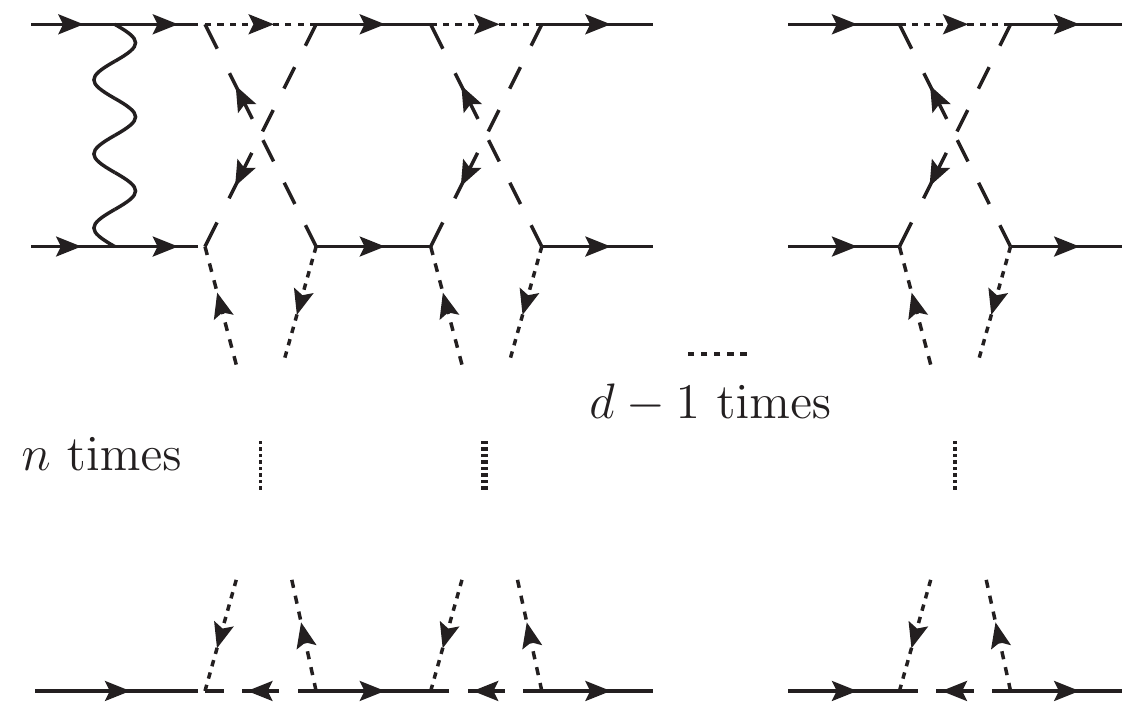}
\end{minipage}
\begin{minipage}{0.5\linewidth}
\begin{equation}\small\label{MultiDavydichevNLO}
\begin{split}
&\!\!\!\!\!\!\!\!\!\!\!\!\!\!\!\!\!\!\!\!\!\!\!\!\!\!\!\!\!\!=- \frac{n^{d} }{(16\pi^2)^{nd+1}} \binom{2(n+1)}{n+1}
\frac{{\color{red}\zeta}_{2n+1}}{(4\pi^2 x^2)^n}\left[\binom{2n}{n}\frac{{\color{red}\zeta}_{2n-1}}{n}\right]^{d-1}\\
&\!\!\!\!\!\!\!\!\!\!\!\!\!\!\!\!\!\!\!\!\!\!\!\!\!\!\!\!\!\!\equiv n^{d+1} N(n) L_{d-1}(n)\frac{1}{x^{2n}}
\end{split}
\end{equation}
\end{minipage}
\\[0.4cm]
Then the total result for $G_{\vec n}^{(q,1,d+1)}\big|_{\mathrm{NLO}}$ should account for the multiple combinations of insertions of the vector $V$, which generates a complicated coupling dependence. At least the idea is pretty similar to the previous cases: the spacetime factor arising from the integral is independent on the insertion of the vector $V$, and the full NLO correlator is a sum over all the possible insertion of $V$. Schematically:
\begin{align}
G_{\vec n}^{(q,1,d+1)}\big|_{\mathrm{NLO}} = \sum_{\mathrm{All~possible~insertions~of~V}}\parbox[c]{.45\textwidth}{\includegraphics[width = .45\textwidth]{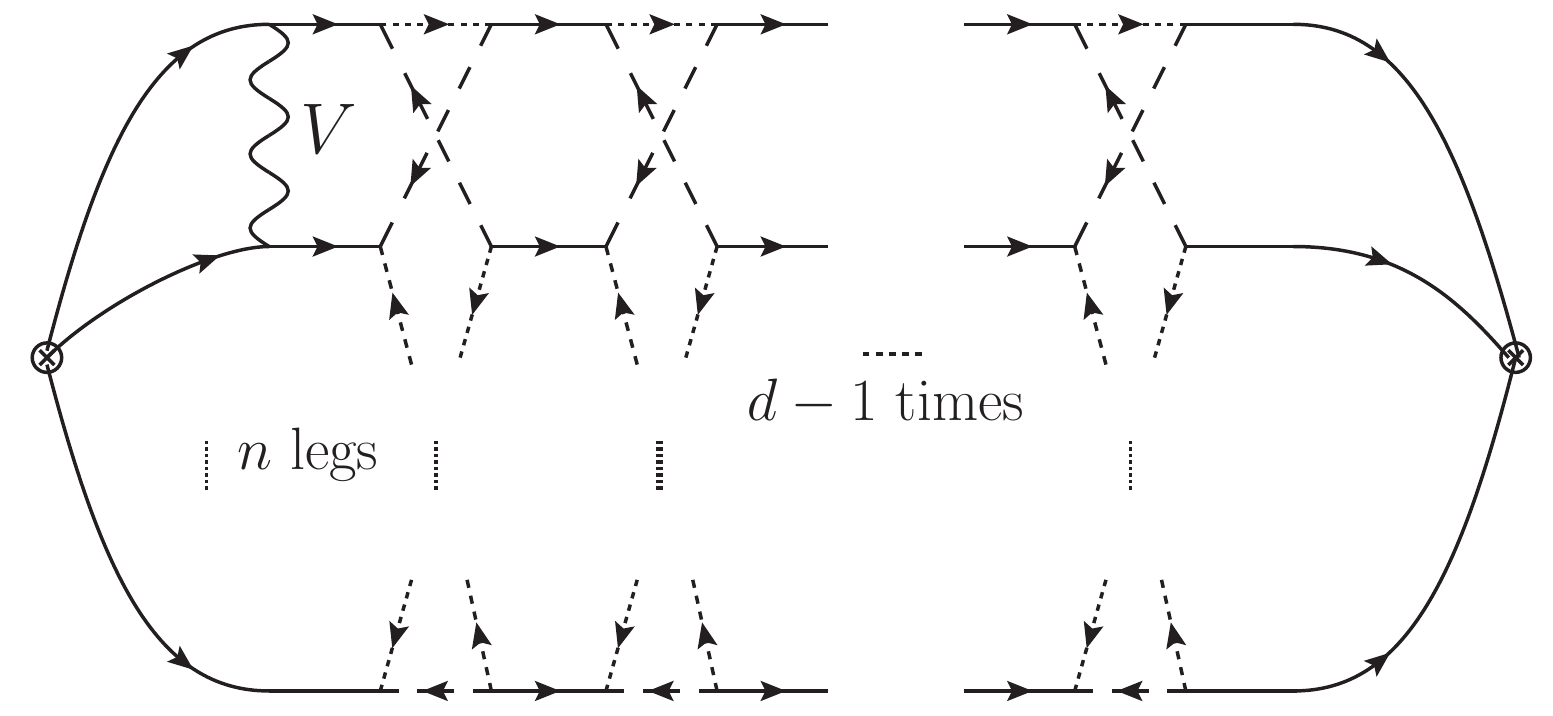}}.
\end{align}
Again the situation becomes more involved in the maximal distance case $d=q/2$ but the general idea is preserved.
The crucial point is that high degree of symmetry for these observables reduces the complexity of computing the Feynman diagrams to a combination of simple building blocks.

\vskip 1.5cm
\noindent {\large {\bf Acknowledgments}}
\vskip 0.2cm
\noindent
We really thank E. Pomoni for several contributions at the early stage of this project and for many useful comments during its realisation. 
We thank M. Bill\`o, A. Lerda and K. Zarembo for 
many useful discussions and suggestions.

\noindent
F.G. work is supported by a grant from the Swiss National Science Foundation, as well as via the NCCR SwissMAP.
The work of M.P. work is supported by the grant ”Exact Results in Gauge and String Theories” from the Knut and Alice Wallenberg foundation.
%
\vskip 1cm

\begin{appendix}

\section{Notations and conventions}
\label{App:Notations}

Chiral and anti-chiral spinor have $\alpha, \beta,\cdots$ and $\dot\alpha,\dot\beta,\cdots$ indices respectively and the they contract as follows \begin{equation}
\psi \,\chi =\psi^\alpha\chi_\alpha~,\qquad
\bar\psi \,\bar\chi=\bar\psi_{\dot{\alpha}}\bar\chi^{\dot{\alpha}}~.
\end{equation}
where we lower and raise indices with 
\begin{equation}
\psi^\alpha=\varepsilon^{\alpha\beta}\,\psi_\beta~,\qquad
\bar\psi_{\dot\alpha}= \varepsilon_{\dot\alpha\dot\beta}\,\bar\psi^{\dot\beta}~,
\end{equation}
where $\varepsilon^{12}=\varepsilon_{21}
=\varepsilon^{\dot 1\dot 2}=\varepsilon_{\dot 2\dot 1}=+1$.

\noindent
In Euclidean space, chiral and anti-chiral spinors satisfy the pseudoreality conditions
\begin{equation}
\psi_\alpha^\dagger = \psi^\alpha~,\qquad
\bar\psi_{\dot{\alpha}}^\dagger =\bar\psi^{\dot{\alpha}}~.
\label{hermitean}
\end{equation}
According to this, the chiral superspace coordinates $y^\mu=x^\mu+\ii\,\theta\sigma^\mu\bar\theta$
are invariant under this conjugation: $y^{\mu\,\dagger}=y^\mu$.
Similarly, the anti-chiral coordinates $\bar y^\mu=x^\mu-\ii\,\theta\sigma^\mu\bar\theta$
are such that $\bar y^{\mu\,\dagger}=\bar y^\mu$.
These coordinates satisfy 
\begin{equation}
\bar D_{\dot\alpha} y^\mu=0~,\qquad D_\alpha \bar y^\mu=0~,
\end{equation}
where $D_\alpha$ and $\bar D_{\dot\alpha}$ are the covariant spinor derivatives, defined as 
\begin{align}
	\label{covspinder}
		D_\alpha=\partial_\alpha+\ii\,(\sigma^\mu)_{\alpha\,\dot{\alpha}}\,
		\bar\theta^{\dot{\alpha}}\,\partial_\mu
		\quad\mbox{and}\quad
		\bar D_{\dot{\alpha}}=-\bar\partial_{\dot{\alpha}}-
		\ii\,\theta^\alpha\,(\sigma^\mu)_{\alpha\,\dot{\alpha}}\,\partial_\mu~.
\end{align}
The integration over Grassmann variables is defined such that
\begin{equation}
\int d^2\theta\,\theta^2 =1~,\qquad \int d^2\bar\theta\,\bar\theta^2 =1~.
\label{inteGr}
\end{equation}

\noindent
The $\sigma$-matrices  are defined by
\begin{equation}
\label{sigmas}
\sigma^\mu =
(\vec\tau,-\ii\mathbf{1})~,\qquad
\bar\sigma^\mu =
-\sigma_\mu^\dagger = (-\vec\tau,-\ii\mathbf{1})~,
\end{equation}
where $\vec\tau$ are the ordinary Pauli matrices and they satisfy the Clifford algebra
\begin{equation}
\label{cliff4}
\begin{aligned}
&\sigma_\mu\bar\sigma_\nu + \sigma_\nu\bar\sigma_\mu =
-2\delta_{\mu\nu}\,\mathbf{1}~.
\end{aligned}
\end{equation}
Using these matrices and the above rules, we can prove 
\begin{equation}\label{A7}
\psi\,\sigma^\mu\bar\psi\,\,\psi\,\sigma^\nu\bar\psi
\,=\,-\frac{1}{2}\,\psi\psi\,\bar\psi\bar\psi\,\delta^{\mu\nu}~.
\end{equation}
Besides, traces of multiple combinations of sigma matrices define a recursive relation:
\begin{equation}\begin{split}\label{tracesolve}
\Tr(\sigma^\mu\bar{\sigma}^\nu\sigma^\rho\bar{\sigma}^{\tau}\dots\sigma^\eta\bar{\sigma}^{\lambda})=\delta^{\mu\nu}&\Tr(\sigma^\rho\bar{\sigma}^{\tau}\dots\sigma^\eta\bar{\sigma}^{\lambda})-\delta^{\mu\rho}\Tr(\sigma^\nu\bar{\sigma}^{\tau}\dots\sigma^\eta\bar{\sigma}^{\lambda})\\
&+\delta^{\nu\rho}\Tr(\sigma^\mu\bar{\sigma}^{\tau}\dots\sigma^\eta\bar{\sigma}^{\lambda})-\epsilon^{\mu\nu\rho\kappa}\Tr(\sigma_\kappa\bar{\sigma}^{\tau}\dots\sigma^\eta\bar{\sigma}^{\lambda})~,
\end{split}\end{equation}
with
\begin{equation}\label{trace4sigma}
\begin{split}
\Tr(\sigma^\mu\bar{\sigma}^\nu\sigma^\rho\bar{\sigma}^{\tau})=2(\delta^{\mu\nu}\delta^{\rho\tau}-\delta^{\mu\rho}\delta^{\nu\tau}+\delta^{\mu\tau}\delta^{\nu\rho}-\epsilon^{\mu\nu\rho\tau})~.
\end{split}
\end{equation}

\section{Feynman rules}\label{App:Feynman}
We derive the Feynman rules from the actions \eqref{Sgauge1} and \eqref{Smatter1}, by expanding all the superfields in terms of the generators of the gauge group. In particular the fields belonging to the gauge part of the action transform in the adjoint representation of each $SU(N)_I$:
\begin{equation}
V_I = V_I^a (T_a)^u_{~v}~,~~~~~~~~~\Phi_I = \Phi_I^a (T_a)^u_{~v}~,
\end{equation}
 the fields of the matter part transform in the bifundamental of $SU(N)_I \times SU(N)_J$:
 \begin{equation}
 Q=Q^A(B_A)^u_{~\hat{v}}~,~~~~~~~~~\tilde{Q} = \tilde{Q}_A(B^A)^{\hat{u}}_{~v}~.
 \end{equation}
The indices $A$ are (anti-)bifundamental indices, $u,v,\hat u,\hat v = 1,\dots N$ are fundamental indices. These matrices obey the following relations:
\begin{align}\label{matrixRel}
\comm{T^a}{T^b} = \ii f_{abc}T^c~,~~~~~~ (T_a)^u_{~v}(T_a)^w_{~z} = \delta^i_z\delta^w_v-\frac{1}{N} \delta^u_v\delta^w_z~,~~~~~~(B_A)^u_{~\hat{v}}(B^A)^{\hat{u}}_{~v} = \delta^u_v \delta^{\hat u}_{\hat v}~.
\end{align}
Expanding \eqref{Sgauge1} and \eqref{Smatter1} the explicit expressions of the propagators and the vertices are derived.  
The explicit expressions for the propagators in the configuration space read:
\begin{align}\label{propagators}
		\Phi^\dagger\Phi-\mathrm{propagator}\parbox[c]{.2\textwidth}{\includegraphics[width = .2\textwidth]{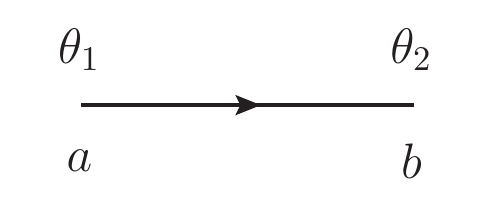}} &= \delta_{ab} ~\rme^{\parenth{-\theta_1 \sigma\,\bar{\theta}_1-\theta_2 \sigma\,\bar{\theta}_2+2\theta_1 \sigma\,\bar{\theta}_2}\cdot \ii \partial_{x_1}}\frac{1}{4\pi^2 x_{12}^2} ~, \notag \\
	VV-\mathrm{propagator}	\parbox[c]{.2\textwidth}{\includegraphics[width = .2\textwidth]{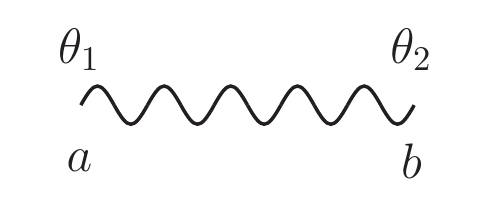}} &=-\frac{\delta_{ab}}{2}\frac{\theta_{12}^2\,\bar \theta_{12}^2}{4\pi^2 x_{12}^2}~, \notag \\
	Q^\dagger Q-\mathrm{propagator}	\parbox[c]{.2\textwidth}{\includegraphics[width = .2\textwidth]{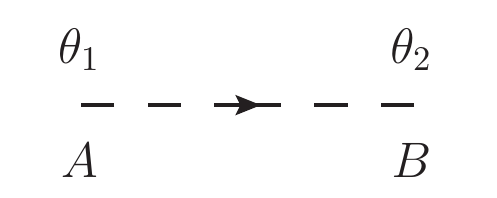}} &= \delta_{AB} ~\rme^{\parenth{-\theta_1 \sigma\,\bar{\theta}_1-\theta_2 \sigma\,\bar{\theta}_2+2\theta_1 \sigma\,\bar{\theta}_2}\cdot \ii \partial_{x_1}}\frac{1}{4\pi^2 x_{12}^2}~,  \notag \\
\tilde Q\tilde Q^\dagger-\mathrm{propagator}	\parbox[c]{.2\textwidth}{\includegraphics[width = .2\textwidth]{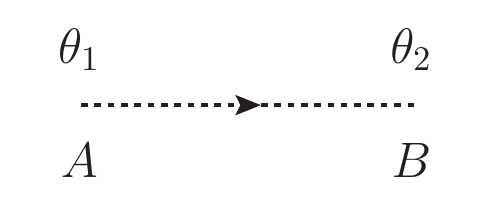}} &= \delta_{AB} ~\rme^{\parenth{-\theta_1 \sigma\,\bar{\theta}_1-\theta_2 \sigma\,\bar{\theta}_2+2\theta_1 \sigma\,\bar{\theta}_2}\cdot \ii \partial_{x_1}}\frac{1}{4\pi^2 x_{12}^2}~. 
\end{align}
where $\theta_{12}^2 = (\theta_1-\theta_2)^2$ and all the combinations in the exponents stand for 
\begin{equation}
\parenth{\theta^\a (\sigma^\mu)_{\a\dot\b} \bar{\theta}^{\dot{\b}}} \ii \frac{\partial}{\partial x_1^\mu}~.
\end{equation}
We write the vertices of the theory, only showing those we are going to use for present purposes.
\begingroup
\allowdisplaybreaks
\begin{align}\label{vertices}
V_I\,\Phi_I^\dagger\Phi_I-\mathrm{vertex}\parbox[c]{.2\textwidth}{\includegraphics[width = .2\textwidth]{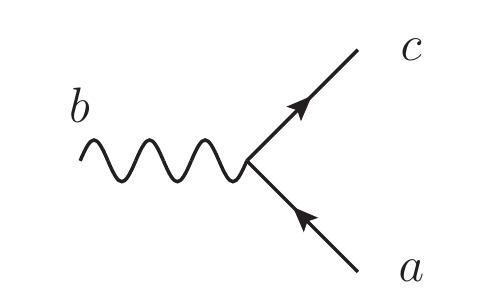}} &= 4 g_I \tr \parenth{T^a \comm{T^b}{T^c}}~,  \notag \\
V_I\,Q^\dagger Q-\mathrm{vertex}\parbox[c]{.2\textwidth}{\includegraphics[width = .2\textwidth]{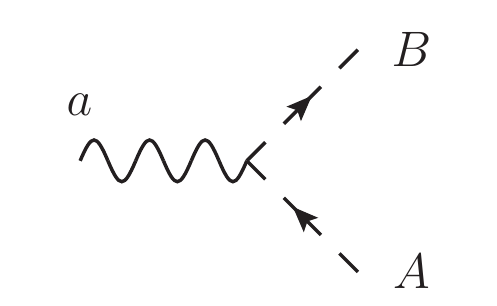}} &= 2 g_I \tr \parenth{T^a B^A B_B}~,  \notag \\
V_I\,\tilde Q\tilde Q^\dagger-\mathrm{vertex}\parbox[c]{.2\textwidth}{\includegraphics[width = .2\textwidth]{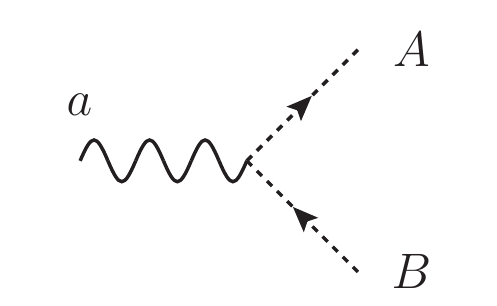}} &= -2 g_I \tr \parenth{T^a B_A B^B}~, \notag \\
\Phi_I\,Q\tilde Q-\mathrm{vertex}\parbox[c]{.2\textwidth}{\includegraphics[width = .2\textwidth]{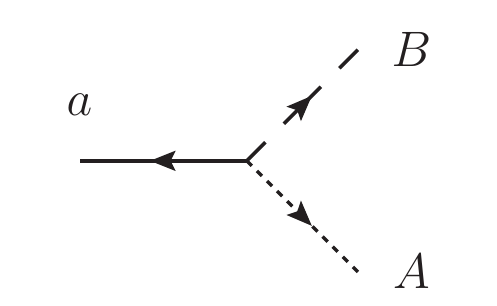}} &= \sqrt{2}\, \ii\, g_I  \tr \parenth{T^a B^A B_B} \bar \theta^2~, \notag \\
\Phi_I^\dagger\,Q^\dagger\tilde Q^\dagger-\mathrm{vertex}\parbox[c]{.2\textwidth}{\includegraphics[width = .2\textwidth]{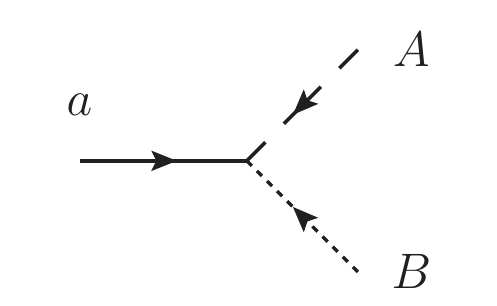}} &= -\sqrt{2}\,\ii\, g_I  \tr \parenth{T^a B^A B_B} \theta^2~.
\end{align}
\endgroup
Using these Feynman rules we can evaluate the chiral/antichiral correlators of $A_{q-1}$ theories at the perturbative level.

\section{Computation of color factors in $A_{q-1}$ theories}\label{App:2}
We compute the color factors of the two-loops correction in the $A_{q-1}$ theory. The notation can be simplified, in this section we use straight lines for adjoint fields, dotted lines for matter fields.\\
First of all we have the SCQCD contributions, where we have a single coupling $g_1 \equiv g$. The two diagrams to be analyzed are the two-loops propagator and the box diagram.

\paragraph{Two-loops propagator in SCQCD}.
The contribution from the vertices is simply $\ii \sqrt{2} g \times (-\ii \sqrt{2} g) \times 2g \times (-2g) = -8 g^4$. 
The color factor can be obtained using the method of the diagrammatic difference between $\cN=2$ and $\cN=4$  \cite{Billo:2017glv,Billo:2018oog,Billo:2019job,Billo:2019fbi}\footnote{See Chapter 2 of \cite{Galvagno:2020imh} for a thorough discussion about this method.}. The idea is to swap the computation of diagrams only involving fields of the vector multiplet with the computation of an ''improved`` color factor of diagrams involving matter fields. Such improved color factor consists in a combination $(\cN=2)-(\cN=4)$. \\ In this case it becomes:
\begin{equation}\label{B1}
\parbox[c]{.25\textwidth}{\includegraphics[width = .25\textwidth]{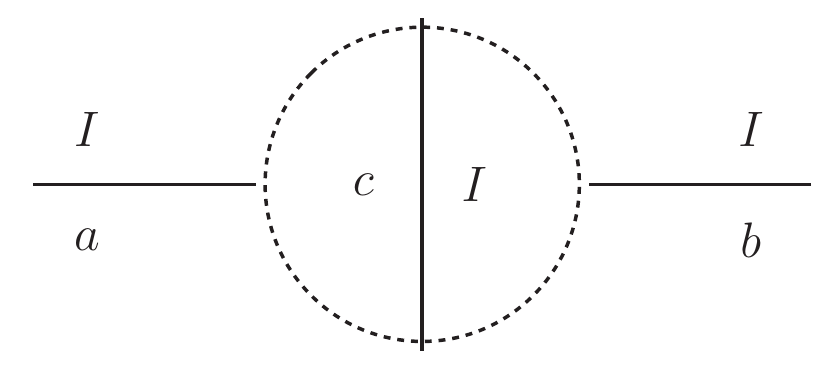}} =-8 g^4\parenth{2N \tr- \Tr_{\mathrm{Adj}}} T^a T^c T^b T^c~,
\end{equation}
which can be evaluated using the fusion/fission identities, and precisely returns:
\begin{equation}\label{B2}
C_2^{ab}(g,N) = 4 g^4 (N^2+1) \delta^{ab}~.
\end{equation}

\paragraph{Box diagram in SCQCD}
The color factor of the box diagram can be computed using the same technique.
\begin{align}\label{B3}
&\parbox[c]{.25\textwidth}{\includegraphics[width = .25\textwidth]{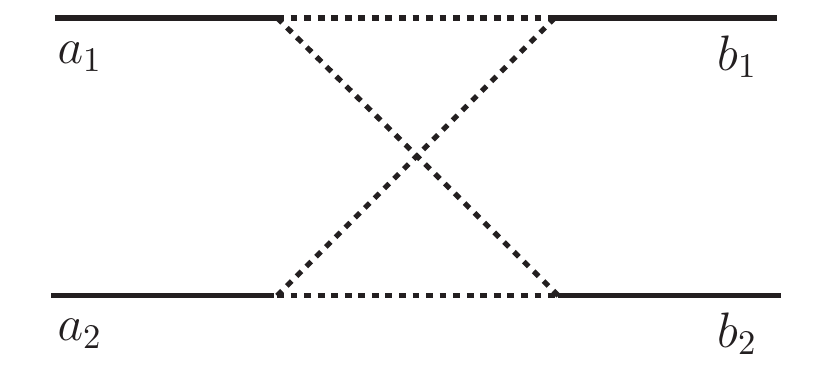}} =2g^4 \parenth{2N \tr - \Tr_{\mathrm{Adj}}} T^{a_1}T^{a_2}T^{b_1}T^{b_2}\notag \\&=-g^4 \parenth{\delta^{a_1b_1}\delta^{a_2b_2}+\delta^{a_1b_2}\delta^{a_2b_1}+\delta^{a_1a_2}\delta^{b_1b_2}}~.
\end{align}
Notice the same trace combination as \eqref{B1}, coming from the diagrammatic difference method explained above. This diagram must be contracted with the trace structure of chiral/antichiral operators. 

\paragraph{Additional color contributions in $A_{q-1}$ theory}

In $A_{q-1}$ theory the Feynman rules allow  additional possibilities for the two-loops propagator, we can insert a vector multiplet in the vertical line from the neighboring nodes ($J=2$ and $J=q$).
We explicitly compute such contribution using the Feynman rules above.
\begin{align}\label{B4}
&\parbox[c]{.25\textwidth}{\includegraphics[width = .25\textwidth]{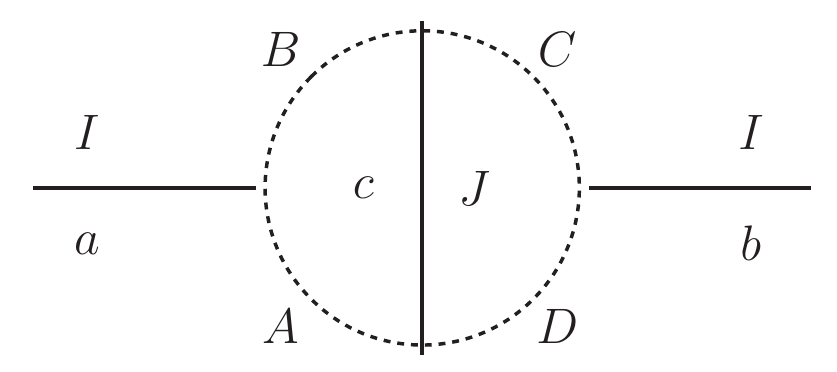}} = \ii \sqrt{2} g_I  (-\ii \sqrt{2} g_I) 2g_J (-2g_J) \times \notag \\
&~~~(T^a)^{u}_{~v}(B^A)^{v}_{~\hat v}(B^B)^{\hat v}_{~u}~(T^c)^{\hat w}_{~\hat y}(B^B)^{\hat y}_{~y}(B^C)^{y}_{~\hat w}~(T^b)^{m}_{~n}(B^C)^{m}_{~\hat n}(B^D)^{\hat n}_{~m}~(T^c)^{\hat o}_{~\hat p}(B^D)^{\hat p}_{~p}(B^C)^{p}_{~\hat o} \notag \\[0.2cm]
&\hspace{3.8cm} = -2 g_I^2 g_J^2 (N^2-1)\delta^{ab}~,
\end{align}
where we used \eqref{matrixRel} in the last step.
The total contribution for the two-loops color propagator in $A_{q-1}$ theory is given by the sum of the SCQCD contribution \eqref{B2} with a contribution \eqref{B4} for both the $J=2$ and $J=q$ cases:
\begin{align}\label{C5}
&\parbox[c]{.2\textwidth}{\includegraphics[width = .2\textwidth]{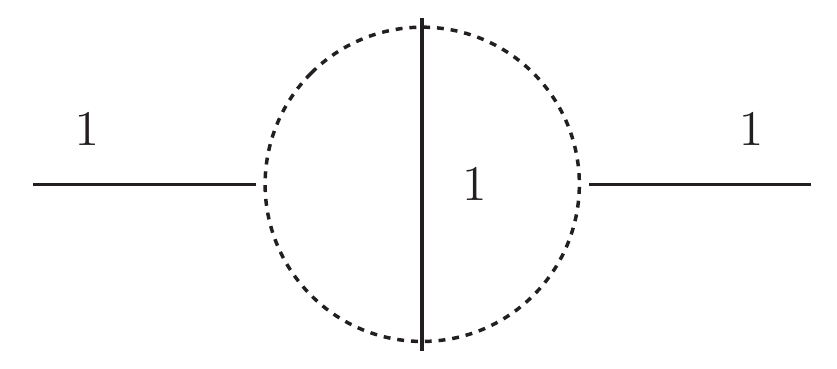}}+\parbox[c]{.2\textwidth}{\includegraphics[width = .2\textwidth]{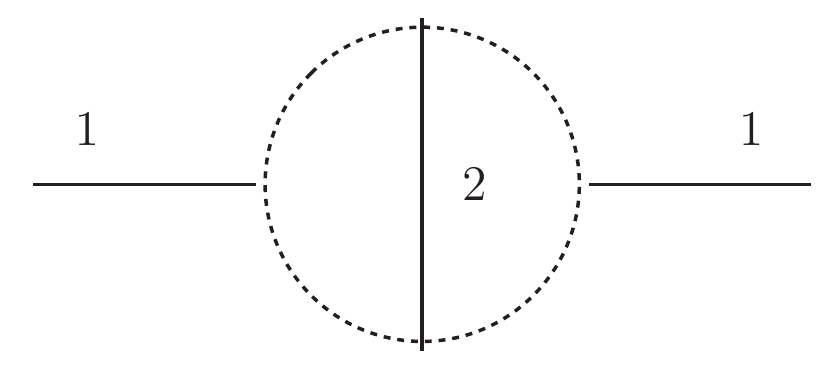}}+\parbox[c]{.2\textwidth}{\includegraphics[width = .2\textwidth]{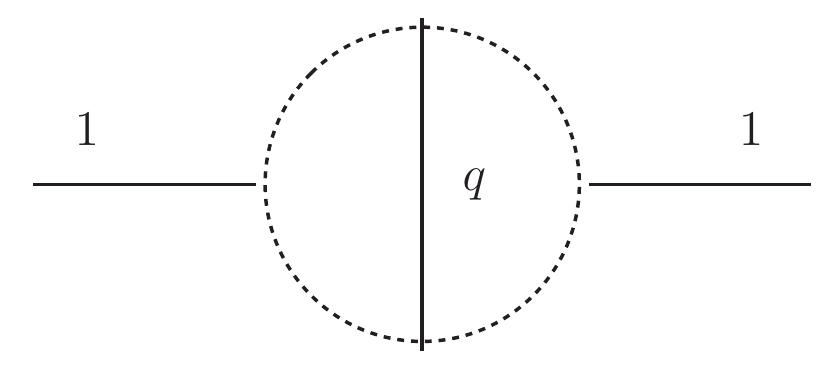}} = \notag \\
&~~~~~2g_1^2 \big[ 2g_1^2(N^2+1)-(g_2^2+g_q^2)(N^2-1)  \big]\delta^{ab}~.
\end{align}

\section{Superspace integrals}
\label{App:superspace}
\subsection{Exact ladder integrals}
\begin{figure}[!t]
 \begin{center}
\includegraphics[width=13cm]{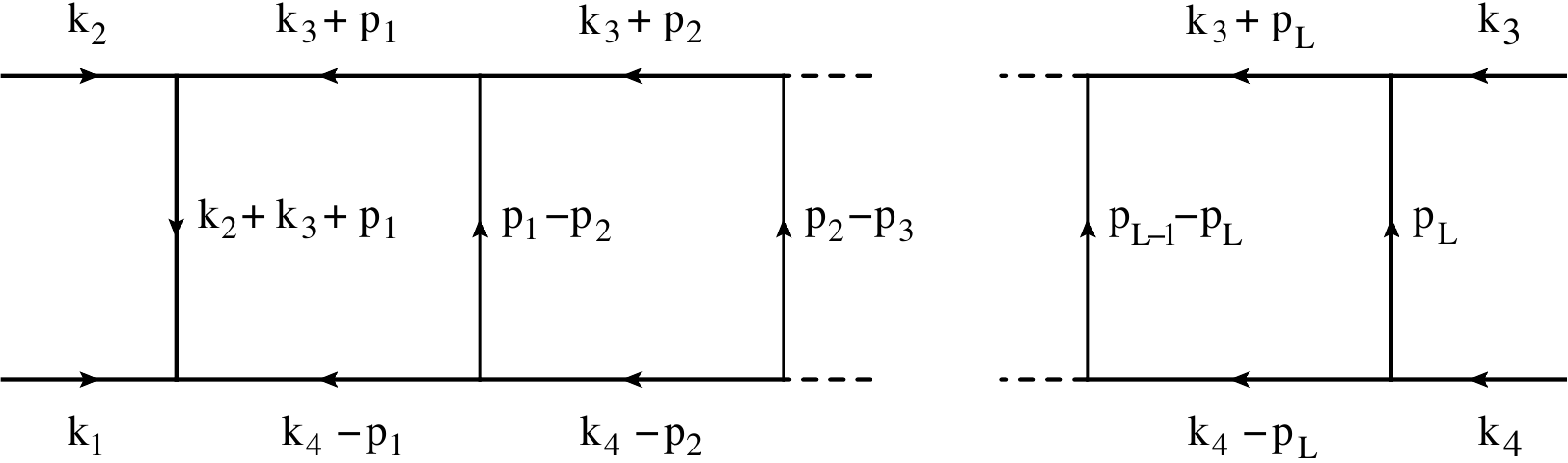}
\end{center}
\caption{Diagrammatic representation of the integral $\tilde{\mathcal{D}}^L$.  }
  \label{fig:ladders}
\end{figure}

We describe the conformal map which allows to exactly compute the space-time integral of the diagram \eqref{Davydichev}.
\begin{figure}[!t]
 \begin{center}
\includegraphics[width=6cm]{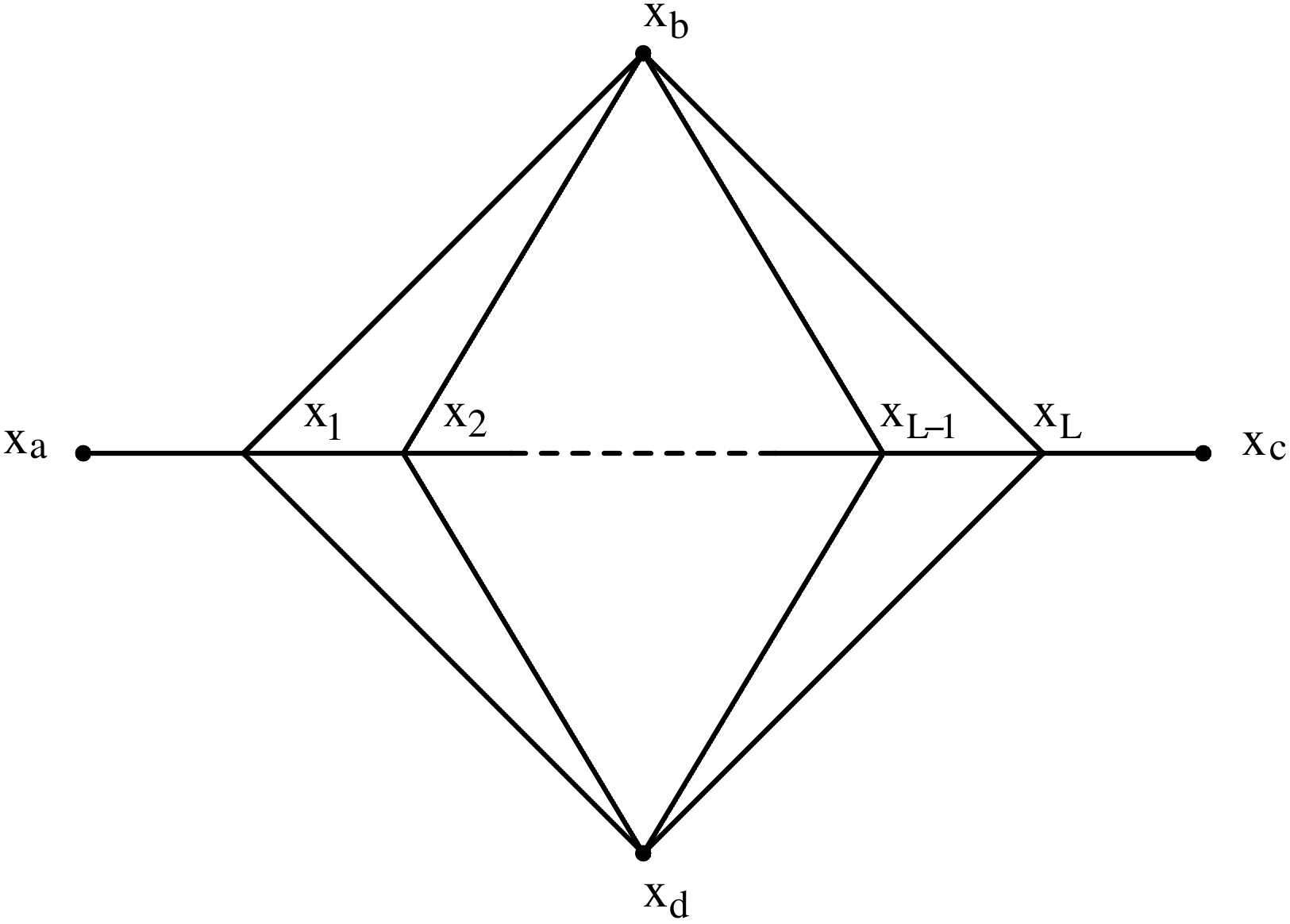}
\end{center}
  \caption{Diagrammatic representation of the integral $\mathcal{D}^L$.  }
  \label{fig:ladderx}
\end{figure}
We are interested in the following class of integrals
 \begin{equation}\label{Dx}
\mathcal{D}^{L}(x_a,x_b,x_c,x_d)=\int \frac{  d^4x_1\ldots   d^4x_L}{ x_{a1}^2\,x_{12}^2\, x_{23}^2\ldots x_{L-1,L}^2 \,x_{Lc}^2 \, {\prod_{i=1}^L} x_{ib}^2x_{id}^2\phantom{\Big|}}~,
\end{equation}
represented in Figure \ref{fig:ladderx}. 
Using the following map
\begin{figure}[H]
 \begin{center}
\includegraphics[width=6cm]{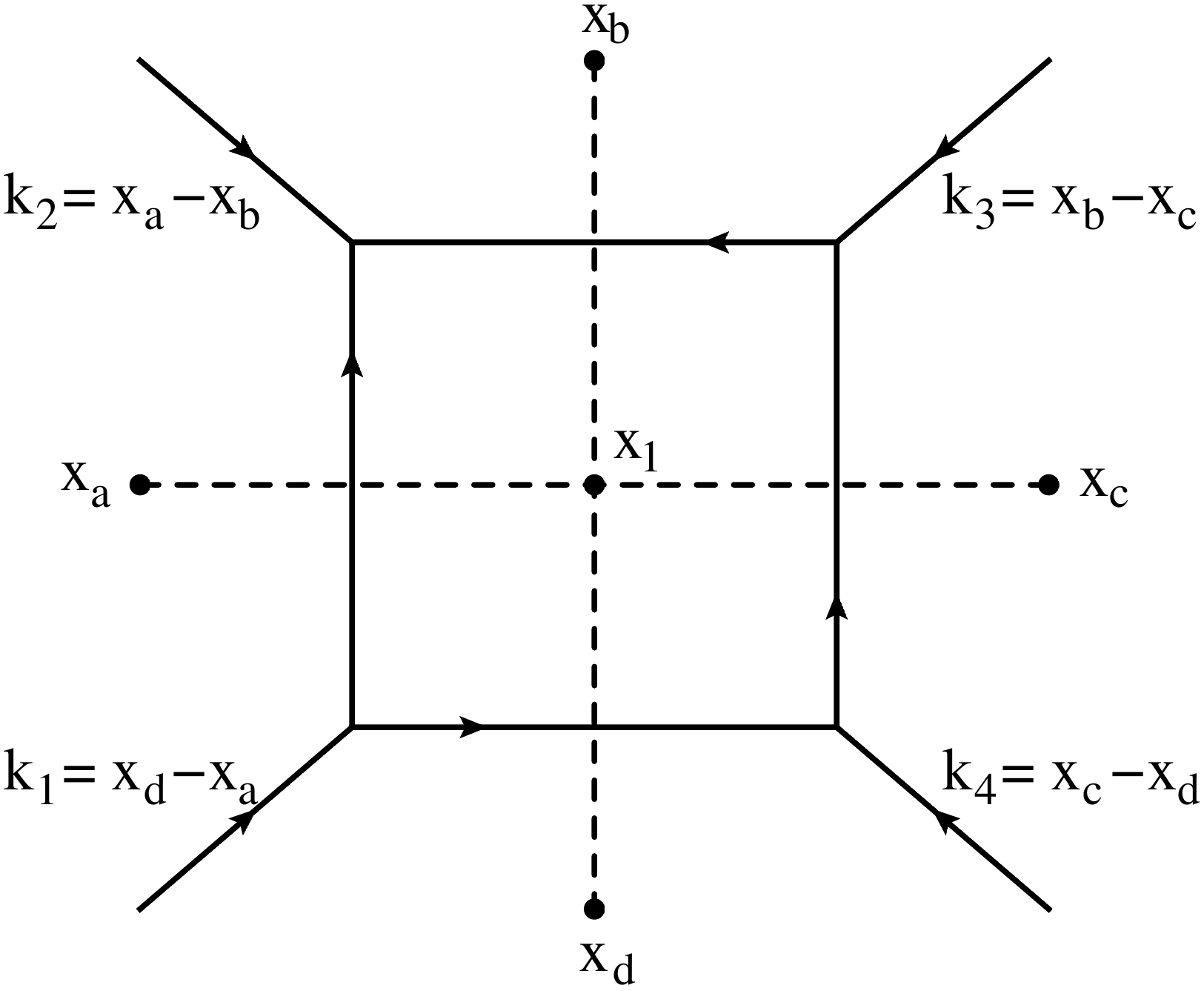}
\end{center}
  \label{fig:dual}
\end{figure}

\noindent
such that $\sum_{i}k_i=0$, we can map the integrals \eqref{Dx} in their momentum space representation.
The resulting integrals are the following
\begin{equation}\label{Dp}
\tilde{\mathcal{D}}^{L}(k_i^2;s,t)=\int \frac{  d^4p_1\ldots   d^4p_L}{ (k_2+k_3+p_1)^2\, (p_1-p_2)^2\ldots (p_{L-1}-p_L)^2\,p_L^2\, {\prod_{i=1}^L} (k_3+p_i)^2(k_4-p_i)^2 \phantom{\Big|}}~,
\end{equation}
where, as usual, $s=(k_1+k_2)^2$ and $t=(k_2+k_3)^2$. 
The integrals \eqref{Dp} belong to an infinite sequence computing the 
$L$-loop contribution of ladder diagrams to the four-point function of a $\phi^3$-theory (see Figure \ref{fig:ladders}).  
In \cite{Usyukina:1993ch}, the authors showed that the four-point function $\tilde{\mathcal{D}}^L$ can be written in terms of the corresponding three-point function which is known for arbitrary $L$. Then the integrals \eqref{Dp} can be expressed in the compact form as follows \footnote{This formula holds in Euclidean space. In Minkowski space-time the same formula holds after replacing $\pi^{2L}$ by $(\ii\pi^2)^L$.}
\begin{equation}\label{ccc}
\tilde{\mathcal{D}}^{L}(k_1^2,k_2^2,k_3^2,k_4^2;s,t)= \frac{\pi^{2L}}{s^L t } \,\Phi^{(L)}( u,v)~,
\end{equation}
where
\begin{equation}
u\equiv\frac{k_1^2 k_3^2}{st}~,\qquad\text{and}\qquad v\equiv\frac{k_2^2 k_4^2}{st}~.
\end{equation}
The dimensionless function $\Phi^{(L)}(u,v)$ is defined as
\begin{equation}
\begin{aligned}\label{PhiL}
\Phi^{(L)}(u,v) &= \sum_{j=L}^{2L} \frac{j! \log^{2L-j}(v/u)}{\lambda\,L! (j-L)!(2L-j)! }
 \big[\mathrm{Li}_j(-\rho\, u)+ (-1)^j\mathrm{Li}_j(-\rho\, v)\big] \\
&~+ 
\mathop{\sum_{k,l=0}^L}_{k+l=\mathrm{even}}
\frac{ 2(k+l)! (1-2^{1-k-l} ) }{ \lambda\,  k! \,  l! \, (L-k)! \, (L-l)!  }{\color{red}\zeta}_{k+l} \,
\log^{L-k}(\rho\, u) \log^{L-l}(\rho\, v) ~.
\end{aligned} 
\end{equation}
with
\begin{equation}
\lambda=\sqrt{(1-u-v)^2-4uv } \quad  \mbox{and} \quad \rho = \frac{2}{1-u-v+\lambda}~.
\end{equation}
We are interested to solve integrals \eqref{Dx} in the limit $x_c\to x_a$. This limit, in momentum space representation corresponds to $k_3\rightarrow-k_2$ and $k_4\rightarrow-k_1$ which is equivalent to sending $v\to u$ and then $u\to \infty$.
Using (\ref{PhiL}), the leading contribution in this limit is
 \begin{equation}
 \mathop{\lim_{v\to u}}_{u\to \infty} \Phi^{(L)}  \left(u,v \right)  
 =\frac {(2L)! }{L!^2} \, \frac{{\color{red}\zeta} (2L-1)}{u}~.
 \end{equation}
Then, inserting this in (\ref{ccc}) one obtains
 \begin{equation}\label{Dtilde}
 \tilde{\mathcal{D}}^{L}(k_1^2,k_2^2,k_2^2,k_1^2;s,t) =\frac{ \pi^{2L}}{s^{L-1}\,k_1^2\,k_2^2 } \, 
 \frac{(2L)! }{ L!^2} \, {\color{red}\zeta} (2L-1) ~.
 \end{equation}
The solution of integrals \eqref{Dx} can be read from \eqref{Dtilde} using the dual conformal map again, obtaining
 \begin{equation}\label{SolDx}
 \mathcal{D}^{L}(x_a,x_b,x_a,x_d) =\frac{ \pi^{2L}}{x_{db}^{2L-2}  \,  x_{ab}^2 \, x_{da}^2 } \, 
 \frac{(2L)! }{ L!^2} \, {\color{red}\zeta} (2L-1) ~.
 \end{equation}
 where
 \begin{equation}
 u=\frac{x_{da}^2x_{bc}^2}{x_{ac}^2x_{db}^2}\,,\qquad\text{and}\qquad v=\frac{x_{ab}^2x_{cd}^2}{x_{ac}^2x_{db}^2}~.
 \end{equation}
This integral is fundamental for the computation of the space-time integral $W_4$ and its generalization $W_{2L}$.
\subsection{Computation of $W_4$}
\begin{figure}[!h]
 \begin{center}
  \includegraphics[scale=0.6]{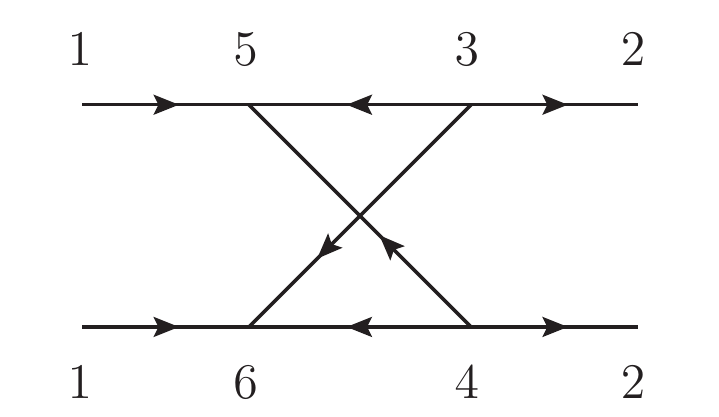}\hspace{1cm}\includegraphics[scale=0.6]{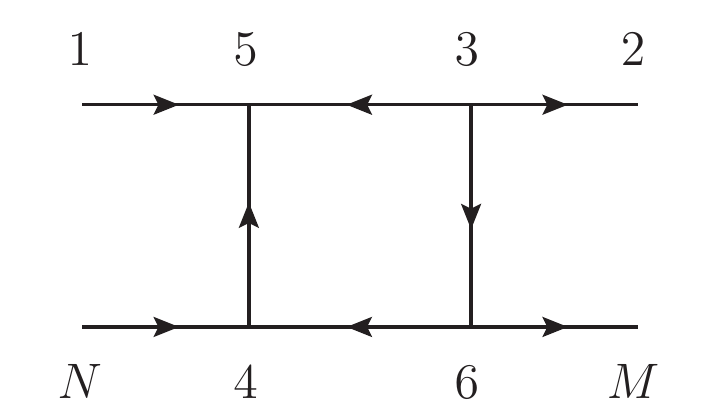}
\end{center}
  \caption{$W_4$ with its point-splitting representation. All the propagators come from chiral fields, so they can be simply represented by plain lines. }
  \label{Fig:W4}
\end{figure}
We represent the space-time integral $W_4$ as in Figure \ref{Fig:W4}. For this computation we only need the superspace part of the chiral propagators, which we shorten as:
\begin{equation}
\braket{ij}:=e^{\left(\xi_{ii}+\xi_{jj}-2\xi_{ij}\right)\cdot\partial_{x_i}}\frac{1}{4\pi^2x_{ij}^2}~,\quad \xi^{\mu}_{ij}:=\ii(\theta\sigma^{\mu}\bar\theta)~.
\end{equation}

The integral $W_4(x_{12})$ is finite but needs to be regularized. The point-splitting regularization is the most natural one (see Figure \ref{Fig:W4}). In this representation the integral can be written as follows
\begin{equation}\label{ID}
W_4(x_{12})\equiv\lim_{x_{M}\to x_1}\lim_{x_{N}\to x_2}I_2(x_1,x_{M},x_2,x_{N})~,
\end{equation}
where the integral $I_2$ is defined as
\begin{equation}
I_2(x_1,x_{M},x_2,x_{N})=\int\prod_{i=3}^6d^4x_id^4\theta_i\braket{15}\braket{M 6}\braket{45}\braket{46}\braket{35}\braket{36}\braket{4 N}\braket{32}\bar{\theta}_3^2\bar{\theta}_4^2\theta_5^2\theta_6^2~,
\end{equation}
where we followed the Feynman rules above (notice the $\theta^2,\bar \theta^2$ factors coming from the vertices).

Taking $\theta_1=\theta_2=0$ we get:
\begin{equation}
I_2=\!\!\int\prod_{i=3}^6d^4x_id^2\theta_3d^2\theta_4d^2\bar{\theta}_5d^2\bar{\theta}_6\frac{\left(e^{-2\xi_{45}\cdot\partial_4}\frac{1}{x_{45}^2}\right)\!\left(e^{-2\xi_{46}\cdot\partial_4}\frac{1}{x_{46}^2}\right)\!\left(e^{-2\xi_{35}\cdot\partial_3}\frac{1}{x_{35}^2}\right)\!\left(e^{-2\xi_{36}\cdot\partial_3}\frac{1}{x_{36}^2}\right)}{(4\pi^2)^8x_{15}^2x_{M 6}^2x_{32}^2x_{4 N}^2}~.
\end{equation}
We expand the exponentials, and using \eqref{psipsi}, \eqref{A5}, \eqref{A7}
we can integrate over $\theta$'s:
\begin{equation}
\begin{aligned}
I_2=&\frac{1}{(4\pi^2)^8}\int\prod_{i=3}^6d^4x_i\frac{1}{x_{15}^2x_{M 6}^2x_{32}^2x_{4 N}^2}\Bigg[\partial_4^2\left(\frac{1}{x_{46}^2x_{36}^2}\right)\partial_3^2\left(\frac{1}{x_{45}^2x_{35}^2}\right)+\partial_3^2\left(\frac{1}{x_{46}^2x_{36}^2}\right)\partial_4^2\left(\frac{1}{x_{45}^2x_{35}^2}\right)\\&-\Tr\left(\sigma_{\rho}\overline{\sigma}_{\mu}\sigma_{\nu}\overline{\sigma}_{\eta}\right)\partial_3^{\rho}\left(\frac{1}{x_{36}^2}\right)\partial_4^{\mu}\left(\frac{1}{x_{46}^2}\right)\partial_4^{\nu}\left(\frac{1}{x_{45}^2}\right)\partial_3^{\eta}\left(\frac{1}{x_{35}^2}\right)\Bigg]~.
\end{aligned}
\end{equation}
We now use the fermionic star-triangle relations  (see \cite{Preti:2018vog,Preti:2019rcq} for a recent review)
\begin{equation}
\sigma_{\mu}\overline{\sigma}_{\nu}\partial_i^{\mu}\partial_j^{\nu}\int\frac{d^4x_a}{x_{ia}^2x_{ja}^2x_{ka}^2}=-4\pi^2\sigma_{\mu}\overline{\sigma}_{\nu}\frac{x_{ik}^{\mu}x_{kj}^{\nu}}{x_{ik}^2x_{jk}^2x_{ij}^2}~,
\end{equation}
we use \eqref{trace4sigma}
and we performe the $x_5,x_6$ integrals. We finally obtain:
\begin{equation}\begin{split}
I_2(x_1,x_M,x_2,x_{N})=&\frac{1}{(4\pi^2)^6}\int d^4x_3d^4x_4\frac{x_{34}^2x_{1 M}^2+2\epsilon_{\rho\mu\nu\eta}x_{3 M}^{\rho}x_{M4}^{\mu}x_{14}^{\nu}x_{31}^{\eta}}{x_{34}^4x_{13}^2x_{M 3}^2x_{32}^2x_{41}^2x_{4 M}^2x_{4 N}^2}\\
=&\frac{x_{1M}^2}{(4\pi^2)^6}\int \frac{d^4x_3d^4x_4}{x_{34}^2x_{13}^2x_{M 3}^2x_{32}^2x_{41}^2x_{4M}^2x_{4N}^2}~,
\end{split}\end{equation}
where in the second line we used the $\epsilon_{\rho\mu\nu\eta}x_{3M}^{\rho}x_{M4}^{\mu}x_{14}^{\nu}x_{31}^{\eta}=0$.
The resulting integral is finite in the limit $x_N\to x_2$ but it is quadratically divergent in the limit $x_M\to x_1$. However, taking into account also the prefactor $x_{1M}^2$ in $I_2$, we find in the end a finite result. After taken the limit we exploit the map of the previous section, and using \eqref{ID}, \eqref{Dx} and \eqref{SolDx} we have: 
\begin{equation}
W_4(x_{12})=\lim_{x_M\to x_1}\frac{x_{1M}^2}{(4\pi^2)^6} \mathcal{D}(x_1,x_M,x_2,x_2)
=\frac{6 {\color{red}\zeta}_3}{2^{12}\pi^8}\frac{1}{x_{12}^4}~.
\end{equation}

\subsection{Computation of $W_{2L}$}
We are now ready to generalize to the $W_{2L}(x_{12})$ case.\\
 Again we perform a point-splitting regularization as in the previous case. Here we have a diagram like Figure \ref{Fig:W4} with $2L$ external legs. In this representation the integral can be written as follows
\begin{equation}\label{PSL}
W_{2L}(x_{12})\equiv\lim_{x_{M_i}\to x_1}\lim_{x_{N_i}\rightarrow x_2}I_L(x_1,\{x_{M_i}\},x_2,\{x_{N_i}\})~,
\end{equation}
where $i=1,2,\dots,L$ and the integral $I_L$ is defined as
\begin{equation}\begin{split}
I_L(x_1,\{x_{M_i}\},x_2,\{x_{N_i}\})\!=\!\!\int\prod_{i=3}^{2L+2} \!\!d^4x_i d^4\theta_i \!&\prod_{j=1}^{L}\bar{\theta}_{2+j}\braket{2\!+j,N_{j-1}}\theta_{2+L+j}\braket{M_{j-1},2\!+L\!+j}\\
&\times\braket{2+j,2L+3-j}\braket{3+j|_{L},2L+3-j}~,
\end{split}\end{equation}
where $x_{M_0}=x_1$ and $x_{N_0}=x_2$ and where $j|_{L}$ stands for $j$ modulo $L$ such that $L|_{L}=0$.
The brackets $\{x_{M_i}\}$ represent the sets of points $\{x_{M_1},\dots,x_{M_{L-1}}\}$ (and the same for $\{x_{N_i}\}$).
We then follow the same steps of the previous section, with the remark that combinations of multiple sigma matrices can be solved using the recursive formula \eqref{tracesolve}. We finally get:
\begin{equation}\begin{split}
W_{2L}(x_{12})&=\lim_{x_{M_1}\to x_1}\frac{(-1)^L x^2_{1M_1}}{(4\pi^2)^{3L}}\mathcal{D}^L(x_{M_1},x_1,x_{M_1},x_2)\\
&=\parenth{\frac{-1}{16\pi^2}}^L\binom{2L}{L}  {\color{red}\zeta}_{2L-1} \frac{1}{(4\pi x_{12})^{2L}}~.
\end{split}
\end{equation}

\end{appendix}

 \bibliographystyle{nb}

\bibliography{biblio}

\end{document}